\documentclass[%
 reprint,
 amsmath,amssymb,
 aps,
]{revtex4-2}

\usepackage[pdftex]{graphicx}

\graphicspath{
{./Figure/},
{./PhaseDiagrams/},
{./DTSPs/},
{./ADTSPs/}
}

\usepackage{diagbox}
\usepackage{braket}
\usepackage{dcolumn}
\usepackage{bm}
\usepackage{booktabs}
\usepackage{threeparttable}

\begin{document}

\preprint{APS/123-QED}

\title{Densest ternary sphere packings}

\author{Ryotaro Koshoji}
\email{cosaji@issp.u-tokyo.ac.jp}
\affiliation{Institute for Solid State Physics, The University of Tokyo, Kashiwa 277-8581, Japan}
\author{Taisuke Ozaki}%
 \email{t-ozaki@issp.u-tokyo.ac.jp}
\affiliation{Institute for Solid State Physics, The University of Tokyo, Kashiwa 277-8581, Japan}

\date{\today}

\date{\today}

\begin{abstract}

We present our exhaustive exploration of the densest ternary sphere packings (DTSPs) for 45 radius ratios and 237 kinds of compositions,
which is a packing problem of three kinds of hard spheres with different radii, under periodic boundary conditions by a random 
structure searching method. 
To efficiently explore DTSPs we further develop the searching method based on the piling-up and iterative 
balance methods [Koshoji \textit{et al.}, Phys. Rev. E 103, 023307 (2021)]. 
The unbiased exploration identifies diverse 38 putative DTSPs appearing on phase diagrams in which 37 DTSPs of them are discovered 
in the study.
The structural trend of DTSPs changes depending especially on the radius of small spheres. 
In case that the radius of small spheres is relatively small, structures of many DTSPs can be understood as derivatives of densest binary 
sphere packings (DBSPs), while characteristic structures specific to the ternary system emerge as the radius of small spheres 
becomes larger. In addition to DTSPs, we reveal a lot of semi-DTSPs (SDTSPs) which are obtained by excluding DBSPs in the calculation of 
phase diagram, and investigate the correspondence of DTSPs and SDTSPs with real crystals based on the space group, showing a considerable correspondence of SDTSPs having high symmetries with real crystals including $\mathrm{Cu}_2 \mathrm{GaSr}$ and $\mathrm{ThCr}_2 \mathrm{Si}_2$ structures.
Our study suggests that the diverse structures of DBSPs, DTSPs, and SDTSPs can be effectively used as structural prototypes 
for searching complex crystal structures. 

\end{abstract}

\maketitle

\section{INTRODUCTION}

Identifying the densest sphere packings (DSPs) is one of the most difficult mathematical problems. It was proved only in the 2000s that the densest unary sphere packing (DUSPs) is the Barlow packings~\cite{10.2307/20159940}. The densest binary sphere packings (DBSPs) for given compositions are not determined in a mathematically rigorous way, however, there have been several studies that estimate the DBSPs by numerical calculations~\cite{doi:10.1021/jp804953r, PhysRevE.79.046714, doi:10.1021/jp206115p, doi:10.1021/jp1045639, Hudson_2011, doi:10.1063/1.5052478, de_laat_de_oliveira_filho_vallentin_2014}. In 2012, Hopkins \textit{et al.} explored the DBSPs under the restriction that the number of spheres in the unit cell with periodic boundary condition is less than or equal to 12 and constructed the phase diagram for the first time~\cite{PhysRevLett.107.125501, PhysRevE.85.021130}. Following the seminal study, in 2021, Koshoji \textit{et al.} revisited their phase diagram, accordingly identified 12 unknown putative DBSPs in addition to confirming known 
DBSPs with a small correction, and presented an updated phase diagram~\cite{PhysRevE.103.023307}. At the present time, totally the 28 putative DBSPs are known.

The correspondences of the DBSPs with crystals are also discussed in Ref.~\cite{PhysRevE.103.023307}. They showed that a considerable number of crystals can be understood as DBSPs. For example, the crystal structure of $\mathrm{LaH_{10}}$, which is a superhydride material synthesized under high pressure~\cite{doi:10.1002/anie.201709970}, can be understood as the $\mathrm{XY_{10}}$ structure~\cite{PhysRevE.85.021130,PhysRevE.103.023307}, and the crystal structure of $\mathrm{UB}_4$ corresponds to the (16-4) structure~\cite{PhysRevE.103.023307}. Although some DBSPs seem not be associated with crystal structures in crystal database such as Inorganic Crystal Structure Database (ICSD)~\cite{ICSD}, they might be used as structural prototypes in structural exploration for novel materials, especially for high-pressure phases. They may also be used as structural models for other physical systems such as colloids~\cite{doi:10.1021/acsnano.9b04274, doi:10.1021/acs.chemrev.6b00196} and glasses~\cite{doi:10.1063/1.4769981}.

On the other hand, the phase diagram for the densest ternary sphere packings (DTSPs) has not yet been constructed, while some studies researched the ternary system within limited conditions~\cite{WONG2014357,doi:10.1021/ie200765h,YI2012129,ROQUIER2019343}. Wang \textit{et al.}~\cite{doi:10.1063/1.4941262} discussed the two periodic ternary dense structures: FCCTC and HCPTC. The FCCTC (HCPTC) comprises the FCC (HCP) dense structures of large spheres with one small and one medium spheres placed per tetrahedral and octahedral site, respectively, while they did not show whether those structures appear on the phase diagram or not. Although a given composition of three kinds of spheres can be achieved by a non-unique phase separation that is comprised by some of known periodic structures, the previous studies proved that there is at least one phase separation with the densest packing fraction that consists of less than or equal to three structures if every candidate structure is periodic~\cite{PhysRevE.85.021130, PhysRevE.103.023307}. Note that the three densest FCC structures consisting of small, medium, and large spheres belong to the candidate structures for the densest phase separations, and additionally, the DBSPs consisting of small and medium, small and large, and medium and large spheres also belong to the candidate structures for the densest phase separations. Thus, the construction of the ternary phase diagrams requires the knowledge of the DBSPs as well. Moreover, the ternary system has a much larger number of composition and radius ratios than the binary system. The difficulties are what make it very hard to explore the DTSPs.

Despite the dramatic advances in computers, the exhaustive search of the densest packings at fixed composition and radius ratios is still challenging due to a large number of local minima in the energy or enthalpy surface. To infer the considerable increase of the computational complexity, let us consider a simple model that three kinds of spheres are placed at the nodes of the $4 \times 4 \times 4$ grid of a cubic cell. The combinatorial number $C$ of occupations by spheres for the grids is given by
\begin{equation}
C = \frac{64!}{N_A ! \, N_B ! \, N_C ! \, \left(64 - N_A - N_B - N_C \right)!},
\end{equation}
where the number of small, medium, and large spheres are $N_A$, $N_B$, and $N_C$, respectively. When $N_A = N_B = N_C = 1$, $C = 249984$, while when $N_A = N_B = N_C = 4$, $C \simeq 1.138 \times 10^{17}$. The non-deterministic polynomial-time (NP) hardness indicates that an exhaustive search of DTSPs is getting much more difficult along with an increase in the number of spheres per unit cell.

The explosive increase makes it difficult to apply a random structure searching method, however, the unbiased method is powerful in search of the DTSPs, since we have no a priori knowledge about what kind of structures are the DTSPs. To overcome the difficulty in part, Koshoji \textit{et al.} developed an efficient random structure searching method based on two approaches: the \textit{piling-up method}, which randomly generates appropriate initial structures, and the \textit{iterative balance method}, which efficiently optimizes the initial structure~\cite{PhysRevE.103.023307}. With the two methods, 12 unknown putative DBSPs were discovered successfully, resulting in that 28 putative DBSPs have been known in total. It is reasonable to assume that the random structure searching method can be successfully applied to the exploration of DTSPs.

In this study, to explore the DTSPs, we furthermore improve the piling-up method and the structural optimization scheme. The improved piling-up method directly generates multilayered initial structures being close to packing structures by properly choosing the number of 
spheres on each layer with the binomial or uniform distribution while avoiding a large overlap between spheres. 
Since we try to unbiasedly distribute the initial structures in the configuration space for a given composition and radius ratios 
as much as possible, it is expected that the exhaustive exploration can reach the DSP structure if the number of the trials is 
large enough.
Besides, to reduce the computational cost, the optimization scheme is also improved so as to reject sparse structures before the fine iterative balance optimization which optimizes structures with minimizing the volume of the unit cell. The improved methods to explore the DSPs are implemented in our open-source program package \textbf{SAMLAI} (Structure search Alchemy for MateriaL Artificial Invention), 
which is available online~\cite{samlai}. Then, with \textbf{SAMLAI}, we exhaustively explore DTSPs under periodic boundary conditions at 45 kinds of radius ratios and 237 kinds of compositions, and present ternary phase diagrams with discovering of 37 putative DTSPs. We also discuss the characteristic features in structures of the discovered DTSPs. When the radius of the small spheres is relatively 
small, many of DTSPs can be understood as derivative structures of DBSPs, where DTSPs are formed by inserting small spheres into voids of DBSPs. As the radii of small and medium spheres are getting larger, novel DTSP structures, which cannot be understood as derivatives of DBSPs, emerge on the phase diagram. A typical example is the (13-2-1) structure appearing at a radius ratio of $0.45:0.65:1.00$. 
If the slight structural distortion is corrected, 
the (13-2-1) structure with the $Fm\bar{3}m$ symmetry has a well-ordered clathrate structure, where 
large spheres constitute the FCC structure with no contact, a medium sphere is placed in an expanded tetrahedral site, and medium and large spheres 
are surrounded by small spheres constituting semi-regular polyhedrons.
In addition to the DTSPs, we discuss the semi-DTSPs (SDTSPs) that are denser than any phase separation consisting of only the FCC structures and/or the other ternary packings, excluding the DBSPs when the densest phase separation is calculated.
As a result, we discover 65 SDTSPs, and find that some of SDTSPs are highly well-ordered. 
Furthermore, we investigate the correspondence of the DTSPs with real crystals based on the space group. The survey shows
that among 38 DTSPs only the FCCTC can be associated with a real crystal structure: $\mathrm{AlCu}_2 \mathrm{Mn}$ structure, 
while among 65 SDTSPs five SDTSPs can be related to crystal structures such as 
$\mathrm{Cu}_2 \mathrm{GaSr}$ and $\mathrm{ThCr}_2 \mathrm{Si}_2$ structures. 
Although the correspondence of the DTSPs and SDTSPs with real crystals seems to be exceptional, by considering the considerable
correspondence between the DBSPs and crystal structures~\cite{PhysRevE.103.023307}, we expect that 
the DTSPs and SDTSPs might be used effectively as structural prototypes for searching complex crystal structures, especially
for high-pressure phases.

The paper is organized as follows: Sec.~\ref{sec:method_and_implementation} describes our improved methodology to explore the DTSPs; Sec.~\ref{sec:exploration_conditions} details the conditions for the exhaustive exploration of DTSPs; Sec.~\ref{sec:results} presents the phase diagrams of the ternary systems and the discovered 37 putative DTSPs; Sec.~\ref{sec:discussion} discusses the geometric features of the DTSPs, correspondence between the DTSPs and crystals, the discovered SDTSPs, and the efficiency of our method and geometric features of DTSPs. In Sec.~\ref{sec:conclusion}, we summarize this study.

\section{method and implementation}
\label{sec:method_and_implementation}

Koshoji \textit{et al.} developed an efficient algorithm for exploring DSPs, which consists of two methods: The piling-up method and the iterative balance method~\cite{PhysRevE.103.023307}. The former randomly generates initial structures, and the latter optimizes initial structures to packing structures with minimizing the volumes of the unit cells. The algorithm successfully discovered 12 DBSPs~\cite{PhysRevE.103.023307}, but there are still rooms for improvement in two respects. First, generated initial structures are far from packing structures. An initial structure is generated by stacking spheres randomly one by one on top of a randomly generated first layer~\cite{PhysRevE.103.023307}, thereby the generated structure tends to be rather an unnatural structure such as a tower-like structure with a large overlap on the first layer. Although the initial structure can be transformed to a packing structure by expansion and pseudo-annealing steps, the displacements of the fractional coordinates and lattice vectors are too large to assure that some of the initial structures can reach the DSPs. Second, the optimization scheme has room for improvement. As discussed in Ref.~\cite{PhysRevE.103.023307}, when the number of spheres per unit cell is larger, most of the explored structures are sparse compared to the densest structure. The inefficiency can be eliminated by selecting promising structures.

In this section, we discuss the improved algorithms for exploring DSPs.
All the methods presented here are implemented in our open-source program package \textbf{SAMLAI}~\cite{PhysRevE.103.023307,samlai}.

\subsection{Improved piling-up method}
\label{sec:improved_piling-up_method}

As discussed in Ref.~\cite{PhysRevE.103.023307}, any periodic structure can be understood as a multilayered structure if it is expanded in a direction perpendicular to a chosen base plane. In the improved piling-up method, the multilayered structure is generated directly.

First, we randomly choose the number of spheres placed on the first layer. In our methodology, there are two different ways to randomly choose the number, and which way to use is selected randomly per structure. One of the two ways is based on the binomial distribution, and the other is based on the uniform distribution. 
By default, the binomial distribution is used by 70 percent of the trials, since the number of spheres in the first layer of DSPs 
seems to follow broadly the binomial distribution from observation for the structures of DBSPs.
However, first layers having a small or large number of spheres should also be generated with high possibility because some of DSPs may have such kinds of unit cells. In fact, as discussed in Ref.~\cite{PhysRevE.103.023307}, the DBSPs are diverse, for example, the (7-3) structure has a tower-like unit cell, while the (16-4) structure has a thin unit cell. Since the uniform distribution can choose a small or large number with high probability, the combination of the binomial and uniform distributions can realize a suitable distribution 
of the number of spheres on the first layer for the exploration of DSPs.

Second, we randomly choose spheres that are placed on the first layer. Fractional coordinates of the spheres on the first layer are randomly set as $\left(r, r^{\prime}, 0 \right)$, where $r$ and $r^{\prime}$ are random values between $0$ and $1$.

Third, all of the remaining spheres are placed on layers that are generated above the first layer. The step is executed as follows: If there are remaining spheres, one layer is generated above the top layer. The number of spheres placed on the generated layer is randomly chosen by only the uniform distribution, where the maximum number is set to be the number of the remaining spheres unless 
that of the remaining spheres exceeds that of the first layer. In the latter case, the maximum number is set to be 
the number of spheres on the first layer instead. 
The treatment guarantees generation of diverse initial structures including cubic structure, 
tower-like structure, and flattened structure with modest overlaps of spheres.
Note that we cannot determine the fractional coordinates for $z$ components in this step because we do not set the lattice vectors yet.
Therefore, the provisional fractional coordinates of the spheres are randomly set as $\left(r, r^{\prime}, 0 \right)$, where $r$ and $r^{\prime}$ are random values between $0$ and $1$. 
The step for arranging spheres is repeated until the number of remaining spheres becomes zero.

Fourth, we randomly set the two lattice vectors $\bm{a}_1$ and $\bm{a}_2$ that are spanned on the first layer as follows: The length $l$ of the two lattice vectors is set as $l = \sqrt{c}$, where $c$ is calculated by 
\begin{equation}
c = \max_{j} \sum_{i_{j} = 1}^{N_{j}} 2 r_{i_j} ^2
\end{equation}
with $N_j$ being the number of spheres on the $j$-th layer and $i_j$ being the label of the spheres placed on the $j$-th layer. The angle $\theta$ of the two lattice vectors are also chosen randomly to be between $60^{\circ}$ and $120^{\circ}$. In short, the two lattice vectors are set as
\begin{align}
\bm{a}_1 &= \left(l, 0, 0 \right), \\
\bm{a}_2 &= \left(l \cos \theta, l \sin \theta, 0 \right).
\end{align}
By the construction, the area of the base plane by $\bm{a}_1$ and $\bm{a}_2$ changes depending on the arrangement of spheres placed 
on the layers, resulting in avoidance of a large overlap between spheres. 

Fifth, we randomly set the third lattice vector $\bm{a}_3$ as follows: 
the $z$ component $l_z$ is determined as the sum of the maximum radii, where the maximum radius for each layer is taken to be 
the maximum one among spheres placed on the corresponding layer. 
The setting corresponds to the operation of a piling-up of each layer, where the distance between two layers is the average 
of the maximum radii assigned to the two layers. On the other hand, the $x$ and $y$ components of $\bm{a}_3$ are set randomly 
by linear combination of $\bm{a}_1$ and $\bm{a}_2$, thus we have 
\begin{equation}
\bm{a}_3 = r \bm{a}_1 + r^{\prime} \bm{a}_2 + \left(0, 0, l_z \right),
\end{equation}
where $r$ and $r^{\prime}$ are randomly chosen to be between $-0.5$ and $0.5$.

Finally, we determine the third fractional coordinates of spheres placed on the second and later layers. The third fractional coordinates of each layer can be determined uniquely because each layer is piled up one by one as the distance between two layers is the average of the maximum radii assigned to the two layers.

As described above, the improved piling-up method directly generates multilayered structures with modest overlaps. In many cases, the structural optimization, which will be discussed in the next subsection, does not significantly transform the initial structures. In fact, if the spheres are placed uniformly in each layer, only lattice vectors are expanded uniformly until the overlaps become zero, and finally the structure is distorted slightly so as to increase the packing fraction. 
So, one can assume that initial structures generated by the improved piling-up method are close to some packing structures 
from the beginning, and that they can be smoothly optimized for the packing structure. 
If the initial structures are unbiasedly distributed in the configuration space for a given composition and radius ratio, 
it is expected that the exhaustive exploration can reach the DSP structure if the number of the trials is 
large enough.

Note that overlaps are necessary to let the unit cell have the freedom to change to some extent. 
Since an initial structure is optimized by the pseudo-annealing and the iterative balance optimization, 
lattice vectors can change to some extent when the spheres are placed non-uniformly. 
For example, if there are large overlaps in the $x$ or $y$ direction on some layers, 
the lattice vectors are expanded in that direction. Such an occasional displacement gives the structural diversity.

\subsection{Improved optimization scheme}
\label{improved_optimization_scheme}

\begin{figure}
\centering
\includegraphics[width=\columnwidth]{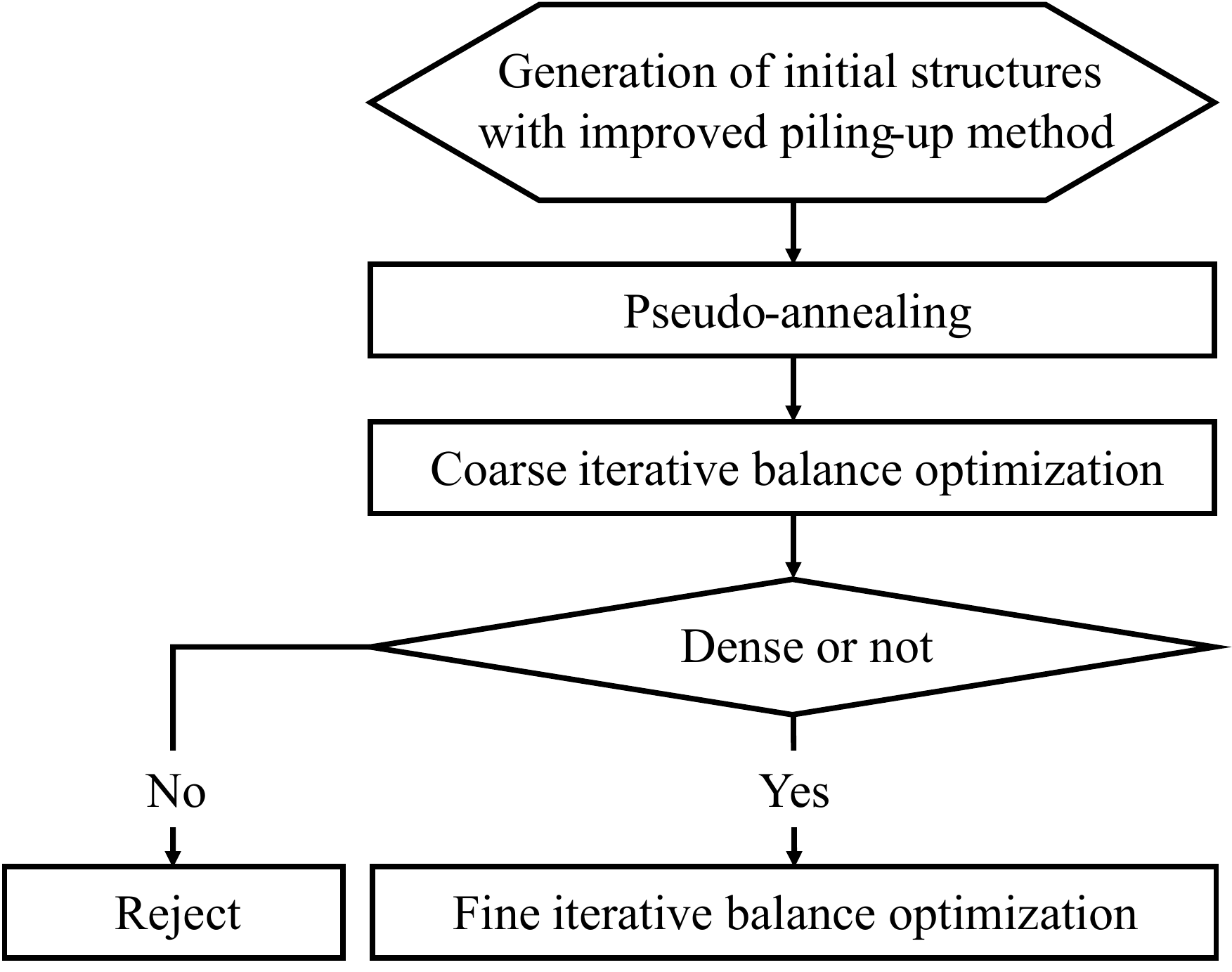}
\caption{The flow chart of the generation steps.}
\label{fig:flowChart}
\end{figure}

As discussed in Ref.~\cite{PhysRevE.103.023307}, when the number of spheres per unit cell is large, most of generated structures have much lower packing fractions than the maximum. In addition, the optimization by the iterative balance method occupies most of the computational time in the whole optimization step. Therefore, the computational cost can be reduced significantly if unpromising structures are rejected properly before the fine optimization.

The improved scheme of optimization step comprises four steps as depicted in the flow chart of Fig.~\ref{fig:flowChart}: (1) pseudo-annealing; (2) coarse iterative balance optimization; (3) selection of promising structures; (4) fine iterative balance optimization. In this subsection, we discuss the operation and importance of the four steps.

\subsubsection{Pseudo-annealing}
\label{sec:pseudo-annealing}

First, an initial structure is annealed by the steepest descent method with the hard-sphere potential~\cite{PhysRevE.103.023307}, defined by
\begin{equation}
U \left(\left|\bm{r}_j + \bm{T} - \bm{r}_i \right| \right) \equiv \begin{cases}
-z _{i j} ^{\left(\bm{T}\right)} & \text{$z _{i j} ^{\left(\bm{T}\right)} \le 0$} \\
0 & \text{$0 < z _{i j} ^{\left(\bm{T}\right)}$} \end{cases} \label{eq:potential}
\end{equation}
with
\begin{equation}
z _{i j} ^{\left(\bm{T}\right)} \equiv \left|\bm{r} _{j} + \bm{T} - \bm{r}_{i} \right| - \left(c_i + c_j \right),
\end{equation}
where $\bm{r}_i$ and $c_i$ is the position vector and the radius of sphere $i$, respectively. 
Note that all values we discuss in the paper are dimensionless.
The enthalpy $H$ per unit cell is defined as
\begin{equation}
H = \frac{1}{2} \sum_{\bm{T}} \sum_{i=1}^N \sum_{j=1} ^N U \left(\left|\bm{r}_j + \bm{T} - \bm{r}_i \right| \right) + PV,
\end{equation}
where $N$ is the number of spheres per unit cell, $P$ is the pressure, and $V$ is the volume of the unit cell. The pressure is necessary to reduce the volume of the unit cell. The fractional coordinates $\{\bm{q} \}$ and lattice vectors $\bm{a}_1$, $\bm{a}_2$, $\bm{a}_3$ are updated by the steepest descent method as
\begin{equation}
\Delta \bm{u} = -k_1 \frac{\partial H}{\partial \bm{u}}
\end{equation}
with
\begin{equation}
\bm{u} \equiv \left(q_{11}, q_{12}, q_{13}, \cdots , q_{N3}, a_{11}, a_{12}, a_{13}, \cdots , a_{33} \right),
\end{equation}
where $k_1$ is a constant. Through our optimization step, we do not directly set the $k_1$, but we scale the $\Delta \bm{u}$ so that the maximum displacement in the lattice vectors and the position vectors of the spheres is equal to the value $D_{\mathrm{max}}$ that is set independently in the first annealing, second annealing, coarse iterative balance optimization, and fine iterative balance optimization, where the displacement of each sphere is calculated by Cartesian coordinate.

The pseudo-annealing step is aimed at optimizing the initial structure for a packing structure with overlaps, depending on the sphere arrangement. The operation induces repetitions of collision and repulsion between spheres which lets void in the unit cell filled with spheres by reducing the volume of unit cell, distorting the unit cell, and optimizing the sphere arrangement.

Since the pressure never balances with the repulsive forces by the hard-sphere potential, no initial structure converges to any packing structure. However, the structure are sometimes trapped at a local minimum by circulating around the local minimum in the configuration space. In that case, only a change of maximum displacement $D_{\mathrm{an}, \mathrm{max}}$ is the driving force to escape the local minimum. Therefore, in our algorithm, the pseudo-annealing step is applied twice with different maximum displacement, $D_{\mathrm{an1}, \mathrm{max}}$ and $D_{\mathrm{an2}, \mathrm{max}}$, whose default values are shown in Sec.~\ref{sec:default_values_for_optimization_parameters_in_exhaustive_search} together with those of the numbers of steepest descent steps $N_{\mathrm{an1}} = N_{\mathrm{an2}}$.

\subsubsection{Coarse optimization by the iterative balance method}

To select promising structures which may have the densest packing fraction, the annealed structure is coarsely optimized for a packing structure by the coarse iterative balance optimization. In the coarse optimization, the maximum displacement $D_{\mathrm{cItr}, \mathrm{max}}$ is gradually decreased by the factor $d_{\mathrm{cItr}}$ in each optimization step. 
As discussed in Ref.~\cite{PhysRevE.103.023307}, 
if the decreasing factor $d_{\mathrm{cItr}}$ is chosen to be small and the number of optimization steps $N_{\mathrm{cItr}}$ is set to be large, the maximum displacement is converged to nearly zero, resulting in that any structure converges to a packing structure whose unit cell is locally minimized and in which as many spheres contact each other as possible. However, this step is only aimed at estimating the packing fractions to select promising structures, therefore, a large number of optimization steps is unnecessary and we do not need the maximum displacement converged to zero. If we properly choose a suitable parameter set: $D_{\mathrm{cItr}, \mathrm{max}}$, $d_{\mathrm{cItr}}$, and $N_{\mathrm{cItr}}$, it becomes possible to efficiently estimate the packing fractions precisely enough to choose promising structures, and accordingly the computational cost will be largely reduced. The default values which work well for most cases
are given in Sec.~\ref{sec:default_values_for_optimization_parameters_in_exhaustive_search}.

The initial maximum displacement $D_{\mathrm{cItr}, \mathrm{max}}$ is different from that for the second pseudo-annealing $D_{\mathrm{an2}, \mathrm{max}}$, so the difference sometimes induces a structural change by breaking out of a local minimum. Furthermore, the maximum displacement is gradually decreased in each step. The reduction also induces breaking out of a local minimum. The coarse optimization is necessary not only for the selection of promising structures but also for the optimization for a denser structure.

We checked the distributions of the error of packing fractions estimated by the coarse optimization at some radius ratios and compositions, where the error is defined as the difference between packing fractions estimated by the coarse optimization and the accurate packing fractions calculated by the fine iterative balance optimization. The result shows that most of the error are less than 0.03, indicating that a packing fraction estimated by the coarse optimization is smaller than the accurate value by 0.03 to 0.04. The error is accurate enough to select promising structures.

\subsubsection{Selection of promising structures}

Since a packing fraction estimated by the coarse optimization is about 0.03 less than the accurate value, we can efficiently select dense structures before the fine iterative balance optimization. As discussed in Ref.~\cite{PhysRevE.103.023307}, in the (16-4) and (14-5) systems, most of obtained structures are much sparser than the densest structure; the differences between their packing fractions and the densest one is much larger than 0.03. Therefore, it might be possible to assume for other systems containing a large number of spheres that most of packing fractions obtained by the exploration process are also much smaller than the densest one.
To reject sparse structures, we need to determine a threshold value to screen dense structures. Referring to the discussion in the previous subsection, we set the threshold value of the packing fraction as the provisional maximum packing fraction minus about $0.03$. 
More precisely, the threshold value at each radius ratio and composition is determined in our algorithm based on statistical data 
as follows: 
First, at the beginning of exploration all of the generated initial structures are optimized completely to make the distribution of error. Second, we take average of the absolute values of the error with a weight which becomes larger with increase in the absolute value of error. Then, we set the initial threshold value using the weighted average error. If a packing fraction is less than the threshold value, the structure is rejected. The threshold value is refined whenever the error of packing fraction is calculated by the fine optimization. In most cases, the threshold value is in between $0.03$ and $0.04$.
The selection significantly reduces the computational cost. In case the number of spheres per unit cell is large, the rate of promising structures is about only 1 to 5 percent, and the screening reduces 85-90 percent of computational cost.

\subsubsection{Fine optimization by the iterative balance method}

As discussed in Ref.~\cite{PhysRevE.103.023307}, in the iterative balance method 
the fractional coordinates and lattice vectors are optimized by the steepest descent method to minimize the enthalpy of system, 
where the maximum displacement $D_{\mathrm{fItr}, \mathrm{max}}$ is gradually decreased by the factor $d_{\mathrm{fItr}}$ in each optimization step. The number of optimization steps $N_{\mathrm{fItr}}$ is set to be so large that the maximum displacement is reduced to zero so gradually enough to let as many spheres contact each other as possible, and accordingly the volume of the unit cell is minimized with slight distortion. 
The default values of $D_{\mathrm{fItr}, \mathrm{max}}$, $d_{\mathrm{fItr}}$, and $N_{\mathrm{fItr}}$ are given in Sec.~\ref{sec:default_values_for_optimization_parameters_in_exhaustive_search}.

\subsubsection{Default parameters for exploration of DSPs}
\label{sec:default_values_for_optimization_parameters_in_exhaustive_search}

In Secs.~\ref{sec:improved_piling-up_method} and \ref{improved_optimization_scheme}, we discussed the way to generate initial structures and the optimization step. The operations are applied to the exhaustive search of the DTSPs. The default parameters such as maximum displacement are given in Table~\ref{table:samlai-optimization-parameters}. $M_{\mathrm{an1}}$, $M_{\mathrm{an2}}$, $M_{\mathrm{cItr}}$, and $M_{\mathrm{fItr}}$ are the interval to update the neighbor list in the first and second pseudo-annealing steps, the coarse and 
fine optimization steps by the the iterative balance method, respectively. The neighbor list records which pairs of spheres possibly overlap each other in the optimization step, which lets us significantly reduce computational cost. 
This is because fractional coordinates and lattice vectors are not largely changed after the initial structure is converged 
close to a packing structure.

As discussed in Ref.~\cite{PhysRevE.103.023307}, the efficiency of structural optimization depends on the choice of parameters especially for the maximum displacement, and even the reachable maximum packing fraction is also varied depending on the parameters in case that many local minima are having competitive packing fractions. Therefore, to find the densest packing fraction, the maximum displacement in each optimization step should be randomly changed for each generated initial structure. By default, $D_{\mathrm{an1}, \mathrm{max}}$, $D_{\mathrm{an2}, \mathrm{max}}$, $D_{\mathrm{cItr}, \mathrm{max}}$, and $D_{\mathrm{fItr}, \mathrm{max}}$ for each generated initial structure are determined randomly within the range given in Table~\ref{table:samlai-optimization-parameters}.

\begin{table}
\caption{Default parameters for exploration of DSPs. $D_{\mathrm{an1}, \mathrm{max}}$, $D_{\mathrm{an2}, \mathrm{max}}$, $D_{\mathrm{cItr}, \mathrm{max}}$, and $D_{\mathrm{fItr}, \mathrm{max}}$ are the the maximum displacement in first annealing, second annealing, coarse iterative balance optimization, and fine iterative balance optimization, respectively. 
$N_{\mathrm{an1}}$, $N_{\mathrm{an2}}$, $N_{\mathrm{cIter}}$, and $N_{\mathrm{fIter}}$ are the number of optimization steps in first annealing, second annealing, coarse iterative balance optimization, and fine iterative balance optimization, respectively. $M_{\mathrm{an1}}$, $M_{\mathrm{an2}}$, $M_{\mathrm{cIter}}$, and $M_{\mathrm{fIter}}$ are the interval to update neighbor list in first annealing, second annealing, coarse iterative balance optimization, and fine iterative balance optimization, respectively. $d_{\mathrm{cItr}}$ and $d_{\mathrm{fItr}}$ are the decreasing factor of maximum displacement in coarse iterative balance optimization, and fine iterative balance optimization, respectively.}
\label{table:samlai-optimization-parameters}
\begin{ruledtabular}
\begin{tabular}{cc}
Optimization parameter & default value \\
$D_{\mathrm{an1}, \mathrm{max}}$, $D_{\mathrm{an2}, \mathrm{max}}$ & $0.03 \le D_{\mathrm{an1}, \mathrm{max}}, D_{\mathrm{an2}, \mathrm{max}} \le 0.10$ \\
$N_{\mathrm{an1}}$, $N_{\mathrm{an2}}$ & 500 \\
$M_{\mathrm{an1}}$, $M_{\mathrm{an2}}$ & 200 \\
$D_{\mathrm{cItr}, \mathrm{max}}$ & $0.02 \le D_{\mathrm{cItr}, \mathrm{max}} \le 0.05$ \\
$N_{\mathrm{cItr}}$ & 2500 \\
$d_{\mathrm{cItr}}$ & 0.9995 \\
$M_{\mathrm{cItr}}$ & 300 \\
$D_{\mathrm{fItr}, \mathrm{max}}$ & $0.02 \le D_{\mathrm{fItr}, \mathrm{max}} \le 0.05$ \\
$N_{\mathrm{fItr}}$ & 40000 \\
$d_{\mathrm{fItr}}$ & 0.9997 \\
$M_{\mathrm{fItr}}$ & 300
\end{tabular}
\end{ruledtabular}
\end{table}

\subsection{Reoptimization}

Some of packing structures have so complex local minima that tens and thousands of trials of reoptimization with slight fluctuation are necessary to find the highest packing fraction from the large number of local maxima. Besides, as discussed in Ref.~\cite{PhysRevE.103.023307}, a unit cell sometimes places a limit on the maximum packing fraction that can be reached. Therefore, a large number of reoptimization of several {\it isomorphic} structures with small fluctuation is necessary to determine the maximum packing fraction, where the term {\it isomorphic} means the equality between two structures that two structures are identical when the Cartesian coordinates and 
lattice vectors of one of the two structures are displaced in a range of a small threshold value.

In our algorithm, after the exhaustive exploration for DSPs, some of the candidate structures, one of which will have the densest packing fraction, are reoptimized. The default values for the number of candidate structures $N_{\mathrm{d}}$, which are applied for reoptimization, and the number of the reoptimization step $N_{\mathrm{reo}}$ are given in Table~\ref{table:default_parameters_for_reoptimization}. During the exhaustive exploration, we pick the top $N_{\mathrm{d}}$ structures up with respect to the packing fraction, and we perform the reoptimization $N_{\mathrm{reo}}$-th times for the selected structures as explained below.

The reoptimization process comprises three steps: (a) random fluctuation, (b) pseudo-annealing, (c) iterative balance optimization. 
The random fluctuation of the Cartesian coordinates and lattice vectors of seed structures is necessary to find the densest packing fraction from a large number of local minima. The default maximum value of the fluctuation $\Delta f$ is given in Table~\ref{table:default_parameters_for_reoptimization}.

The pseudo-annealing and the iterative balance optimization are aimed at optimizing the fluctuated structures to the densest packing. The range of maximum displacement both in pseudo-annealing $D_{\mathrm{rean}, \mathrm{max}}$ and in the iterative balance optimization $D_{\mathrm{reitr}, \mathrm{max}}$ are set to be larger than those for the exhaustive exploration so that the reachable packing fractions can be diverse. Besides, if we need to calculate packing fractions accurately, the number of iterative balance optimization steps $N_{\mathrm{reitr}}$ should be larger than that for the exhaustive exploration, because overlaps between spheres make packing fractions a little larger. The default parameters for the pseudo-annealing and the iterative balance optimization are also given in Table~\ref{table:default_parameters_for_reoptimization}.
In most cases, the default parameters given in Table~\ref{table:default_parameters_for_reoptimization} are effective enough to find the densest packing fractions at each radius ratio and composition. 
The effectiveness of our method to search DSPs will be discussed in Sec.~\ref{sec:discussion}.

\begin{table}
\caption{Default parameters for reoptimization. $N_{\mathrm{d}}$ is the number of candidate structures which are applied for reoptimization. $N_{\mathrm{reo}}$ is the number of the reoptimization step per candidate structure. $\Delta f$ is the maximum value of the fluctuation. $D_{\mathrm{rean}, \mathrm{max}}$ and $D_{\mathrm{reitr}, \mathrm{max}}$ are the range of maximum displacement  in pseudo-annealing and in iterative balance optimization, respectively. 
$N_{\mathrm{rean}}$ and $N_{\mathrm{reitr}}$ are the number of iterative balance optimization steps in pseudo-annealing and in iterative balance optimization, respectively. $M_{\mathrm{rean}}$, and $M_{\mathrm{reitr}}$ are the interval to update the neighbor list in pseudo-annealing and in iterative balance optimization, respectively. $d_{\mathrm{reitr}}$ is the decreasing factor of maximum displacement in iterative balance optimization.}
\label{table:default_parameters_for_reoptimization}
\begin{ruledtabular}
\begin{tabular}{cc}
Reoptimization parameter & default value \\
$N_{\mathrm{d}}$ & 20 \\
$N_{\mathrm{reo}}$ & 5000 \\
$\Delta f$ & 0.03 \\
$D_{\mathrm{rean}, \mathrm{max}}$ & $0.003 \le D_{\mathrm{rean}, \mathrm{max}} \le 0.05$ \\
$N_{\mathrm{rean}}$ & 500 \\
$M_{\mathrm{rean}}$ & 200 \\
$D_{\mathrm{reitr}, \mathrm{max}}$ & $0.001 \le D_{\mathrm{reitr}, \mathrm{max}} \le 0.02$ \\
$N_{\mathrm{reitr}}$ & 80000 \\
$d_{\mathrm{reitr}}$ & 0.9997 \\
$M_{\mathrm{reitr}}$ & 1000 \\
\end{tabular}
\end{ruledtabular}
\end{table}

\section{exploration conditions}
\label{sec:exploration_conditions}

With our open-source program package \textbf{SAMLAI}~\cite{PhysRevE.103.023307,samlai},
we have exhaustively explored DTSPs at 45 kind of radius ratios and 237 kinds of compositions. 
In this section, we detail the conditions for the exhaustive exploration of DTSPs including the choice of radius ratios and compositions.

In this study, a ternary composition that contains $n_1$ small, $n_2$ medium, and $n_3$ large spheres per unit cell is named as $n_1$-$n_2$-$n_3$ system. The radii of small and medium spheres are $\alpha_1$ and $\alpha_2$, respectively, where the radius of the large spheres is 
always set to be $1.0$. The composition ratios of small, medium, and large spheres are denoted by $x_1$, $x_2$, and $x_3$, respectively, which are defined as
\begin{equation}
x_i \equiv \frac{n_i}{N}, \label{eq:definition-of-composition-ratio-xi}
\end{equation}
where $N$ is the total number of spheres per unit cell:
\begin{equation}
N \equiv n_1 + n_2 + n_3.
\end{equation}

\subsection{Optimization parameters}

All parameters of the exhaustive exploration and reoptimization for DTSPs are set to the default values given 
in Tables~\ref{table:samlai-optimization-parameters} and \ref{table:default_parameters_for_reoptimization}, respectively.

\subsection{Exhaustive search conditions}

In this study, $\alpha_1$ and $\alpha_2$ are the radii of small and medium spheres, respectively, where we fix the radius of large spheres to be $1.0$. We have chosen 13 values for $\alpha_1$ and $\alpha_2$ as $0.20$, $0.25$, $\cdots$, $0.70$. Besides, we impose a restriction as $\alpha_1 + 0.1 \le \alpha_2$. The number of radius ratios which satisfy those conditions is 45.

The number of spheres per unit cell are set to be from 3 to 20:
\begin{equation}
3 \le N \le 20, \label{eq:total-number-of-n1-n2-n3}
\end{equation}
with the constraint:
\begin{equation}
n_3 \le n_2 \le n_1. \label{eq:constraint-of-n1-n2-n3}
\end{equation}
The number of compositions which satisfy Eqs.~(\ref{eq:total-number-of-n1-n2-n3}) and (\ref{eq:constraint-of-n1-n2-n3}) is 237.

At each composition and radius ratio, we generate a large number of initial structures by the improved piling-up method so as to exhaustively search the DTSPs. The number of the generated structures are different depending on the number of spheres per unit cell as follows: For large basis as $13 \le N \le 20$, the number of generated structures is set to be 20,000,000. For medium basis as $9 \le N \le 12$, the number of generated structures is set to be 10,000,000. For small basis as $3 \le N \le 8$, the number of generated structures is set to be 1,000,000.

In addition to the 237 compositions determined by Eqs.~(\ref{eq:total-number-of-n1-n2-n3}) and (\ref{eq:constraint-of-n1-n2-n3}), 
we have additionally calculated compositions having a larger number of spheres in the unit cell as follows:
At the radius ratios of $0.20:0.30:1.00$ and $0.20:0.35:1.00$, we have studied at the additional composition ratios including 19-1-1, 20-1-1, 21-1-1, 22-1-1, 18-2-1, 19-2-1, 20-2-1, and 21-2-1 systems. At the radius ratio of $0.20:0.45:1.00$, we have studied at the additional composition ratios including 10-6-6, 12-6-6, 13-6-6, 14-6-6, 15-6-6, 16-6-6, 17-6-6, and 18-6-6 systems. At the radius ratios of $0.20:0.65:1.00$, $0.20:0.70:1.00$, $0.25:0.65:1.00$, $0.25:0.70:1.00$, $0.30:0.60:1.00$, $0.30:0.65:1.00$, $0.30:0.70:1.00$, we have studied at the additional composition ratios including 14-6-2, 15-6-2, 16-6-2, 17-6-2, 18-6-2, 19-6-2, 20-6-2, 21-6-2, and 22-6-2 systems. At the radius ratio of $0.25:0.45:1.00$, we have studied at the additional composition ratios including 9-6-6, 10-6-6, 11-6-6, 12-6-6, 13-6-6, 14-6-6, 15-6-6, and 16-6-6 systems. At the radius ratio of $0.25:0.50:1.00$, we have studied at the additional composition ratios including 19-1-1, 20-1-1, 21-1-1, and 22-1-1 systems. Finally, at the radius ratio of $0.35:0.45:1.00$, we have studied at the additional composition ratios including 17-2-2, 18-2-2, 19-2-2, 20-2-2, 21-2-2, and 22-2-2 systems.

As discussed in previous studies~\cite{PhysRevE.85.021130, PhysRevE.103.023307}, for a given composition of three kinds of spheres, the densest packing fraction can be achieved by the phase separation which consists of less than or equal to three structures when every candidate structure is periodic. Note that the densest FCC structures consisting of small, medium, and large spheres, and additionally the DBSPs consisting of small and medium, small and large, and medium and large spheres are the candidate structures for the densest phase separations. Therefore, there may be composition ratios for which the densest packing fraction is achieved by the phase separation which consists of only FCC structures and/or DBSPs. In fact, we found many systems whose densest phase separations do not contain any DTSPs.
We have identified the densest phase separations at many composition ratios $\left(x_1, x_2, x_3 \right)$, where $x_i$ is calculated as Eq.~(\ref{eq:definition-of-composition-ratio-xi}) with $n_1$, $n_2$, and $n_3$ that satisfy Eq.~(\ref{eq:total-number-of-n1-n2-n3}) and the constraint:
\begin{equation}
n_3 \le n_1 + n_2. \label{eq:constraint-of-n1-n2-n3-2}
\end{equation}
The number of composition ratios which satisfy those conditions is 779. 
At the 779 composition ratios we evaluate densest phase separations using information of 237 compositions of
the generated ternary packings, DBSPs, and FCCs. So, a phase diagram at each radius ratio has 779 points for which the densest phase separations are evaluated
as shown in Sec.~\ref{sec:results}.

\subsection{Phase separation}
\label{sec:phase_separation}

To reduce the computational cost to identify the densest phase separations, the denser ternary sphere packings are picked up as follows: Any composition ratio of ternary systems can be realized by a phase separation consisting of only the densest FCC structures and/or DBSPs. If a ternary packing is not denser than the phase separation consisting of only the densest FCC structures and/or DBSPs, the ternary structure cannot contribute the densest phase separations, so we can exclude it from the candidates. Furthermore, if the packing fractions of the remaining ternary structures are less than or equal to phase separations consisting of the other structures, the ternary structures
do not necessarily make the densest phase separations, so the structures can also be excluded from the candidates. The exclusion is important not only for the computational cost but also for unique determination of the densest phase separations.

Finally, we identify the densest phase separations at each composition ratio by exhaustive calculation: First, we make all the feasible phase separations which can be realized from candidate structures. Second, we calculate their packing fractions. Finally, we pick the densest phase separations up.

\begin{table*}
\caption{DTSPs discovered at radius ratios, $\alpha_1$ and $\alpha_2$, for small and medium sphere relative to the radius of large sphere
and the packing fraction, where {\it none} and '-' mean that we did not find any DTSPs and that the result is the same as that of $\alpha_2$ and $\alpha_1$, respectively.}
\label{table:PF-dtsps}
\begin{ruledtabular}
\begin{tabular}{c|ccccccccc}
\diagbox{$\alpha_1$}{$\alpha_2$} & $0.30$ & $0.35$ & $0.40$ & $0.45$ & $0.50$ & $0.55$ & $0.60$ & $0.65$ & $0.70$\\
\hline
    &(18-2-1) &(18-1-1) &(16-1-1) &(14-1-1) &(10-2-1) &(4-2-1)  &(4-2-1)  &(22-6-2) &none\\
    &0.823226 &0.812846 &0.815179 &0.813881 &0.806565 &0.791468 &0.774420 &0.782561 &\\
    &(10-4-1) &(12-1-1) &FCCTC    &(16-2-2) &(6-2-1)  &         &         &         &\\
    &0.815904 &0.810788 &0.799719 &0.798402 &0.787225 &         &         &         &\\
    &(10-1-1) &(8-1-1)  &         &(14-2-2) &         &         &         &         &\\
    &0.814884 &0.807440 &         &0.794755 &         &         &         &         &\\
0.20&(7-4-1)  &(12-2-2) &         &(10-4-2) &         &         &         &         &\\
    &0.811218 &0.804429 &         &0.784534 &         &         &         &         &\\
    &XYZ$_8$  &XYZ$_4$  &         &(6-4-2)  &         &         &         &         &\\
    &0.807864 &0.795924 &         &0.774288 &         &         &         &         &\\
    &         &         &         &(18-6-6) &         &         &         &         &\\
    &         &         &         &0.773209 &         &         &         &         &\\
    &         &         &         &(8-8-4)  &         &         &         &         &\\
    &         &         &         &0.769156 &         &         &         &         &\\
\hline 
    &    -    &(8-4-1)  &none     &(14-6-6) &(18-1-1) &(12-1-1) &none     &none     &none\\
0.25&         &0.788699 &         &0.781847 &0.790979 &0.796895 &         &         &\\
    &         &(9-7-2)  &         &(4-4-2)  &(4-3-1)  &(10-2-2) &         &         &\\
    &         &0.783220 &         &0.778942 &0.777358 &0.788895 &         &         &\\
\hline 
    &    -    &   -     &(13-3-1) &(4-3-1)  &(4-3-1)  &(9-6-3)  &none     &none     &none\\
    &         &         &0.793502 &0.784757 &0.786907 &0.785141 &         &         &\\
0.30&         &         &         &(4-4-2)  &(2-2-2)  &         &         &         &\\
    &         &         &         &0.77569  &0.763813 &         &         &         &\\
    &         &         &         &(2-2-2)  &         &         &         &         &\\
    &         &         &         &0.765616 &         &         &         &         &\\
\hline 
0.35&   -     &   -     &   -     &(16-2-2) &(10-6-3) &none     &none     &none     &none\\
    &         &         &         &0.769450 &0.773180 &         &         &         &\\
\hline 
0.40&   -     &   -     &   -     &    -    &   none  &none     &none     &none     &none\\
\hline 
0.45&   -     &   -     &   -     &    -    &    -    &none     &(13-2-1) &(13-2-1) &none\\
    &         &         &         &         &         &         &0.751416 &0.751922 &\\
\hline 
0.50&   -     &   -     &   -     &    -    &    -    &    -    & none    & none    &(2-1-1)\\
    &         &         &         &         &         &         &         &         &0.765177\\
\hline 
0.55&   -     &   -     &   -     &    -    &    -    &    -    & -       & none    & none\\
\hline 
0.60&   -     &   -     &   -     &    -    &    -    &    -    & -       & -       & none\\
\end{tabular}
\end{ruledtabular}
\end{table*}

\begin{table}
\caption{Packing fractions of DBSPs at the radius ratio of $0.20:0.30$ to $0.30:0.70$. 'none' means that there is no DBSP.}
\label{table:packing_fractions_of_dtsps}
\begin{ruledtabular}
\begin{tabular}{ccc}
Radius ratios & Name of DBSP & Packing fraction \\
\hline
$0.20:0.30$ & none & \\
$0.20:0.35$ & (12-6)~\cite{PhysRevE.103.023307} & $0.778953$ \\
$0.20:0.40$ & $\mathrm{AuTe}_2$~\cite{PhysRevE.79.046714,PhysRevE.85.021130,PhysRevE.103.023307} & $0.758114$ \\
$0.20:0.45$ & (6-6)~\cite{PhysRevE.85.021130,PhysRevE.103.023307} & $0.761084$ \\
$0.20:0.45$ & (14-5)~\cite{PhysRevE.103.023307} & $0.765866$ \\
$0.20:0.50$ & $\mathrm{XY}$~\cite{PhysRevE.85.021130,PhysRevE.103.023307} & $0.787871$ \\
$0.20:0.55$ & $\mathrm{XY}$ & $0.776086$ \\
$0.20:0.60$ & (6-1)~\cite{PhysRevE.85.021130,PhysRevE.103.023307} & $0.779091$ \\
$0.20:0.60$ & $\mathrm{XY}$ & $0.767906$ \\
$0.20:0.65$ & (6-1) & $0.796582$ \\
$0.20:0.65$ & $\mathrm{XY}$ & $0.762051$ \\
$0.20:0.70$ & (6-1) & $0.796002$ \\
$0.20:0.70$ & $\mathrm{XY}$ & $0.757751$ \\
$0.20:1.00$ & $\mathrm{XY}_{12}$~\cite{PhysRevE.103.023307} & $0.811567$ \\
$0.20:1.00$ & (22-1)~\cite{PhysRevE.103.023307} & $0.813313$ \\
$0.25:0.35$ & none & \\
$0.25:0.40$ & $\mathrm{A}_3 \mathrm{B}$~\cite{doi:10.1021/jp206115p,PhysRevE.85.021130,PhysRevE.103.023307} & $0.743409$ \\
$0.25:0.45$ & (12-6) & $0.779389$ \\
$0.25:0.50$ & $\mathrm{AuTe}_2$ & $0.758114$ \\
$0.25:0.55$ & (6-6) & $0.753138$ \\
$0.25:0.55$ & (16-4)~\cite{PhysRevE.103.023307} & $0.759674$ \\
$0.25:0.55$ & $\mathrm{HgBr}_2$~\cite{PhysRevE.79.046714,PhysRevE.85.021130,PhysRevE.103.023307} & $0.758764$ \\
$0.25:0.60$ & (6-6) & $0.789961$ \\
$0.25:0.65$ & $\mathrm{XY}$ & $0.782611$ \\
$0.25:0.70$ & $\mathrm{XY}$ & $0.774212$ \\
$0.25:1.00$ & (10-1)~\cite{PhysRevE.85.021130,PhysRevE.103.023307} & $0.797510$ \\
$0.25:1.00$ & $\mathrm{XY}_4$~\cite{PhysRevE.85.021130,PhysRevE.103.023307} & $0.786761$ \\
$0.30:0.40$ & none & \\
$0.30:0.45$ & none & \\
$0.30:0.50$ & $\mathrm{AlB}_2$~\cite{PhysRevE.79.046714,PhysRevE.85.021130,PhysRevE.103.023307} & $0.757492$ \\
$0.30:0.55$ & (12-6) & $0.780148$ \\
$0.30:0.60$ & $\mathrm{AuTe}_2$ & $0.758114$ \\
$0.30:0.65$ & (8-4) & $0.758913$ \\
$0.30:0.65$ & (16-4) & $0.757125$ \\
$0.30:0.70$ & (6-6) & $0.776184$
\end{tabular}
\end{ruledtabular}
\end{table}
\begin{table}
\caption{Packing fractions of DBSPs at the radius ratio of $0.30:1.00$ to $0.70:1.00$. 'none' means that there is no DBSP.}
\label{table:packing_fractions_of_dtsps_2}
\begin{ruledtabular}
\begin{tabular}{ccc}
Radius ratios & Name of DBSP & Packing fraction \\
\hline
$0.30:1.00$ & (6-1) & $0.797969$ \\
$0.30:1.00$ & $\mathrm{XY}$ & $0.760473$ \\
$0.35:0.45$ & none & \\
$0.35:0.50$ & none & \\
$0.35:0.55$ & $\mathrm{A}_3 \mathrm{B}$ & $0.744921$ \\
$0.35:0.60$ & $\mathrm{AlB}_2$ & $0.772940$ \\
$0.35:0.65$ & (12-6) & $0.780599$ \\
$0.35:0.70$ & $\mathrm{AuTe}_2$ & $0.758114$ \\
$0.35:1.00$ & (6-1) & $0.767159$ \\
$0.35:1.00$ & $\mathrm{XY}$ & $0.772229$ \\
$0.40:0.50$ & none & \\
$0.40:0.55$ & none & \\
$0.40:0.60$ & none & \\
$0.40:0.65$ & $\mathrm{AlB}_2$ & $0.745713$ \\
$0.40:0.70$ & (12-6) & $0.778953$ \\
$0.40:1.00$ & $\mathrm{XY}$ & $0.787871$ \\
$0.45:0.55$ & none & \\
$0.45:0.60$ & none & \\
$0.45:0.65$ & none & \\
$0.45:0.70$ &$\mathrm{A}_3 \mathrm{B}$ & $0.746390$ \\
$0.45:1.00$ & (6-6) & $0.756568$ \\
$0.45:1.00$ & (8-4) & $0.758885$ \\
$0.45:1.00$ & (14-5) & $0.759972$ \\
$0.45:1.00$ & (16-4) & $0.760246$ \\
$0.50:0.60$ & none & \\
$0.50:0.65$ & none & \\
$0.50:0.70$ & none & \\
$0.50:1.00$ & $\mathrm{AuTe}_2$ & $0.758114$ \\
$0.55:0.65$ & none & \\
$0.55:0.70$ & none & \\
$0.55:1.00$ & (12-6) & $0.779943$ \\
$0.60:0.70$ & none & \\
$0.60:1.00$ & $\mathrm{AlB}_2$ & $0.757492$ \\
$0.65:1.00$ & $\mathrm{A}_3 \mathrm{B}$ & $0.746541$ \\
$0.70:1.00$ & none & 
\end{tabular}
\end{ruledtabular}
\end{table}

\section{results}
\label{sec:results}

In this section, we present the phase diagrams for the ternary systems, and describe the discovered 37 putative DTSPs. 
It is not apparent whether there is at least one DTSP in the densest phase separation at any radius and composition ratio, 
since the densest packing fraction may consist of only the FCC structures and/or DBSPs. 
The discovery of 37 putative DTSPs presented in the section implies that many DTSPs actually have relatively large packing 
fractions beyond those of DBSPs. 

\subsection{Overview}

We have constructed the phase diagrams for ternary systems at 45 kinds of radius ratios, where information of 237 compositions of
the ternary packings, DBSPs, and FCCs have been used to evaluate densest phase separations.
As a result, we have successfully discovered novel 37 putative DTSPs and confirmed a known ternary dense structure, FCCTC \cite{doi:10.1063/1.4941262}, as one of DTSPs. 
All the DTSPs we found are shown in Table~\ref{table:PF-dtsps} together with the packing fractions.
Most of the discovered DTSPs are named as ($l$-$m$-$n$) structure. A ($l$-$m$-$n$) structure is constituted by $l$ small, $m$ medium, and $n$ large spheres. On the other hand, if large spheres in a DTSP constitute the FCC structure with contact, the DTSP is named as $\mathrm{X}_n \mathrm{Y}_m \mathrm{Z}_l$ structure, which consists of $n$ large, $m$ medium, and $l$ small spheres. 

\begin{figure}
\centering
\includegraphics[width=\columnwidth]{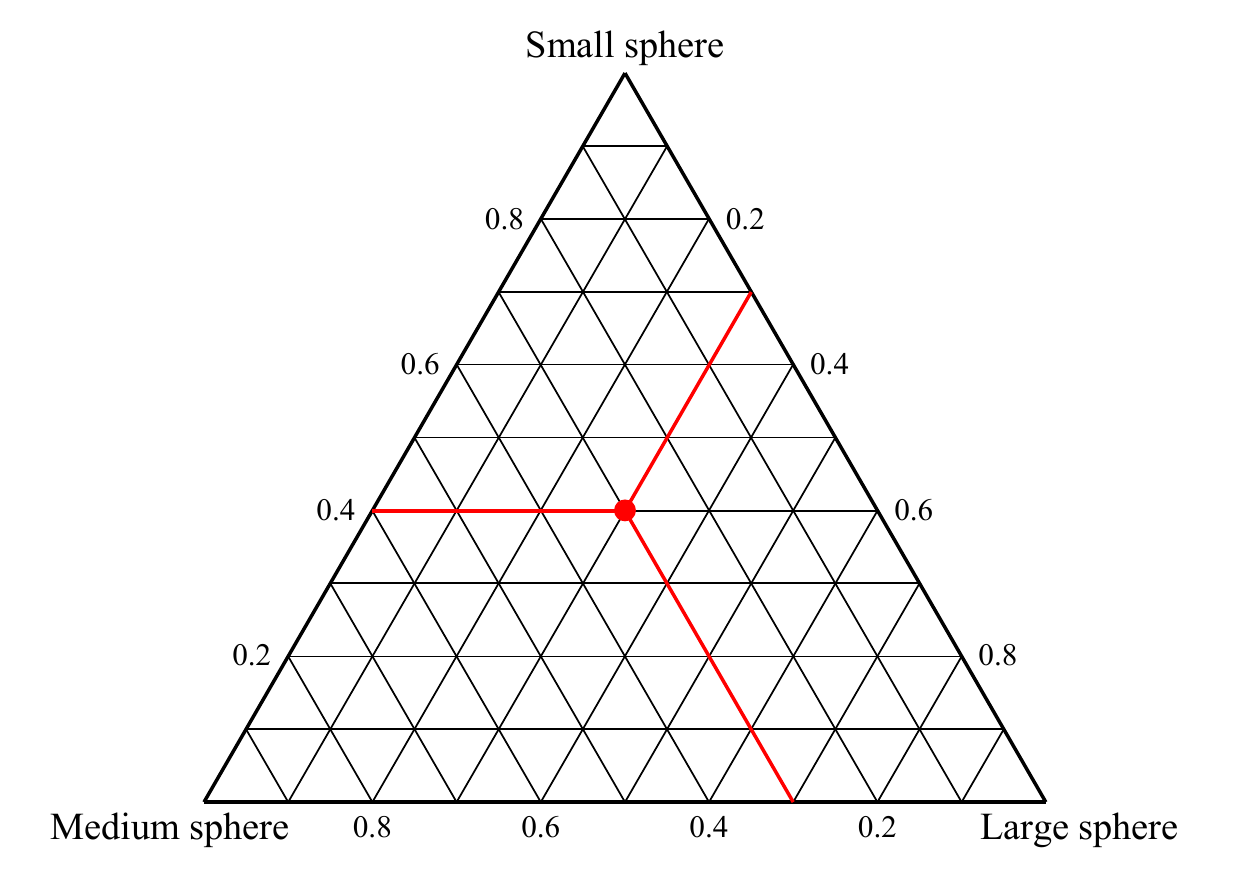}
\caption{The way to read the phase diagram. 
The composition ratio of small, medium, and large spheres at the red point is $0.40:0.30:0.30$.}
\label{fig:how-to-read}
\end{figure}

As discussed in Sec.~\ref{sec:phase_separation}, the highest packing fractions can be achieved by phase separations consisting of less than or equal to three structures. The densest FCC structures consisting of small, medium, and large spheres, and additionally the DBSPs consisting of small and medium, small and large, and medium and large spheres belong to the candidate structures for the densest phase separations. On our phase diagrams,  black disks are plotted at the compositions where the highest packing fractions are achieved by the phase separations consisting of only the FCC structures and/or DBSPs. On the other hand, if the densest phase separation consists of a few DTSPs at a composition, the symbols that correspond to the DTSPs are plotted together. How to read the phase diagram is given in Fig.~\ref{fig:how-to-read}.
In the following subsections, we present the phase separations at 45 kinds of radius ratios and all of the discovered DTSPs.

\subsection{DBSPs for phase separations}
\label{sec:DBSPs_for_phase_separation}

As discussed in Sec.~\ref{sec:phase_separation}, the densest phase separations generally contain DBSPs. In the Tables~\ref{table:packing_fractions_of_dtsps} and \ref{table:packing_fractions_of_dtsps_2}, we show the DBSPs which are comprised by small and medium, small and large, or medium and large spheres, and their packing fractions, at each radius ratio. At some radius ratios listed in the tables, there is no DBSP. 
The case is referred to as {\it none} in the column of the name of DBSPs, and leave the column of the packing fraction blank.

\begin{figure*}
\centering
\includegraphics[width=2\columnwidth]{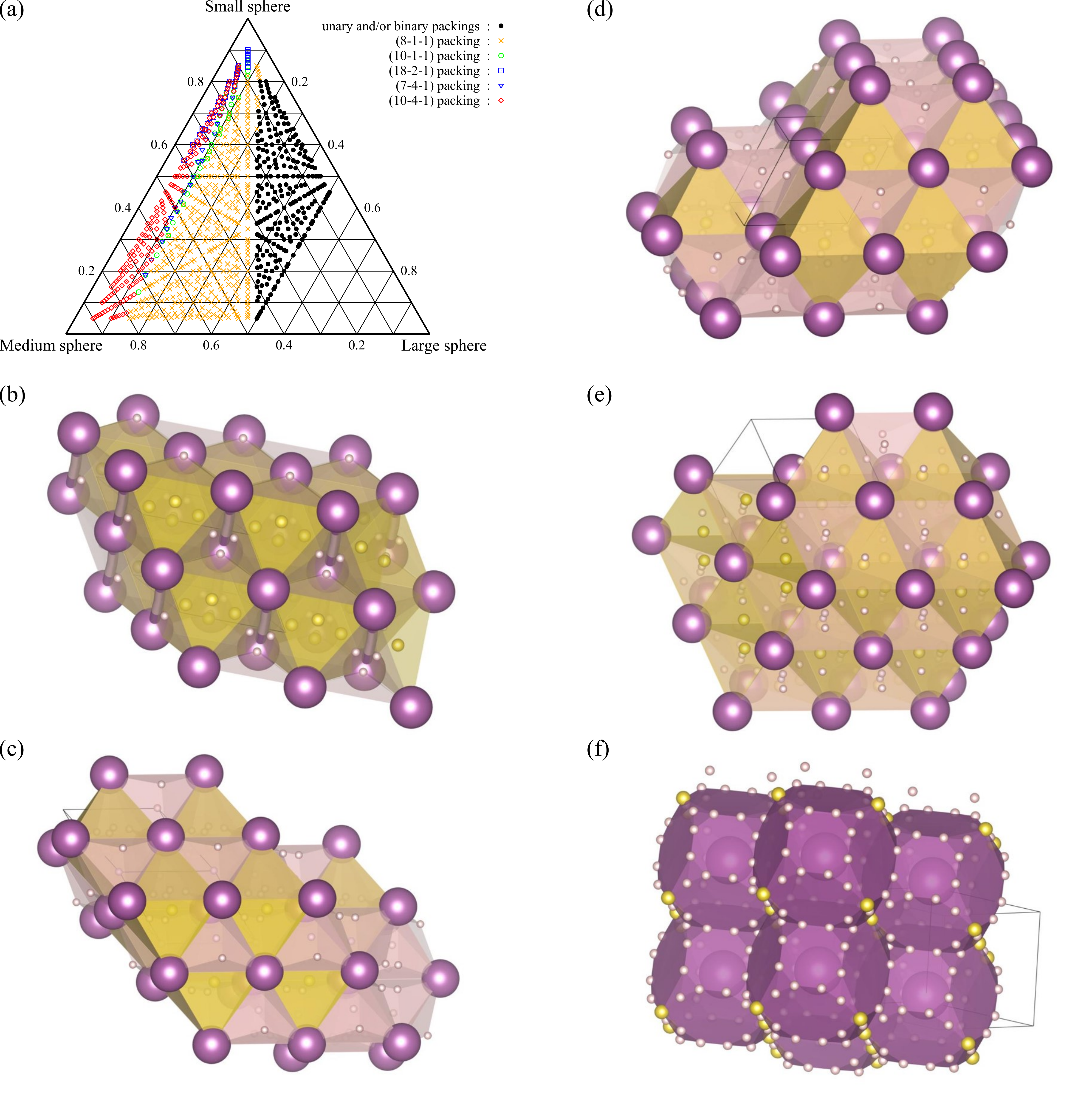}
\caption{The results at the radius ratio of $0.20:0.30:1.00$. (a) The phase diagram. Black disks are plotted at the composition ratios where the densest phase separations consist only of the densest FCC structures and/or DBSPs. On the other hand, at every composition, if the densest phase separation consists of a few DTSPs, the symbols that correspond to the DTSPs are plotted together. (b) The (7-4-1) structure. Figures are generated by VESTA~\cite{Momma:db5098}. (c) The $\mathrm{XYZ}_8$ structure. (d) The (10-1-1) structure. (e) The (10-4-1) structure. (f) The (18-2-1) structure.}
\label{fig:020-030-100}
\end{figure*}
\begin{figure*}
\centering
\includegraphics[width=2\columnwidth]{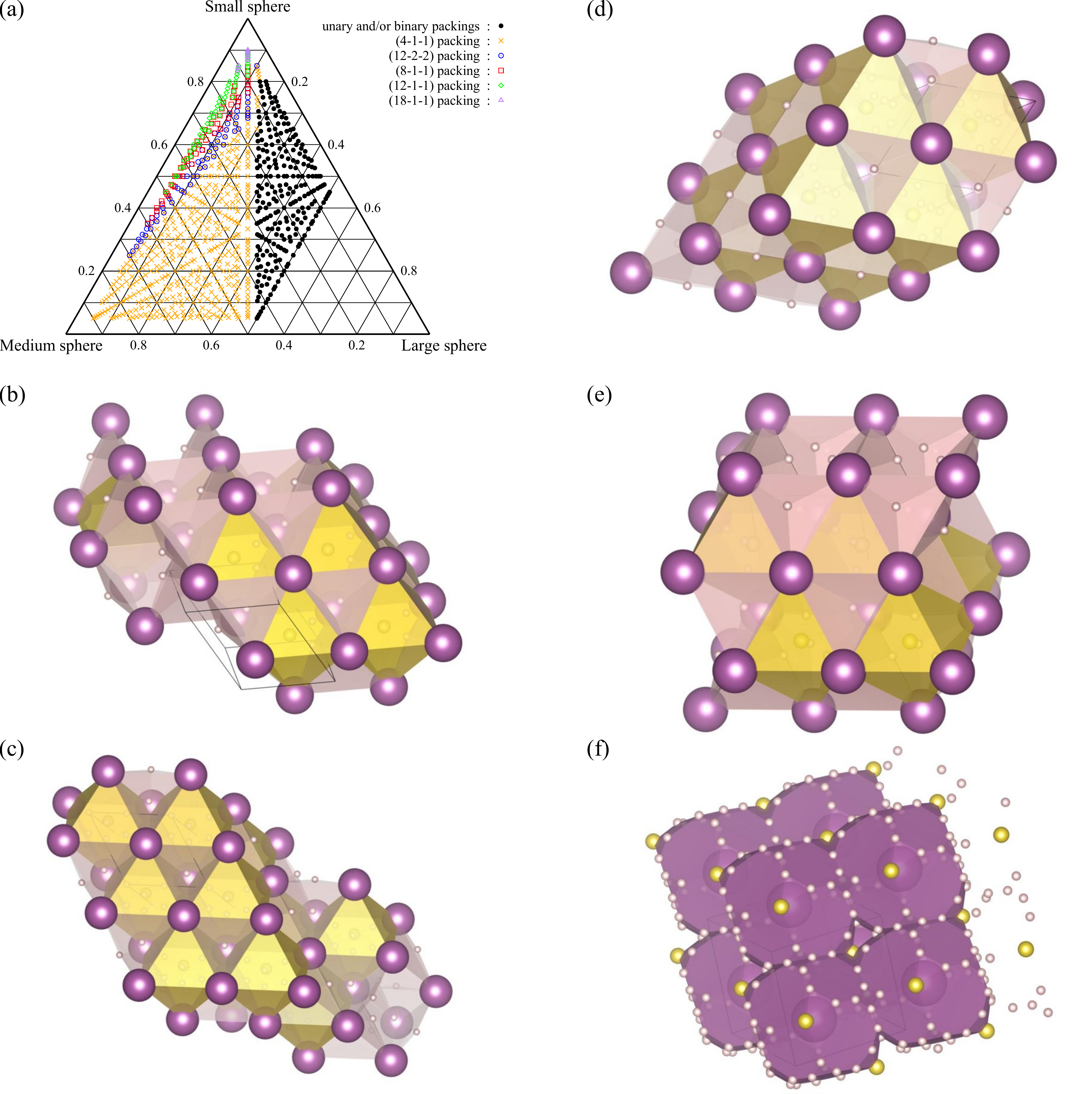}
\caption{The results at the radius ratio of $0.20:0.35:1.00$. (a) The phase diagram. (b) The $\mathrm{XYZ}_4$ structure. (c) The (8-1-1) structure. (d) The (12-1-1) structure. (e) The (12-2-2) structure. (f) The (18-1-1) structure.}
\label{fig:020-035-100}
\end{figure*}
\begin{figure}
\centering
\includegraphics[width=\columnwidth]{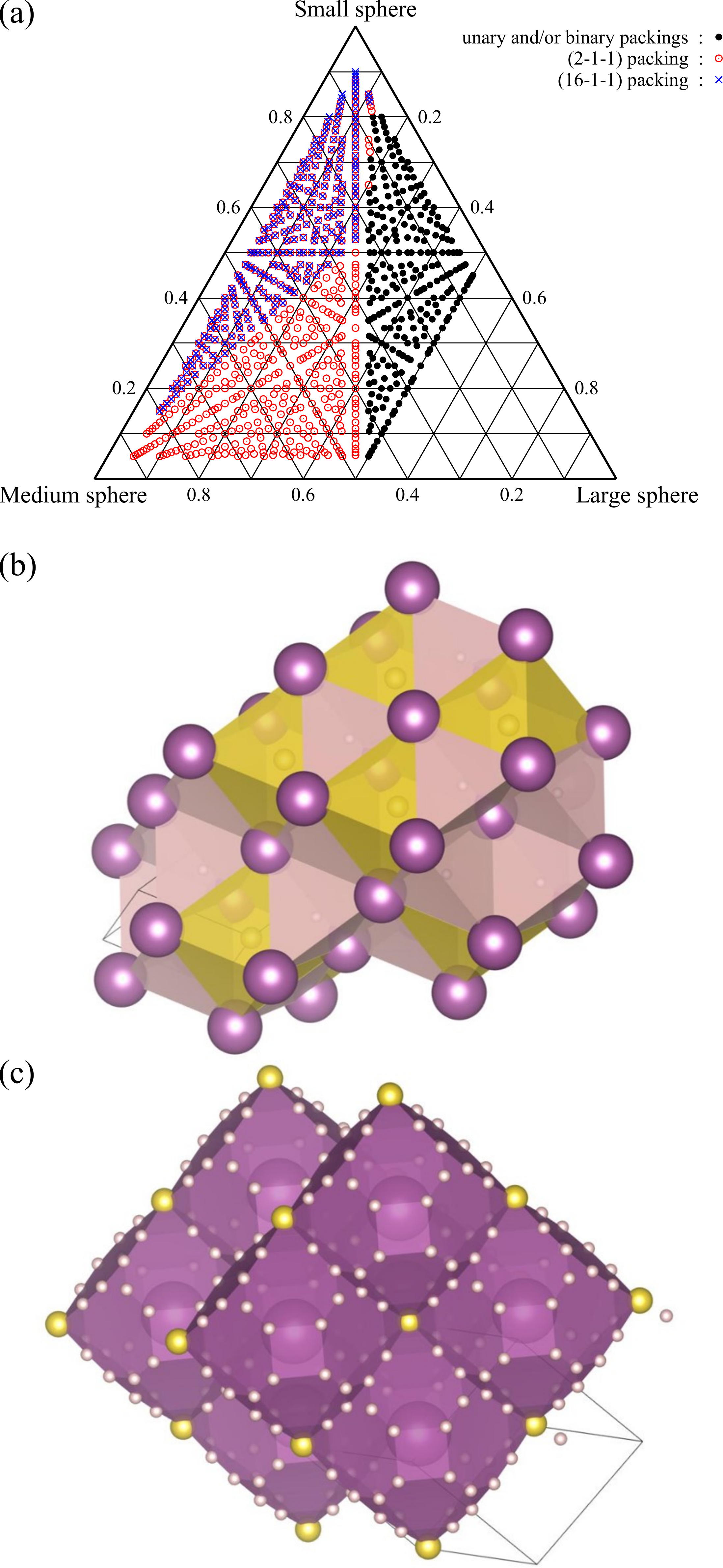}
\caption{The results at the radius ratio of $0.20:0.40:1.00$. (a) The phase diagram. (b) The FCCTC. (c) The (16-1-1) structure.}
\label{fig:020-040-100}
\end{figure}
\begin{figure*}
\centering
\includegraphics[width=2\columnwidth]{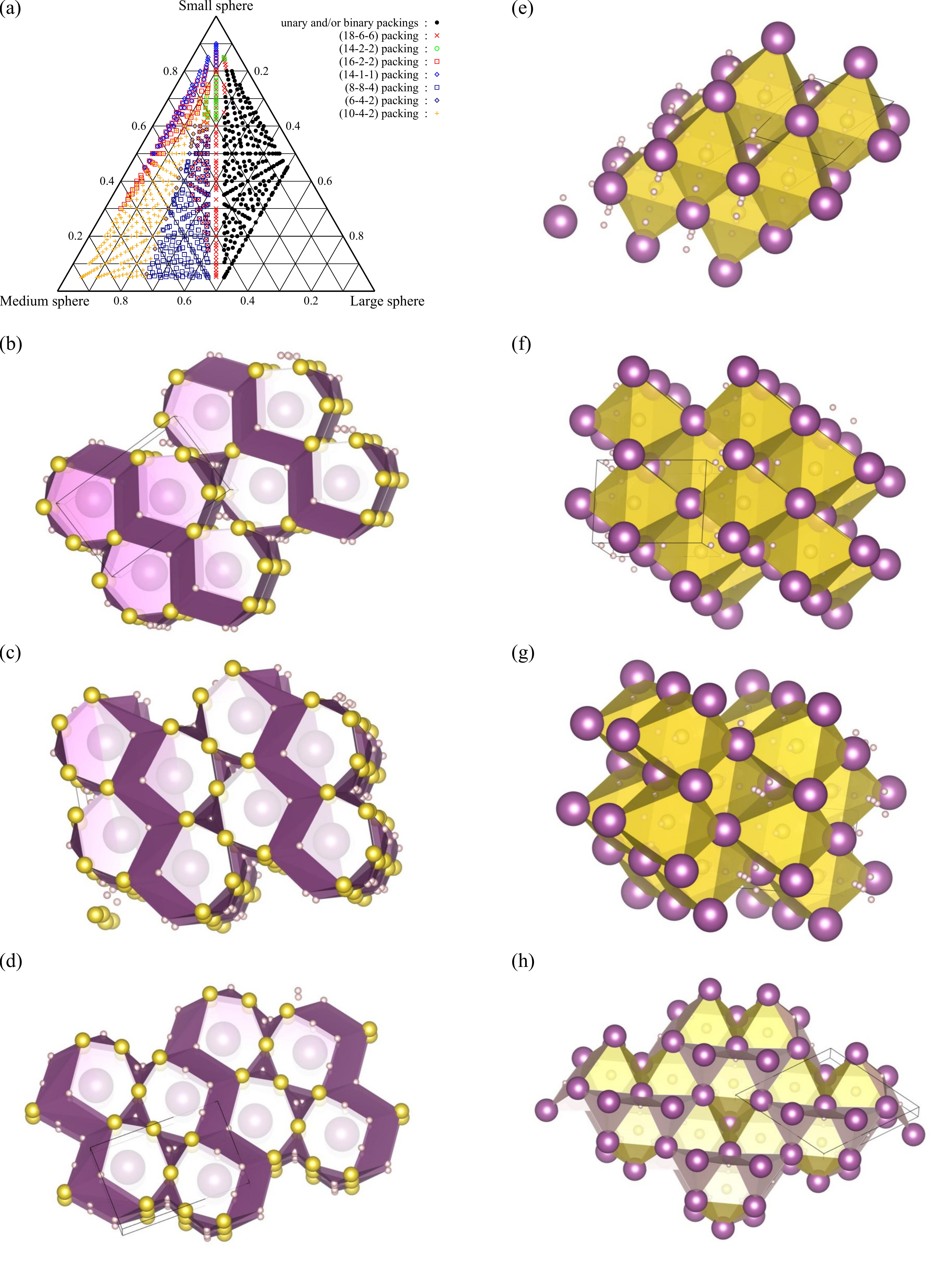}
\caption{The results at the radius ratio of $0.20:0.45:1.00$. (a) The phase diagram. (b) The (6-4-2) structure. (c) The (8-8-4) structure. (d) The (10-4-2) structure. (e) The (14-1-1) structure. (f) The (14-2-2) structure. (g) The (16-2-2) structure. (h) The (18-6-6) structure.}
\label{fig:020-045-100}
\end{figure*}
\begin{figure}
\centering
\includegraphics[width=\columnwidth]{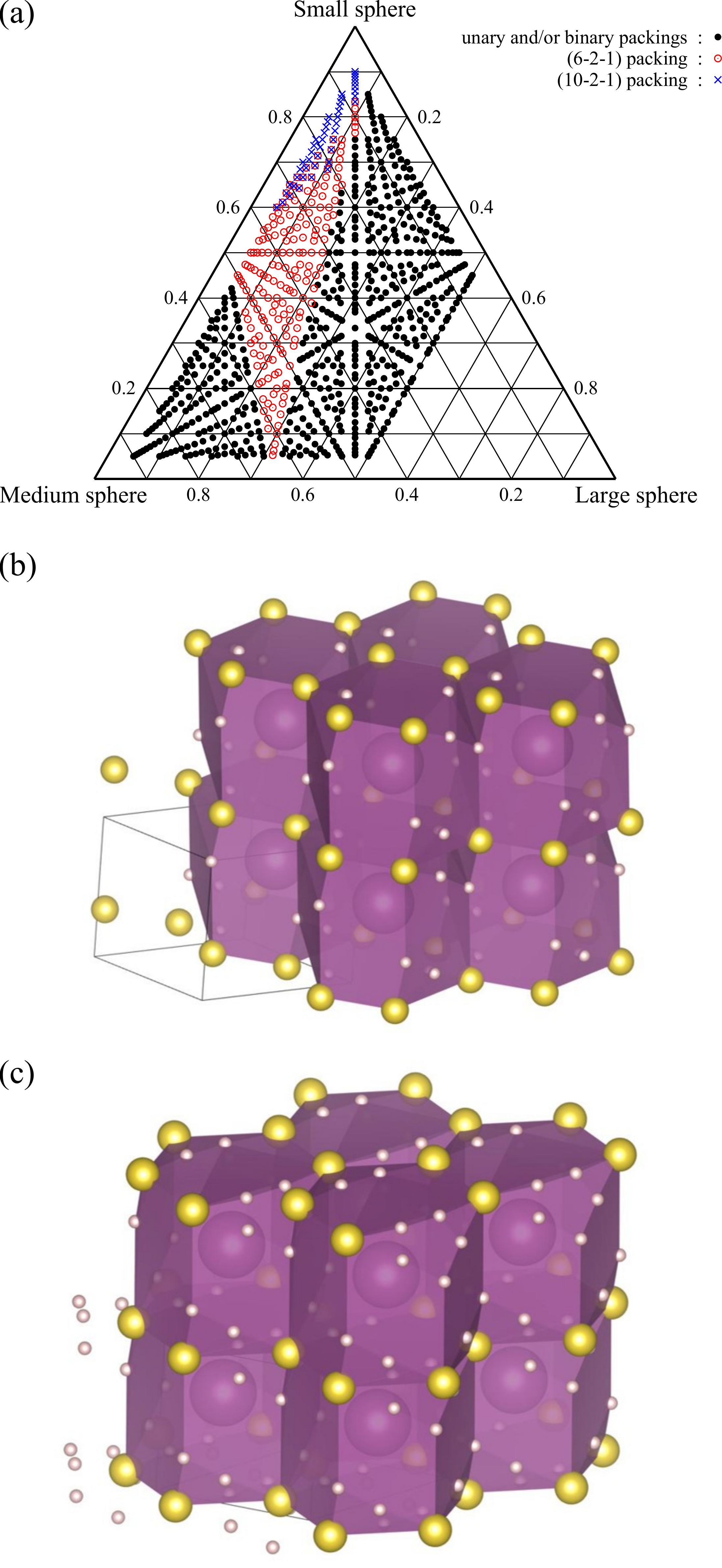}
\caption{The results at the radius ratio of $0.20:0.50:1.00$. (a) The phase diagram. (b) The (6-2-1) structure. (c) The (10-2-1) structure.}
\label{fig:020-050-100}
\end{figure}
\begin{figure}
\centering
\includegraphics[width=0.98\columnwidth]{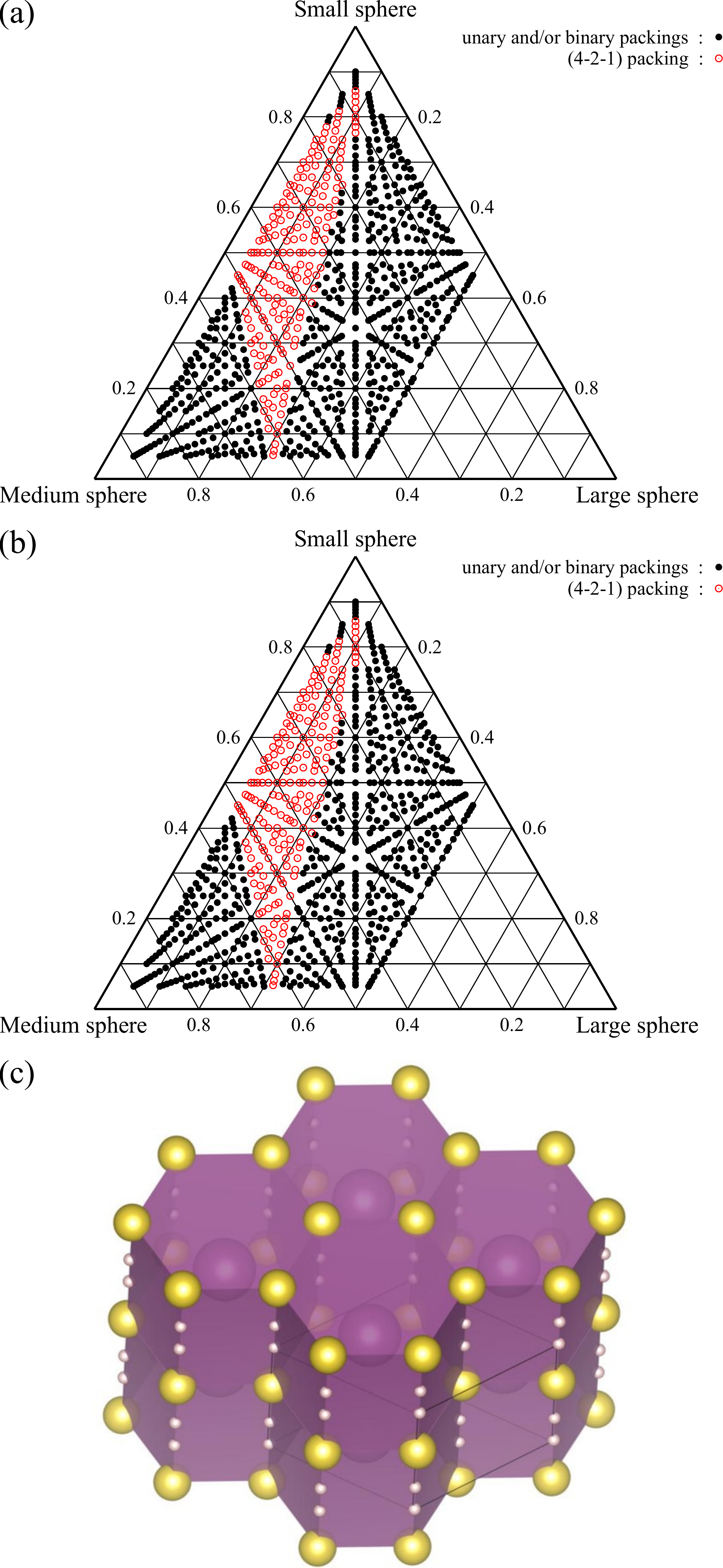}
\caption{The results at the radius ratio of $0.20:0.55:1.00$ and $0.20:0.60:1.00$. (a) The phase diagram at $0.20:0.55:1.00$. (b) The phase diagram at $0.20:0.60:1.00$. (c) The (4-2-1) structure.}
\label{fig:020-055060-100}
\end{figure}
\begin{figure}
\centering
\includegraphics[width=\columnwidth]{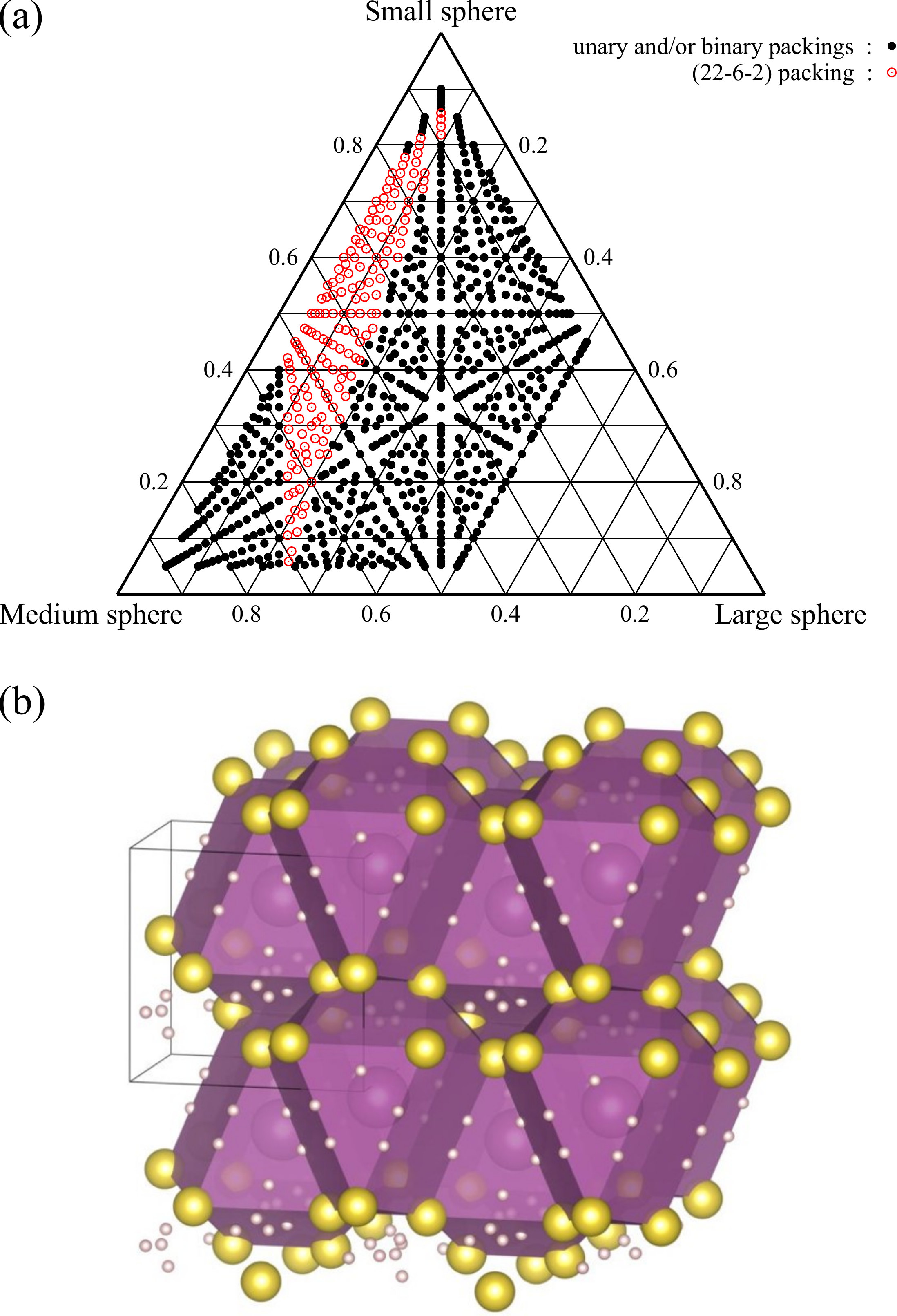}
\caption{The results at the radius ratio of $0.20:0.65:1.00$. (a) The phase diagram. (b) The (22-6-2) structure.}
\label{fig:020-065-100}
\end{figure}
\begin{figure}
\centering
\includegraphics[width=\columnwidth]{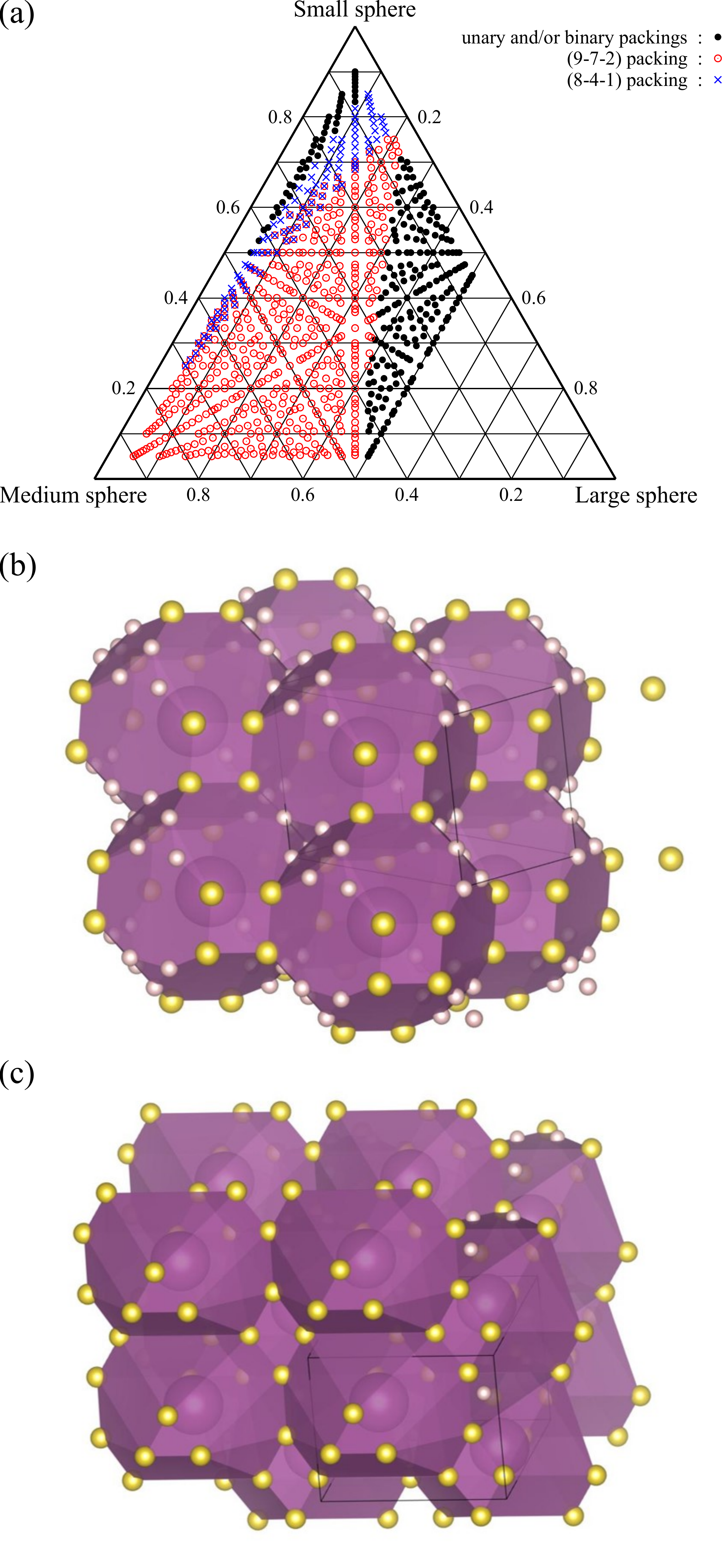}
\caption{The results at the radius ratio of $0.25:0.35:1.00$. (a) The phase diagram. (b) The (8-4-1) structure. (c) The (9-7-2) structure.}
\label{fig:025-035-100}
\end{figure}
\begin{figure}
\centering
\includegraphics[width=\columnwidth]{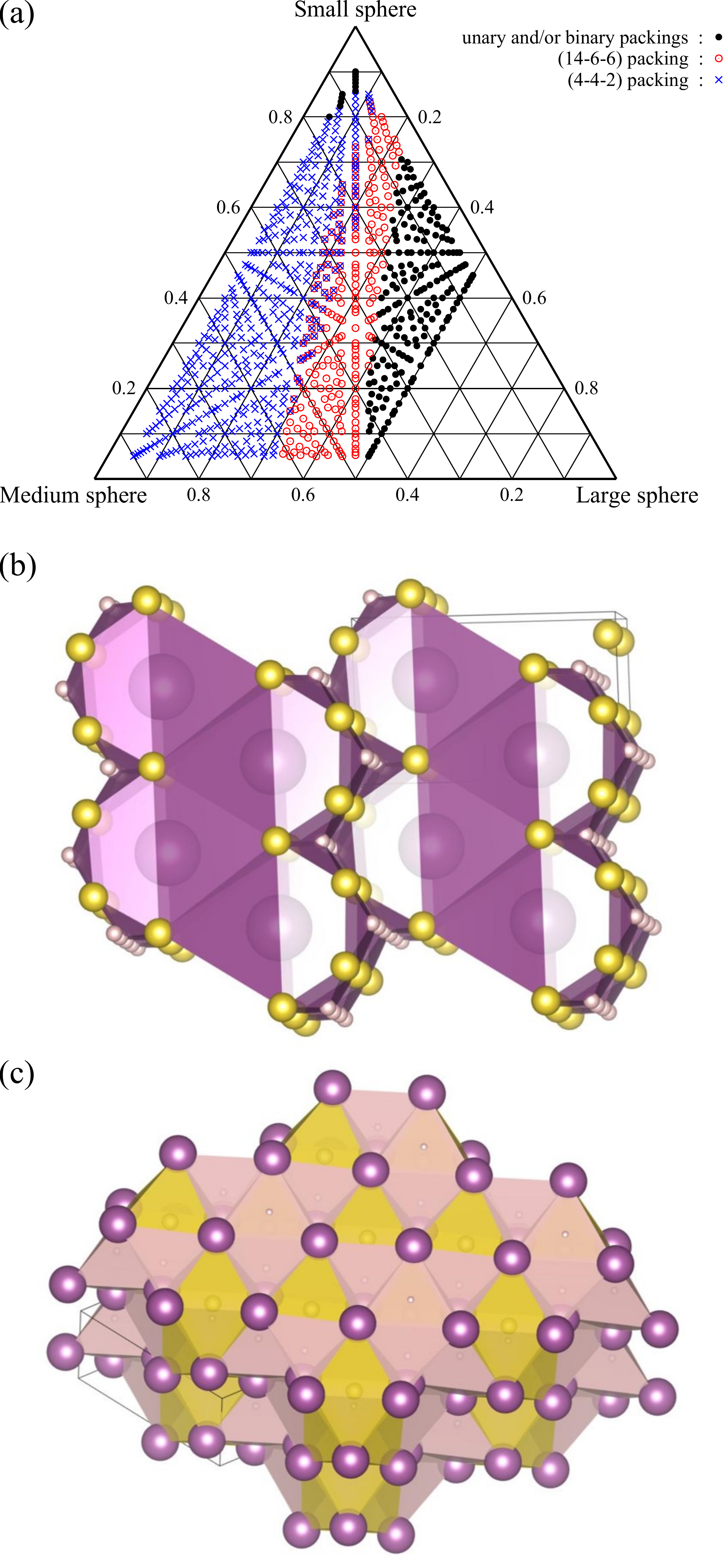}
\caption{The results at the radius ratio of $0.25:0.45:1.00$. (a) The phase diagram. (b) The (4-4-2) structure. (c) The (13-6-6) structure.}
\label{fig:025-045-100}
\end{figure}
\begin{figure}
\centering
\includegraphics[width=\columnwidth]{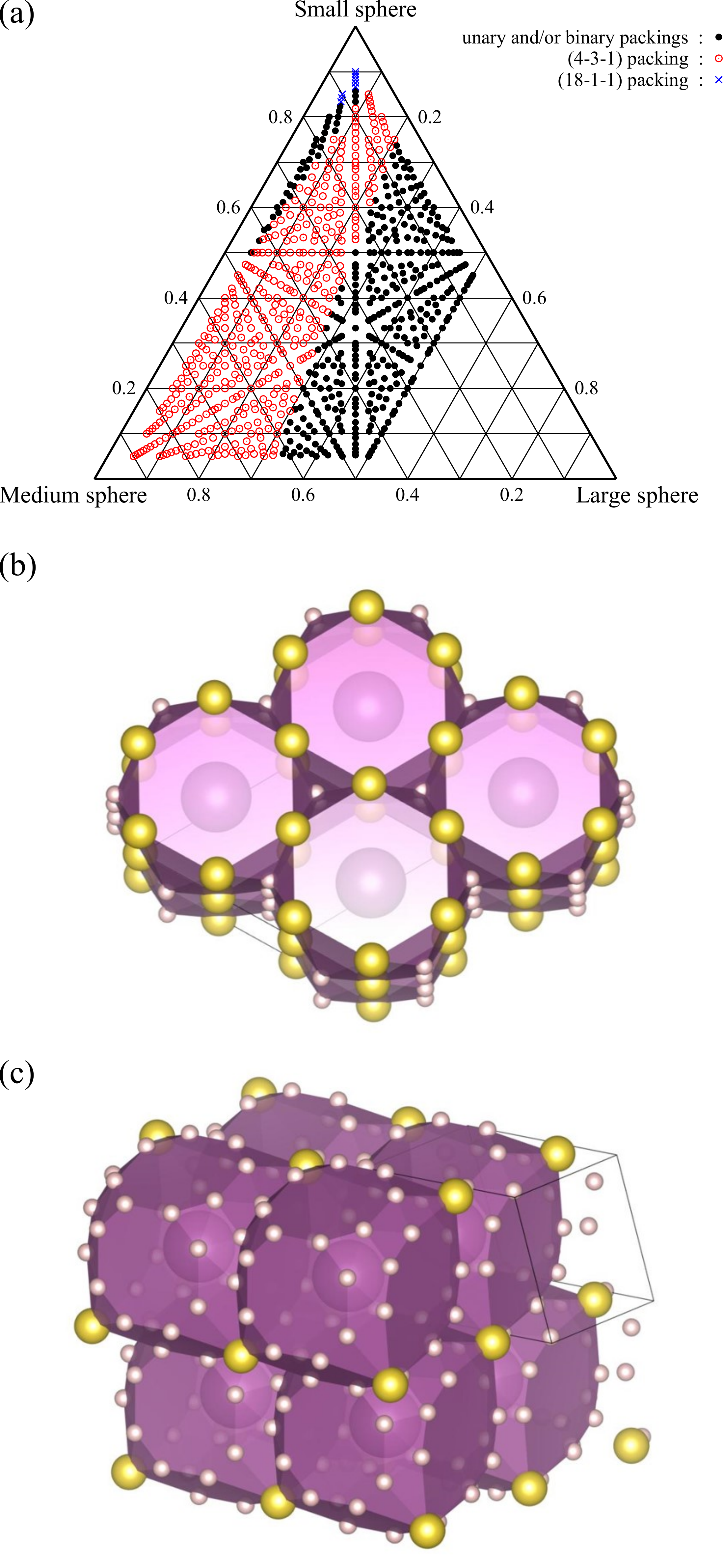}
\caption{The results at the radius ratio of $0.25:0.50:1.00$. (a) The phase diagram. (b) The (4-3-1) structure. (c) The (18-1-1) structure.}
\label{fig:025-050-100}
\end{figure}
\begin{figure}
\centering
\includegraphics[width=\columnwidth]{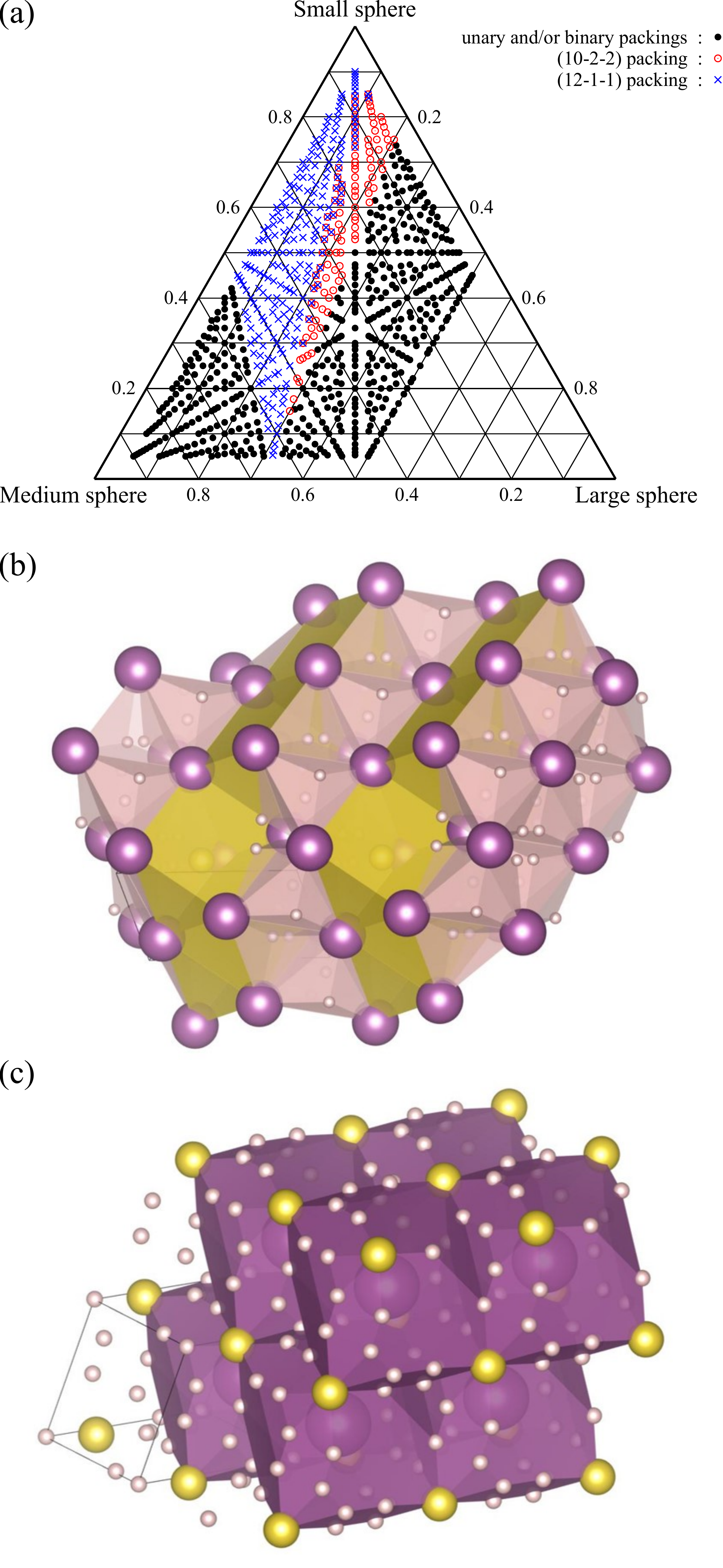}
\caption{The results at the radius ratio of $0.25:0.55:1.00$. (a) The phase diagram. (b) The (10-2-2) structure. (c) The (12-1-1) structure.}
\label{fig:025-055-100}
\end{figure}
\begin{figure}
\centering
\includegraphics[width=\columnwidth]{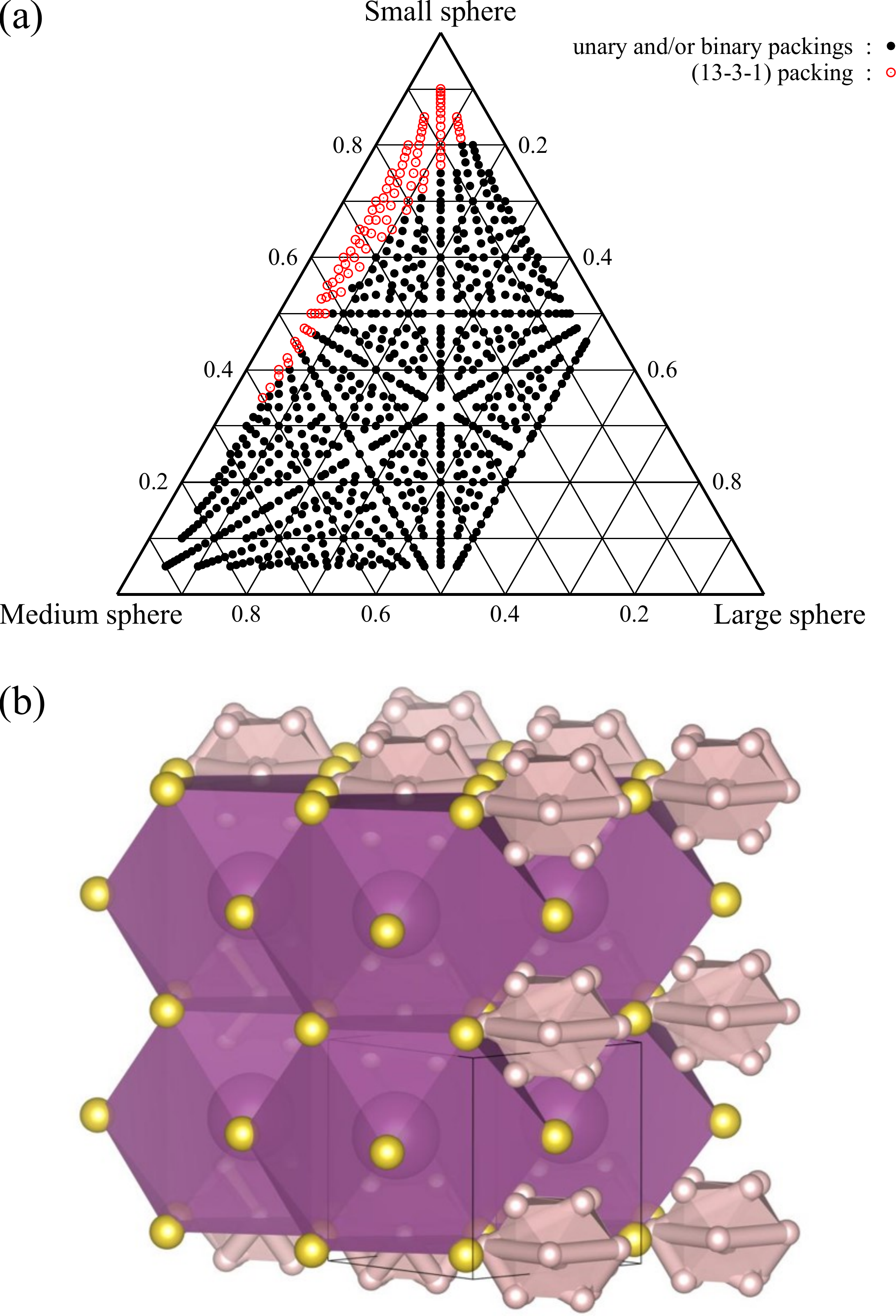}
\caption{The results at the radius ratio of $0.30:0.40:1.00$. (a) The phase diagram. (b) The (13-3-1) structure.}
\label{fig:030-040-100}
\end{figure}
\begin{figure*}
\centering
\includegraphics[width=2\columnwidth]{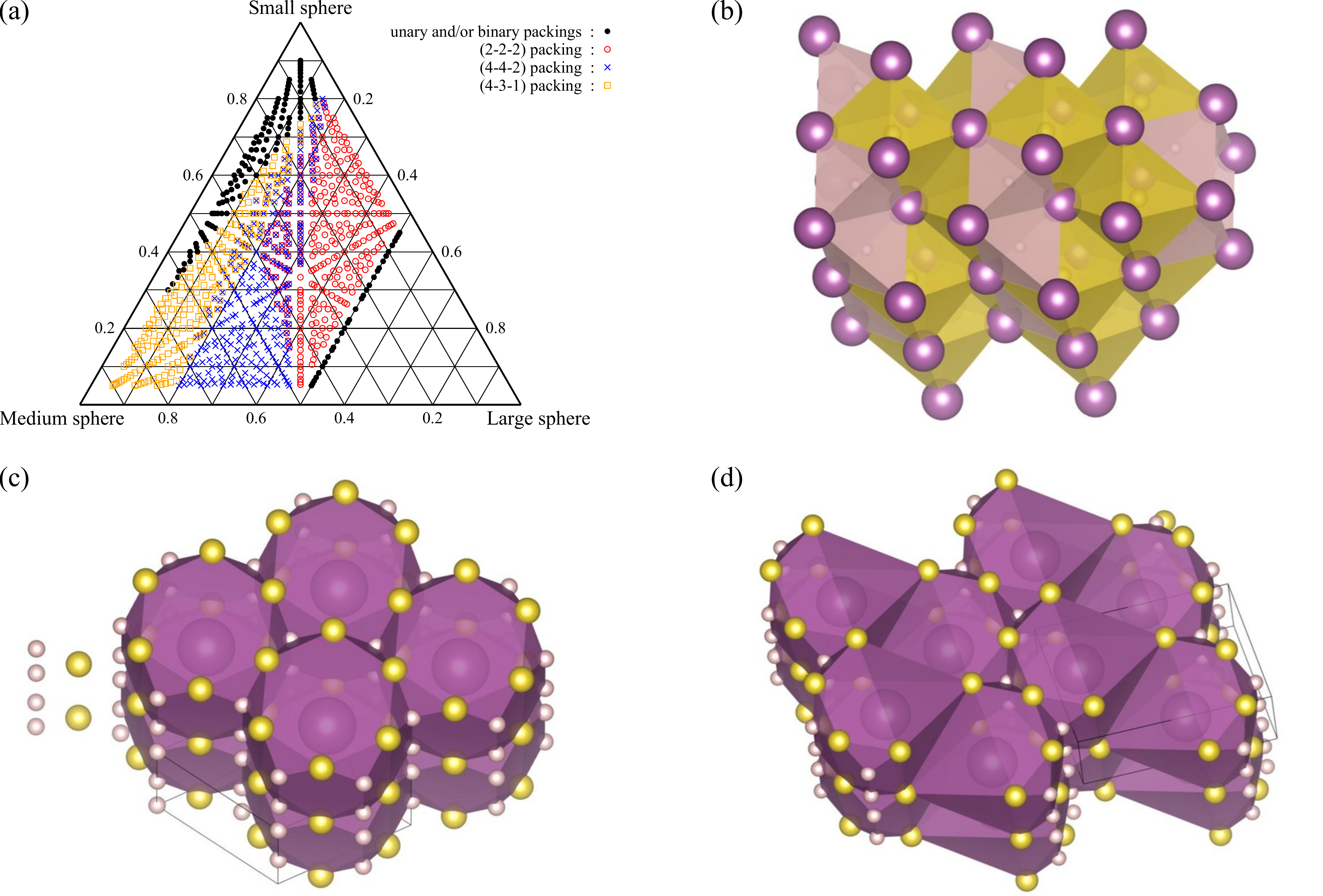}
\caption{The results at the radius ratio of $0.30:0.45:1.00$. (a) The phase diagram. (b) The (2-2-2) structure. (c) The (4-3-1) structure. (d) The (4-4-2) structure.}
\label{fig:030-045-100}
\end{figure*}
\begin{figure}
\centering
\includegraphics[width=\columnwidth]{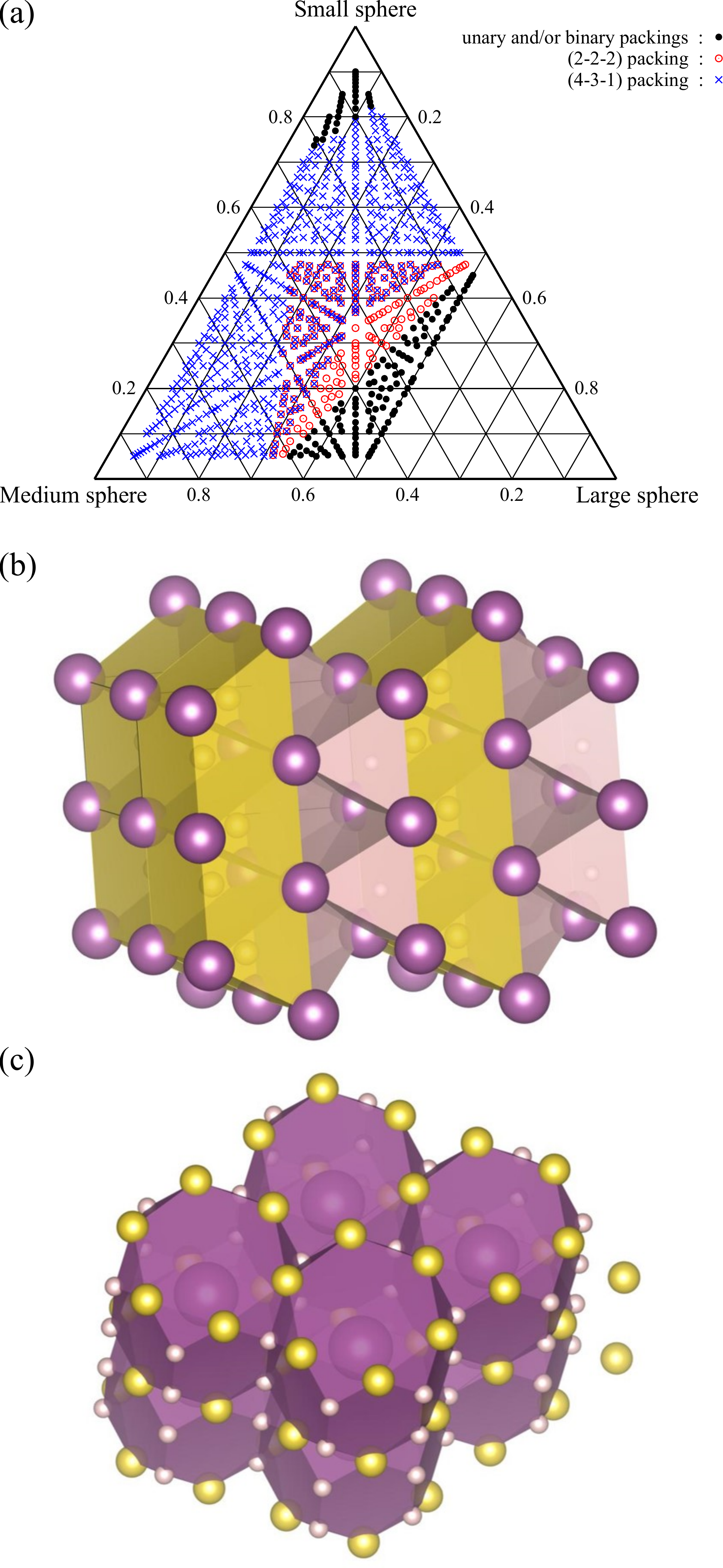}
\caption{The results at the radius ratio of $0.30:0.50:1.00$. (a) The phase diagram. (b) The (2-2-2) structure. (c) The (4-3-1) structure.}
\label{fig:030-050-100}
\end{figure}
\begin{figure}
\centering
\includegraphics[width=\columnwidth]{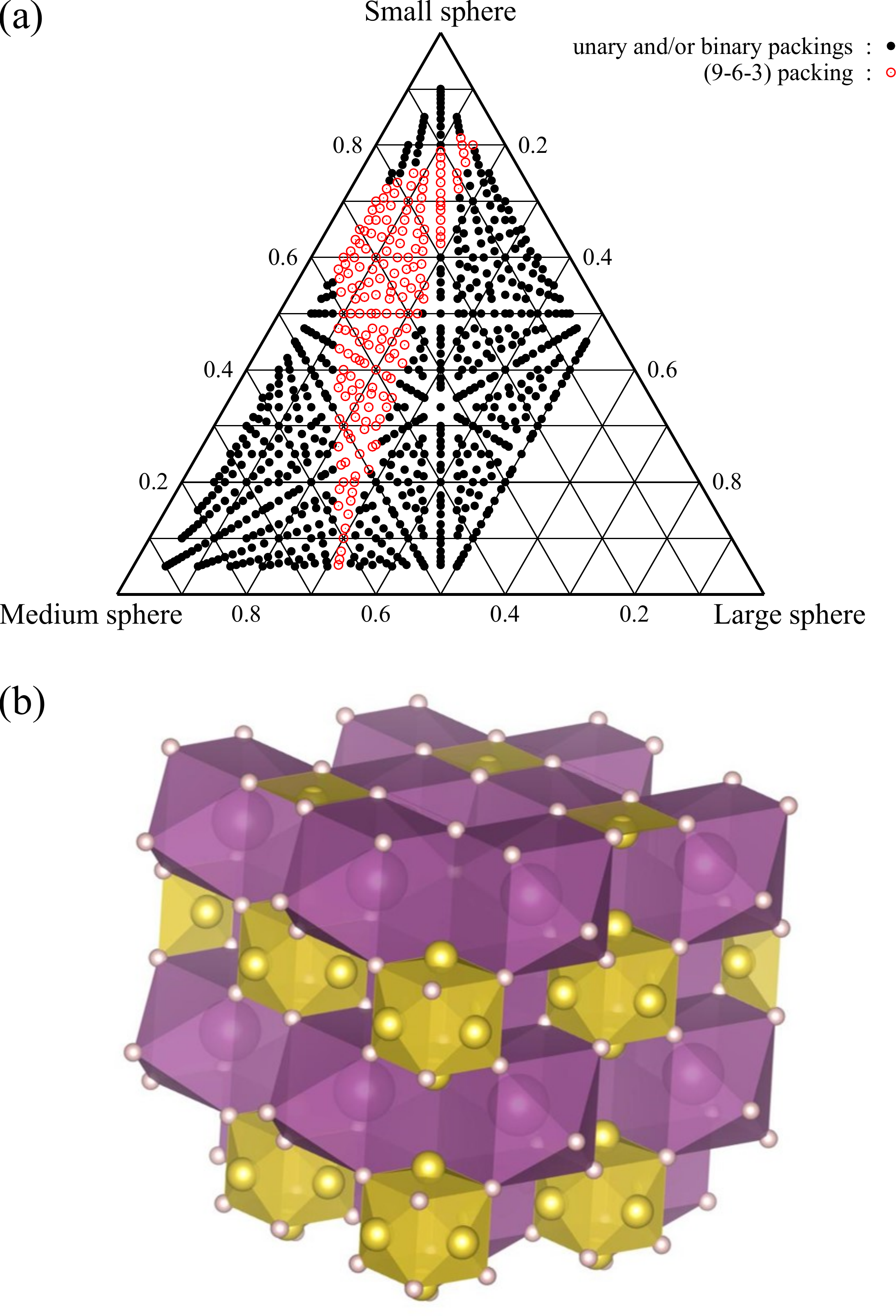}
\caption{The results at the radius ratio of $0.30:0.55:1.00$. (a) The phase diagram. (b) The (9-6-3) structure.}
\label{fig:030-055-100}
\end{figure}
\begin{figure}
\centering
\includegraphics[width=\columnwidth]{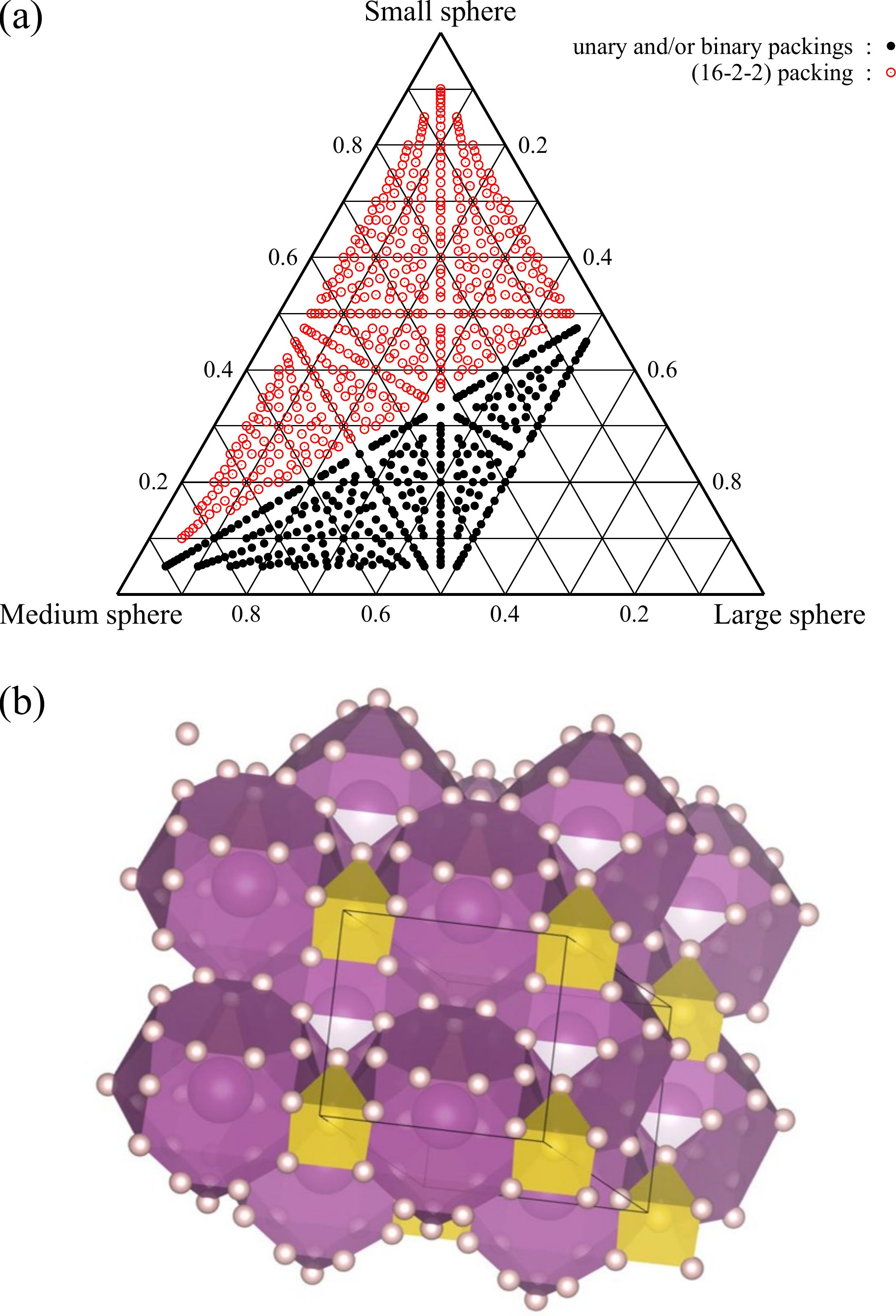}
\caption{The results at the radius ratio of $0.35:0.45:1.00$. (a) The phase diagram. (b) The (16-2-2) structure.}
\label{fig:035-045-100}
\end{figure}
\begin{figure}
\centering
\includegraphics[width=\columnwidth]{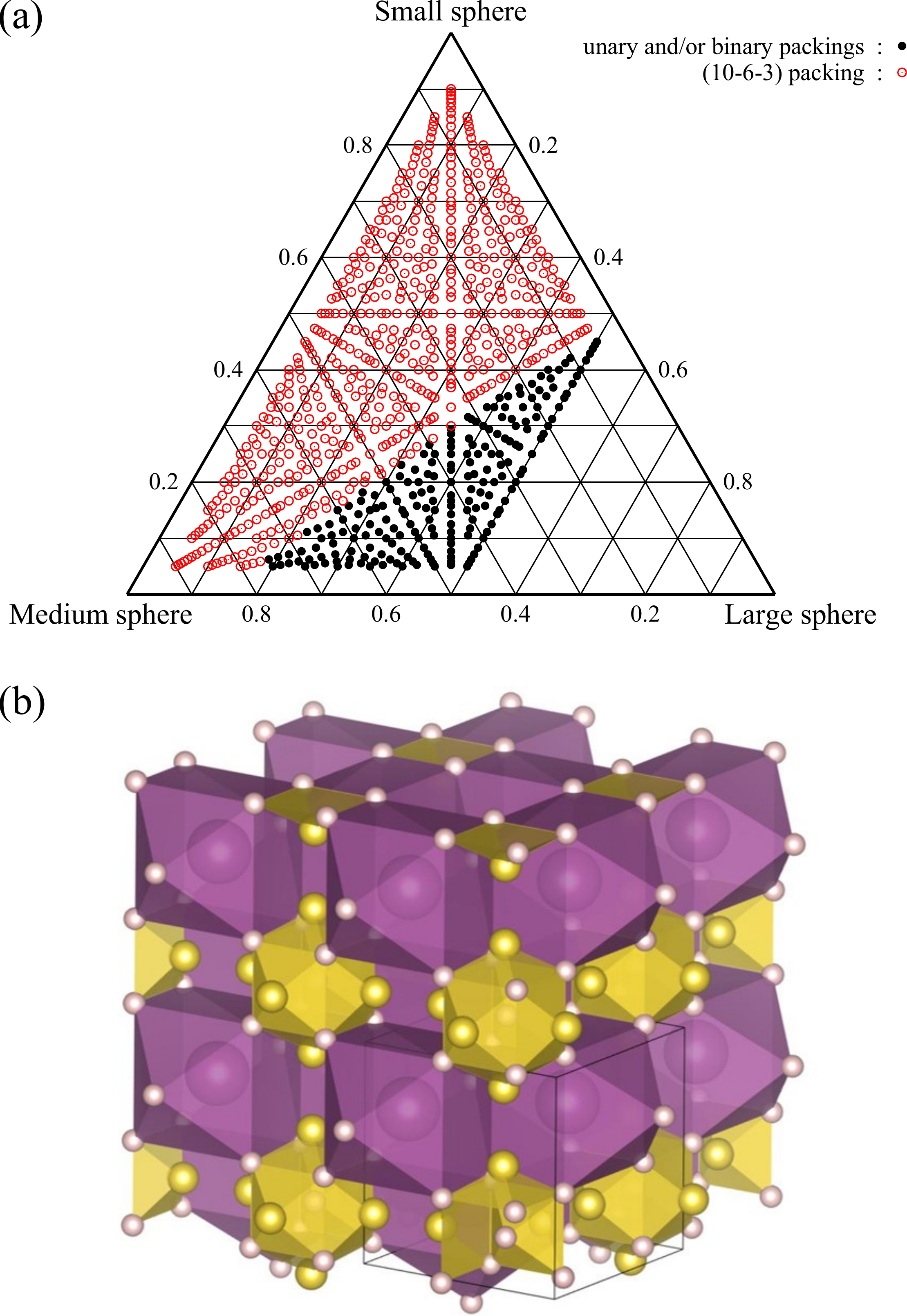}
\caption{The results at the radius ratio of $0.35:0.50:1.00$. (a) The phase diagram. (b) The (10-6-3) structure.}
\label{fig:035-050-100}
\end{figure}
\begin{figure}
\centering
\includegraphics[width=\columnwidth]{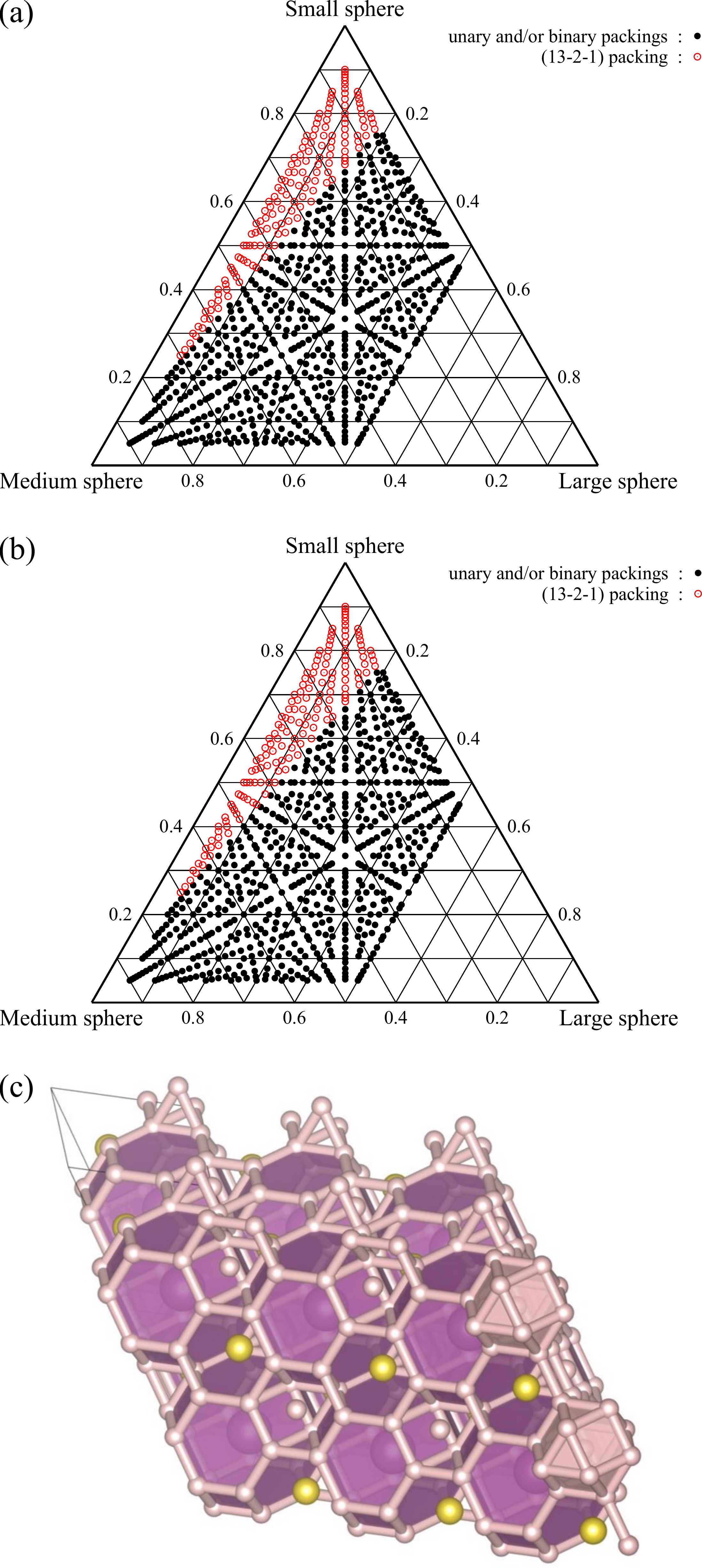}
\caption{The results at the radius ratios of $0.45:0.60:1.00$ and $0.45:0.65:1.00$. (a) The phase diagram at $0.45:0.60:1.00$. (b) The phase diagram at $0.45:0.65:1.00$. (c) The (13-2-1) structure.}
\label{fig:045-060065-100}
\end{figure}
\begin{figure}
\centering
\includegraphics[width=\columnwidth]{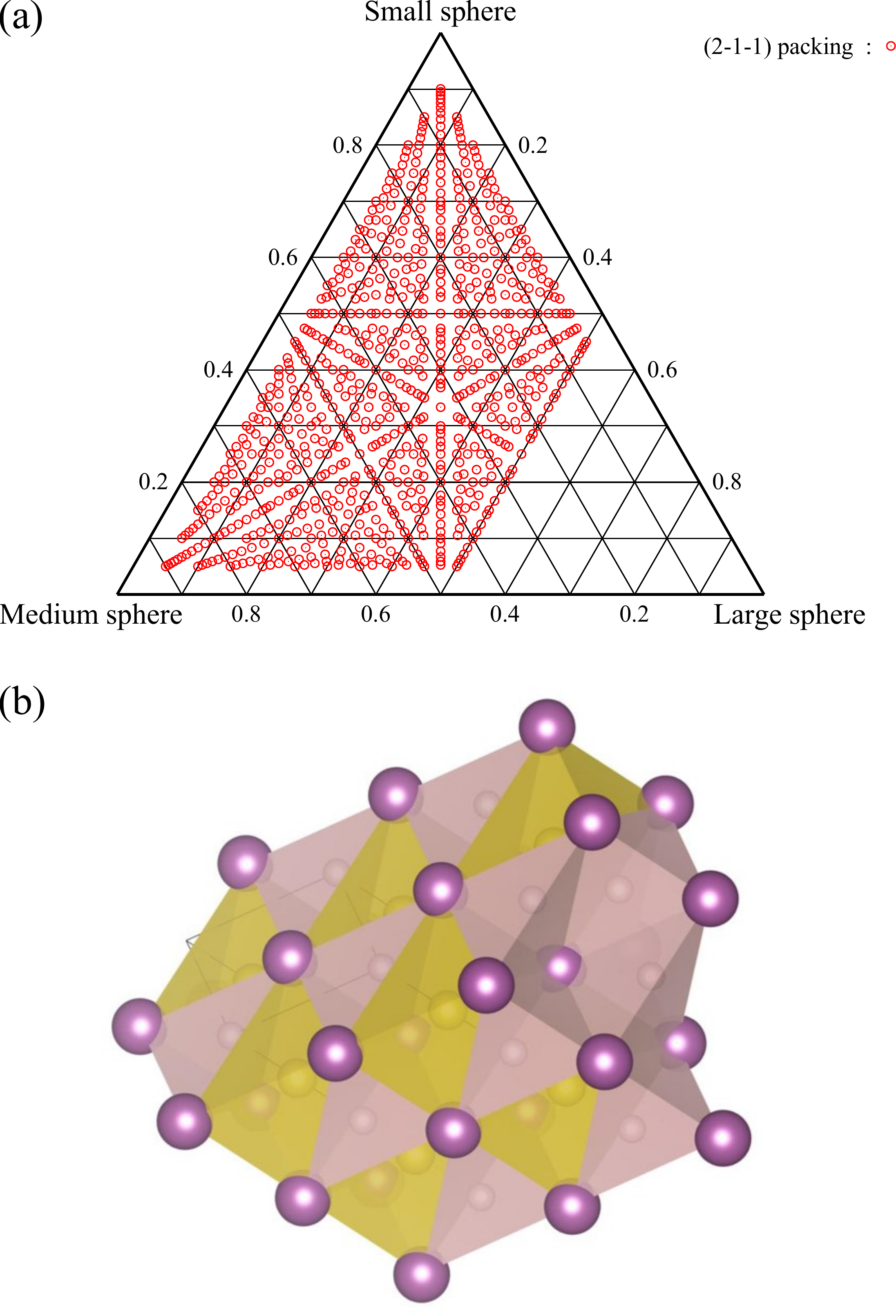}
\caption{The results at the radius ratio of $0.50:0.70:1.00$. (a) The phase diagram. (b) The (2-1-1) structure.}
\label{fig:050-070-100}
\end{figure}

\subsection{Phase diagrams and discovered DTSPs}
\label{sec:phase_diagrams_and_discovered_dtsps}

In this subsection, we show the phase diagrams for ternary systems and the discovered DTSPs.

At the radius ratio of $0.20:0.30:1.00$, we have discovered five putative DTSPs: the (7-4-1) structure, the $\mathrm{XYZ}_8$ structure, the (10-1-1) structure, the (10-4-1) structure, and the (18-2-1) structure, as shown in Fig.~\ref{fig:020-030-100}. The packing fractions are $0.811218$, $0.807864$, $0.814884$, $0.815904$, and $0.823226$, respectively. In the $\mathrm{XYZ}_8$, an octahedral site is occupied by one medium and six small spheres, and a tetrahedral site is occupied by one small sphere. The (10-1-1) structure is similar to the $\mathrm{XYZ}_8$, but large spheres constitute an expanded FCC structure. An expanded octahedral site is occupied by one medium and eight small spheres. Likewise, in the (7-4-1) structure, large spheres constitute a more expanded FCC structure, and an expanded octahedral site is occupied by one small and four medium spheres. The four medium spheres constitute a quadrilateral, and the small spheres are placed near the edge of the octahedron. An expanded tetrahedral site is occupied by three small spheres. The (10-4-1) structure is similar to the (7-4-1) structure, but an expanded octahedral site is occupied by only four medium spheres, and an expanded tetrahedral site is occupied by one small sphere. However, in addition to the arrangement, one small sphere is placed at every plane of the tetrahedrons. Finally, the (18-2-1) structure can be regarded as a clathrate structure.

At the radius ratio of $0.20:0.35:1.00$, we have discovered five putative DTSPs: the $\mathrm{XYZ}_4$ structure, the (8-1-1) structure, the (12-1-1) structure, the (12-2-2) structure, and the (18-1-1) structure, as shown in Fig.~\ref{fig:020-035-100}. The packing fractions are $0.795924$, $0.807440$, $0.810788$, $0.804429$, and $0.812846$, respectively. In the $\mathrm{XYZ}_4$, an octahedral site is occupied by one medium and two small spheres, and a tetrahedral site is occupied by one small sphere. The (8-1-1) structure is almost the same as the $\mathrm{XYZ}_8$, but large spheres constitute an expanded FCC structure, and an expanded octahedral site is occupied by one medium and six small spheres. Besides, both the (12-1-1) structure and the (12-2-2) structure can also be regarded as expanded FCC-type structures. Medium spheres are placed in expanded octahedral sites. The numbers of small spheres per expanded octahedral site are ten and four, respectively. The arrangements of large spheres of the (8-1-1), (12-1-1), and (12-2-2) structures are slightly different each other, and accordingly the volumes of their unit cells are also slightly different each other. Finally, the (18-1-1) structure can be regarded as a clathrate structure.

At the radius ratio of $0.20:0.40:1.00$, we have discovered one putative DTSP: the (16-1-1) structure, as shown in Fig.~\ref{fig:020-040-100}. The packing fraction is $0.815179$. The FCCTC~\cite{doi:10.1063/1.4941262} also appears on the phase diagram as DTSP, and the packing fraction is $0.799719$. As discussed in Wang \textit{et al.}~\cite{doi:10.1063/1.4941262}, an octahedral site is occupied by one medium sphere, and a tetrahedral site is occupied by one small sphere. On the other hand, the (16-1-1) structure can be regarded as a clathrate structure.

At the radius ratio of $0.20:0.45:1.00$, we have discovered seven putative DTSPs: the (6-4-2), (8-8-4), (10-4-2), (14-1-1), (14-2-2), (16-2-2), and (18-6-6) structures, as shown in Fig.~\ref{fig:020-045-100}. The packing fractions are $0.774288$, $0.769156$, $0.784534$, $0.813881$, $0.794755$, $0.798402$, and $0.773209$, respectively. The (6-4-2), (8-8-4), and (10-4-2) structures can be derived from the $\mathrm{HgBr}_2$ structure~\cite{PhysRevE.85.021130,PhysRevE.103.023307} that is one of the DBSPs. Small spheres are placed in the voids of the $\mathrm{HgBr}_2$-type structure. The frameworks of those three structures are almost the same, but the volumes of the unit cells of the (6-4-2) and (10-4-2) structures are slightly different from each other. Note that we also found the (4-4-2) structure which has half as large unit cell as the (8-8-4) structure. Our result shows that the volume of the unit cell of the (4-4-2) structure is the same as that of the (6-4-2) structure, but slightly more than half as large as the volume of the (8-8-4) structure. On the other hands, the (14-1-1), (14-2-2), and (16-2-2) structures consist of an expanded FCC structures of large spheres, where small spheres are placed in the voids. The volumes of the unit cells are slightly different from each other. Finally, the (18-6-6) structure can be derived from the (6-6) structure~\cite{PhysRevE.85.021130,PhysRevE.103.023307} that is one of the DBSPs. In the (18-6-6) structure, medium spheres are placed in expanded octahedral sites, and small spheres are placed in the voids of the (6-6)-type structure.

At the radius ratio of $0.20:0.50:1.00$, we have discovered two putative DTSPs: the (6-2-1) and (10-2-1) structures, as shown in Fig.~\ref{fig:020-050-100}. The packing fractions are $0.787225$ and $0.806565$, respectively. Both the (6-2-1) and (10-2-1) structures can be derived from the $\mathrm{AlB}_2$ structure~\cite{PhysRevE.79.046714,PhysRevE.85.021130,PhysRevE.103.023307} that is one of the DBSPs. Small spheres are placed in the voids of a distorted $\mathrm{AlB}_2$ structure. The volume of unit cell of the (6-2-1) structure is slightly different from that of the (10-2-1) structure.

At the radius ratios of $0.20:0.55:1.00$ and $0.20:0.60:1.00$, we have discovered one putative DTSP: the (4-2-1) structure, as shown in Fig.~\ref{fig:020-055060-100}. The packing fractions at the two radius ratios are $0.791468$ and $0.774420$, respectively. The (4-2-1) structure can be derived from the $\mathrm{AlB}_2$ structure~\cite{PhysRevE.79.046714,PhysRevE.85.021130,PhysRevE.103.023307} that is one of the DBSPs. Two small spheres are placed between two medium spheres that constitute a hexagonal cylinder. Especially, at $0.20:0.60:1.00$, the hexagonal cylinder is regular.

At the radius ratio of $0.20:0.65:1.00$, we have discovered one putative DTSP: the (22-6-2) structure, as shown in Fig.~\ref{fig:020-065-100}. The packing fraction is $0.782561$. The (22-6-2) structure can be derived from the $\mathrm{A}_3 \mathrm{B}$ structure~\cite{doi:10.1021/jp206115p,PhysRevE.85.021130,PhysRevE.103.023307} that is one of the DBSPs. Small spheres are placed in voids constituted by medium and/or large spheres. The volume of unit cell is the same as that of $\mathrm{A}_3 \mathrm{B}$ structure.

At the radius ratio of $0.25:0.35:1.00$, we have discovered two putative DTSPs: the (8-4-1) and (9-7-2) structures, as shown in Fig.~\ref{fig:025-035-100}. The packing fractions are $0.788699$ and $0.783220$, respectively. The (8-4-1) structure can be regarded as a clathrate structure. On the other hand, in the (9-7-2) structure, large spheres are surrounded by medium spheres. Small spheres are placed in voids constituted by medium spheres.

At the radius ratio of $0.25:0.45:1.00$, we have discovered two putative DTSPs: the (4-4-2) and (14-6-6) structures, as shown in Fig.~\ref{fig:025-045-100}. The packing fractions are $0.778942$ and $0.781847$, respectively. The (4-4-2) structure can be derived from the $\mathrm{HgBr}_2$ structure~\cite{PhysRevE.85.021130,PhysRevE.103.023307} that is one of the DBSPs. The (4-4-2) structure is similar to the (8-8-4) structure which is one of the DTSPs at the radius ratio of $0.20:0.45:1.00$. In the (4-4-2) structure, medium spheres are placed on rectangular sides of the triangular prisms constituted by large spheres. Small spheres are placed in the small triangular prisms constituted by medium spheres. On the other hand, the (14-6-6) structure can be derived from the (6-6) structure~\cite{PhysRevE.85.021130,PhysRevE.103.023307} that is one of the DBSPs. In the (14-6-6) structure, 
an expanded tetrahedral site constituted by large spheres is occupied by one small sphere. Besides, the one small sphere is placed on the boundary surface between two octahedrons constituted by large spheres. There are six boundaries between octahedrons per unit cell, but the size of triangle is different, so only the two large boundaries are occupied by one small sphere, respectively.

At the radius ratio of $0.25:0.50:1.00$, we have discovered two putative DTSPs: the (4-3-1) and (18-1-1) structures, as shown in Fig.~\ref{fig:025-050-100}. The packing fractions are $0.777358$ and $0.790979$, respectively. In the (4-3-1) structure, medium spheres constitute a kagome lattice. Large spheres are placed in the hexagonal prisms constituted by medium spheres. Small spheres are placed in the triangular prisms constituted by medium spheres.  On the other hand, the (18-1-1) structure can be regarded as a clathrate structure.
Although the (18-1-1) structure also appears at the radius ratio of $0.20:0.35:1.00$ and the structure at $0.20:0.35:1.00$ is 
slightly different from that at $0.25:0.50:1.00$, we regard that they are basically the same type of structure by taking account of
a large fluctuation of the sphere positions in the clathrate structure.

At the radius ratio of $0.25:0.55:1.00$, we have discovered two putative DTSPs: the (10-2-2) and (12-1-1) structures, as shown in Fig.~\ref{fig:025-055-100}. The packing fractions are $0.788895$ and $0.796895$, respectively. In the (10-2-2) structure, there are three polyhedra constituted by large spheres: tetrahedron, square pyramid, and triangular prism. The tetrahedrons are occupied by one small sphere. The square pyramids are occupied by four small spheres. The triangular prisms are occupied by one medium sphere. On the other hand, the (12-1-1) structure can be regarded as a clathrate structure. 
Although the (12-1-1) structure also appears at the radius ratio of $0.20:0.35:1.00$ and the structure at $0.20:0.35:1.00$ is 
slightly different from that at $0.25:0.55:1.00$, we regard that they are basically the same type of structure by taking account of
a large fluctuation of the sphere positions in the clathrate structure.

At the radius ratio of $0.30:0.40:1.00$, we have discovered one putative DTSP: the (13-3-1) structure, as shown in Fig.~\ref{fig:030-040-100}. The packing fraction is $0.793502$. The unit cell of the (13-3-1) structure is cubic. This structure can be derived from a perovskite structure $\mathrm{ABO}_3$. In the (13-3-1) structure, $\mathrm{A}$ or $\mathrm{B}$ site is occupied by a cluster structure constituted by 13 small spheres. The 12 of them constitute a distorted cuboctahedron and the other one is placed in the center of the cuboctahedron.

At the radius ratio of $0.30:0.45:1.00$, we have discovered three putative DTSPs: the (2-2-2), (4-3-1), and (4-4-2) structures, as shown in Fig.~\ref{fig:030-045-100}. The packing fractions are $0.765616$, $0.784757$, and $0.775691$, respectively. In the (2-2-2) structure, large spheres constitute an expanded FCC structure. Medium spheres are placed in the distorted octahedral sites. Small spheres are also placed in the distorted tetrahedral sites, but only one of the two sites per unit cell is occupied by one small sphere. The (4-3-1) structure is the same as that at the radius ratio of $0.25:0.50:1.00$. The (4-4-2) structure is also the same as that at the radius ratio of $0.25:0.45:1.00$.

At the radius ratio of $0.30:0.50:1.00$, we have discovered two putative DTSPs: the (2-2-2) and (4-3-1) structures, as shown in Fig.~\ref{fig:030-050-100}. The packing fractions are $0.763813$ and $0.786907$, respectively. In the (2-2-2) structure, medium spheres are placed in the triangular prisms constituted by large spheres. Small spheres are placed in the square pyramid constituted by large spheres. The (4-3-1) structure is the same as that at the radius ratios of $0.25:0.50:1.00$ and $0.30:0.45:1.00$.

At the radius ratio of $0.30:0.55:1.00$, we have discovered one putative DTSP: the (9-6-3) structure. The packing fraction is $0.785141$, as shown in Fig.~\ref{fig:030-055-100}. The unit cell of the (9-6-3) structure is cubic. This structure can be derived from the FCC structure constituted by large spheres as follows. In the conventional cell of the FCC structure of large spheres, there are four large spheres. In the (9-6-3) structure, one of the four large spheres are replaced by a cluster structure that are constituted by small and medium spheres. In the cluster structure, eight small spheres constitute a distorted cubic, and six medium spheres are placed over the six rectangular sides of the distorted cubic; medium spheres constitute an octahedron. Besides, the octahedral site consisting of large spheres is occupied by one small sphere.

At the radius ratio of $0.35:0.45:1.00$, we have discovered one putative DTSP: the (16-2-2) structure, as shown in Fig.~\ref{fig:035-045-100}. The packing fraction is $0.769450$. The (16-2-2) structure can be regarded as a clathrate structure. In the (16-2-2) structure, large spheres constitute a distorted HCP structure without contact. Large spheres are surrounded by many small spheres. Medium spheres are placed in voids, and they are also surrounded by small spheres. 

At the radius ratio of $0.35:0.50:1.00$, we have discovered one putative DTSP: the (10-6-3) structure, as shown in Fig.~\ref{fig:035-050-100}. The packing fraction is $0.773180$. The unit cell of the (10-6-3) structure is cubic. This structure is almost the same as the (9-6-3) structure. The small difference is the existence of one small sphere in the cluster structure constituted by small and medium spheres.

At the radius ratios of $0.45:0.60:1.00$ and $0.45:0.65:1.00$, we have discovered one putative DTSP: the (13-2-1) structure, as shown in Fig.~\ref{fig:045-060065-100}. The packing fractions are $0.751416$ and $0.751922$, respectively. In the (13-2-1) structure, large spheres constitute the FCC structure without contacts, where the FCC structure is distorted at the radius ratio of $0.45:0.65:1.00$.
A medium sphere is surrounded by small spheres that constitute truncated tetrahedron. A large sphere is also surrounded by small spheres that constitute a truncated octahedron. Besides, a small sphere is surrounded by small spheres that constitute cuboctahedron.

At the radius ratio of $0.50:0.70:1.00$, we have discovered one putative DTSP: the (2-1-1) structure, as shown in Fig.~\ref{fig:050-070-100}. The packing fraction is $0.765177$.  In the (2-1-1) structure, large spheres constitute the expanded FCC structure. Medium spheres are placed in octahedral sites, while small spheres are placed in tetrahedral sites.

\subsection{Phase diagrams at the other radius ratios}
\label{phase_diagrams_at_the_other_radius_ratios}

\begin{figure}
\centering
\includegraphics[width=\columnwidth]{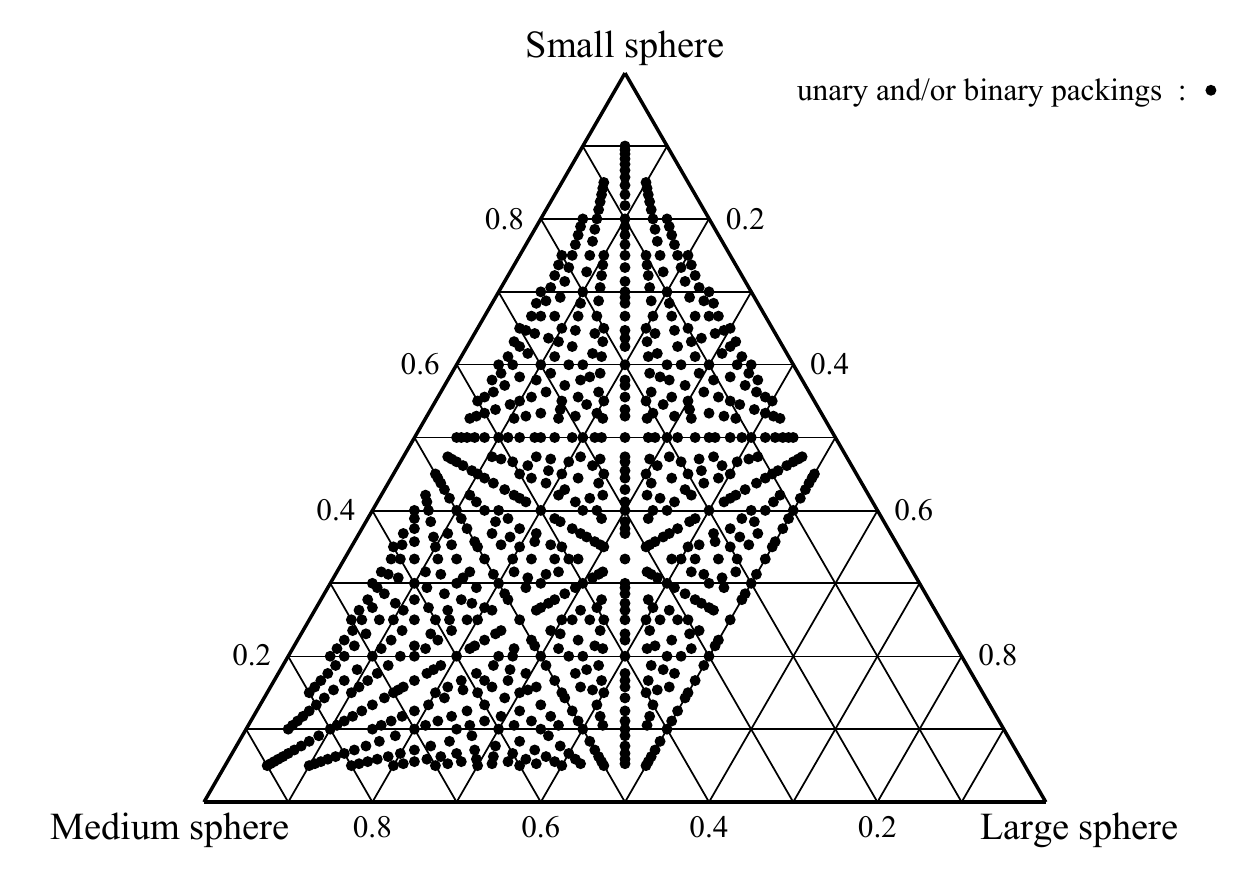}
\caption{The phase diagram at the radius ratios listed in Table~\ref{table:no-dtsps-radius-ratios}. At any composition ratios, densest phase separation consists of only the densest FCC structures and/or DBSPs.}
\label{fig:no-ternary-phaseDiagram}
\end{figure}

At the radius ratios listed in Table~\ref{table:no-dtsps-radius-ratios}, we have not found any DTSP. 
At any composition, the densest packing fractions are achieved by the phase separations consisting of 
only the FCC densest structures and/or DBSPs, as shown in Fig.~\ref{fig:no-ternary-phaseDiagram}.

\begin{table}
\caption{The radius ratios at which there is no DTSP on the corresponding phase diagrams. At the radius ratios shown in this table, densest phase separation consists of only the densest FCC structures and/or DBSPs at any composition ratios.}
\label{table:no-dtsps-radius-ratios}
\begin{ruledtabular}
\begin{tabular}{cccc}
&$0.20:0.70:1.00$ & $0.25:0.40:1.00$  &\\
&$0.25:0.60:1.00$ & $0.25:0.65:1.00$  &\\
&$0.25:0.70:1.00$ & $0.30:0.60:1.00$  &\\
&$0.30:0.65:1.00$ & $0.30:0.70:1.00$  &\\
&$0.35:0.55:1.00$ & $0.35:0.60:1.00$  &\\
&$0.35:0.65:1.00$ & $0.35:0.70:1.00$  &\\
&$0.40:0.50:1.00$ & $0.40:0.55:1.00$  &\\
&$0.40:0.60:1.00$ & $0.40:0.65:1.00$  &\\
&$0.40:0.70:1.00$ & $0.45:0.55:1.00$  &\\
&$0.45:0.70:1.00$ & $0.50:0.60:1.00$  &\\
&$0.50:0.65:1.00$ & $0.55:0.65:1.00$  &\\
&$0.55:0.70:1.00$ & $0.60:0.70:1.00$  & 
\end{tabular}
\end{ruledtabular}
\end{table}

\section{discussion}
\label{sec:discussion}

In this section, we discuss the geometric features of DTSPs, the correspondence between crystals and DTSPs, 
the discovered semi-DTSPs, and the effectiveness of our method.

\subsection{Geometric features of DTSPs with much smaller spheres}
\label{sec:geometric_features_of_dtsps_with_much_smaller_spheres}

When the radius ratio of small spheres is so small as about $0.20$ or $0.25$, many small spheres can be placed in voids of DBSPs constituted by medium and large spheres. In fact, our result suggests that many DTSPs with very small spheres can be derived from DBSPs. 
For example, the (10-1-1), (7-4-1), and (10-4-1) structures discovered at the radius ratio of $0.20:0.30:1.00$ can be derived from the $\mathrm{XY}$-type structure. However, the volumes of the unit cells are different from each other, because a distorted FCC structure constituted by large spheres is expanded with the increase of the number of small spheres. The expansion makes the phase diagrams complex because more structures get to appear on the phase diagrams. Not only at $0.20:0.30:1.00$ but also at $0.20:0.35:1.00$ and $0.20:0.40:1.00$, some DTSPs can be derived from the $\mathrm{XY}$-type structure, and the volumes of their unit cells are slightly different from each other.

Not only the $\mathrm{XY}$-type structure but also the $\mathrm{HgBr}_2$ structure derive some of DTSPs. 
The (6-4-2), (8-8-4), and (10-4-2) structures discovered at the radius ratio of $0.20:0.45:1.00$ can be derived from the $\mathrm{HgBr}_2$ structure. As discussed in Sec.~\ref{sec:phase_diagrams_and_discovered_dtsps}, the volumes of unit cells of the (6-4-2) and (10-4-2) structures are slightly different from each other. On the other hand, the volume of unit cell of the (8-8-4) structure is slightly less than twice as large as that of the (4-4-2) structure, where that of the (4-4-2) structure is the same as that of the (6-4-2) structure. As well as the $\mathrm{XY}$-type DTSPs, these slight differences of the unit cells also make the phase diagrams complex.

Furthermore, the (6-2-1), (10-2-1), and (4-2-1) structures can be derived from the $\mathrm{AlB}_2$ structure. The volume of the unit cell of the (6-2-1) structure is slightly different from that of the (10-2-1) structure. It is also noted that the (14-6-6) structure can be derived from the (6-6) structure.

The (22-6-2) structure, which is the DTSP at the radius ratio of $0.20:0.65:1.00$, can also be derived from the $\mathrm{A}_3 \mathrm{B}$ structure. Interestingly, as discussed in Sec.~\ref{sec:phase_diagrams_and_discovered_dtsps}, the volume of the unit cell of the (22-6-2) structure is the same as that of the $\mathrm{A}_3 \mathrm{B}$ structure. The result indicates two hypotheses. One is that more small spheres may be placed in the (22-6-2) structure without expanding the unit cell. The other is that another DTSP consisting more small spheres with an expanded unit cell may also appear on the phase diagram together with the (22-6-2)-type structure whose volume of the unit cell is the same as that of the $\mathrm{A}_3 \mathrm{B}$ structure, as well as the FCC-type DTSPs such as (10-1-1) structure. On the other hand, the (18-6-6) structure, which is the DTSP at the radius ratio of $0.20:0.45:1.00$, can be derived from the (6-6) structure. In the (18-6-6) structure, a few expanded tetrahedral sites are not occupied by small spheres. Therefore, we guess that more small spheres can be placed in the (18-6-6) structure.

Besides, some of DTSPs such as the (18-2-1) structure can be regarded as clathrate structures. If the radius of small spheres are very small, they do not disturb for medium and large spheres to form a network structure. That is why DTSPs prefer clathrate structures so as to increase packing fractions. However, interestingly, we have discovered a few DTSPs which can also be regarded as clathrate structures, while they consist of relatively large three kinds of spheres, as discussed in the next subsection.

In summary, when the small spheres are much smaller than large spheres, there are two kinds of DTSPs. One can be derived from the DBSPs, and the other can be regarded as clathrate structures. Very small spheres can be placed in small voids constituted by medium and/or large spheres, and that is why many DTSPs can be derived from the DBSPs. Since insertion of more small spheres can also generate dense structures by slightly expanding the unit cell of an original structure, many DTSPs appear on the phase diagrams despite the similarity of DTSPs among them at given radius ratios. In short, the smallness of radius ratio of small spheres makes the phase diagrams more complex. Especially, the phase diagram at the radius ratio of $0.20:0.45:1.00$ is the most complicated, because most of the putative DTSPs have similar structures each other which have almost the same unit cells, and besides the (18-6-6) may contain more small spheres.

\subsection{Geometric features of the other DTSPs}
\label{sec:geometric_features_of_interesting_dtsps}

As the radius of small spheres is getting large, geometric features of DTSPs are getting characteristic, 
which are specific to the ternary system. 

At the radius ratio of $0.30:0.40:1.00$, we have found the (13-3-1) structure. The structure can be derived from the perovskite structure, and the unit cell is cubic. The structure contains a cluster structure which is comprised by 13 small spheres. Similarly, at the radius ratios of $0.30:0.55:1.00$ and $0.35:0.50:1.00$, we have found the (9-6-3) and (10-6-3) structures, respectively. These structure can be derived from the the FCC structure constituted by large spheres, and one of large spheres per conventional cell is replaced by a cluster structure consisting of small and medium spheres. The unit cells of those structures are also cubic. These discoveries indicate that DTSPs prefer containing cluster structures that are constituted by small and/or medium spheres as the radius of small spheres is getting larger. Therefore, we guess that unknown DTSPs may consist of larger unit cells than those used in this study, because cluster structures are generally comprised by many small and/or medium spheres.

At the radius ratios of $0.45:0.60:1.00$ and $0.45:0.65:1.00$, we have discovered the (13-2-1) structure. As discussed in Sec.~\ref{sec:phase_diagrams_and_discovered_dtsps}, medium and large spheres are surrounded by small spheres, and besides the arrangements of small spheres surrounding medium and large spheres are the semi-regular polyhedrons: the truncated tetrahedron and the truncated octahedron, respectively. As discussed in Sec.~\ref{dtsps_and_crystal_structures}, if the slight distortion in the (13-2-1) structure is corrected, the structure has the $Fm\bar{3}m$ symmetry, where large spheres constitute the FCC structure without contact. Surprisingly, a medium sphere is placed in an expanded tetrahedral site despite the large radius ratio of the medium sphere. One small sphere is placed in an expanded octahedral site, and the small sphere is also surrounded by other small spheres. The surrounding structure is also the semi-regular polyhedrons: the cuboctahedron. Accordingly, we can regard the (13-2-1) structure as a highly well-ordered clathrate structure. Similarly, the (16-2-2) structure discovered at the radius ratio of $0.35:0.45:1.00$ can also be regarded as a clathrate structure. These discoveries indicate that even if the radius of small spheres is getting larger, DTSPs also prefer clathrate structures as well as the DTSPs discussed in the previous subsection, and besides some clathrate DTSPs may be ordered very well. Therefore, we also guess that unknown DTSPs may consist of larger unit cells than those used in this study, because many small spheres are necessary to constitute network structures consisting of them.

In addition, at the radius ratio of $0.30:0.50:1.00$, we have found the (2-2-2) structure. In this (2-2-2) structure, small and medium spheres are placed separately in the local structures constituted by large spheres, and the mixture of the different local structures contributes
to the increase of packing fraction in fact. 
In this sense, the combination of local structures is just so a manifestation in unique features of ternary structures. 
On the other hand, the (4-4-2) structure discovered at the radius ratio of $0.30:0.45:1.00$ can be derived from the $\mathrm{HgBr}_2$ structure as well as the (8-8-4) structure discussed in the previous subsection. As discussed in Sec.~\ref{sec:phase_diagrams_and_discovered_dtsps}, in the (4-4-2) structure, medium spheres are placed on rectangular sides of the triangular prisms constituted by large spheres, while small spheres are placed in the small triangular prisms constituted by medium spheres. The combination of local structures 
can also be regarded as a manifestation in unique features of ternary structures as well as the (2-2-2) structure.

Besides, at the radius ratio of $0.50:0.70:1.00$, we have discovered the (2-1-1) structure. As discussed in Sec.~\ref{sec:phase_diagrams_and_discovered_dtsps}, in the (2-1-1) structure, large spheres constitute the expanded FCC structure, and medium spheres are placed in expanded octahedral sites, while small spheres are placed in expanded tetrahedral sites. As discussed in Sec.~\ref{sec:phase_diagrams_and_discovered_dtsps}, when small and medium spheres are relatively small, some of DTSPs can be derived from the $\mathrm{XY}$-type structure, where the medium spheres are placed in octahedral sites. Interestingly, even if small and medium spheres are getting so large, an expanded $\mathrm{XY}$-type structure appears again on the phase diagram.

Finally, at the radius ratios of $0.25:0.50:1.00$, $0.30:0.45:1.00$, and $0.30:0.50:1.00$, we have discovered the (4-3-1) structure. If we exclude all the small spheres from the (4-3-1) structure, the densest packing fractions are achieved by transforming the remaining structure into the $\mathrm{AuTe}_2$ structure~\cite{PhysRevE.79.046714,PhysRevE.85.021130,PhysRevE.103.023307}. In the $\mathrm{AuTe}_2$ structure, medium spheres constitute distorted hexagons. However, it is found in the (4-3-1) structure that medium spheres comprise a kagome lattice, and besides, the comprised hexagons are regular. As discussed in Sec.~\ref{sec:phase_diagrams_and_discovered_dtsps}, small spheres are placed in the triangular prisms constituted by medium spheres. Interestingly, the existence of small spheres transform the $\mathrm{AuTe}_2$ structure into the well-ordered structure.

In summary, as the radius of small spheres is getting large, a lot of unique DTSPs appear on the phase diagrams. Especially, some of DTSPs contain cluster structures and the geometries are ordered well, and besides, a highly well-ordered clathrate structure is discovered, which consists of three kinds of relatively large spheres. The discovery of these two kinds of DTSPs implies that unknown and well-ordered DTSPs might be discovered if the maximum number of spheres per unit cell is set to be larger than those used in this study.

\subsection{DTSPs and Crystal structures}
\label{dtsps_and_crystal_structures}

\begin{figure}
\centering
\includegraphics[width=\columnwidth]{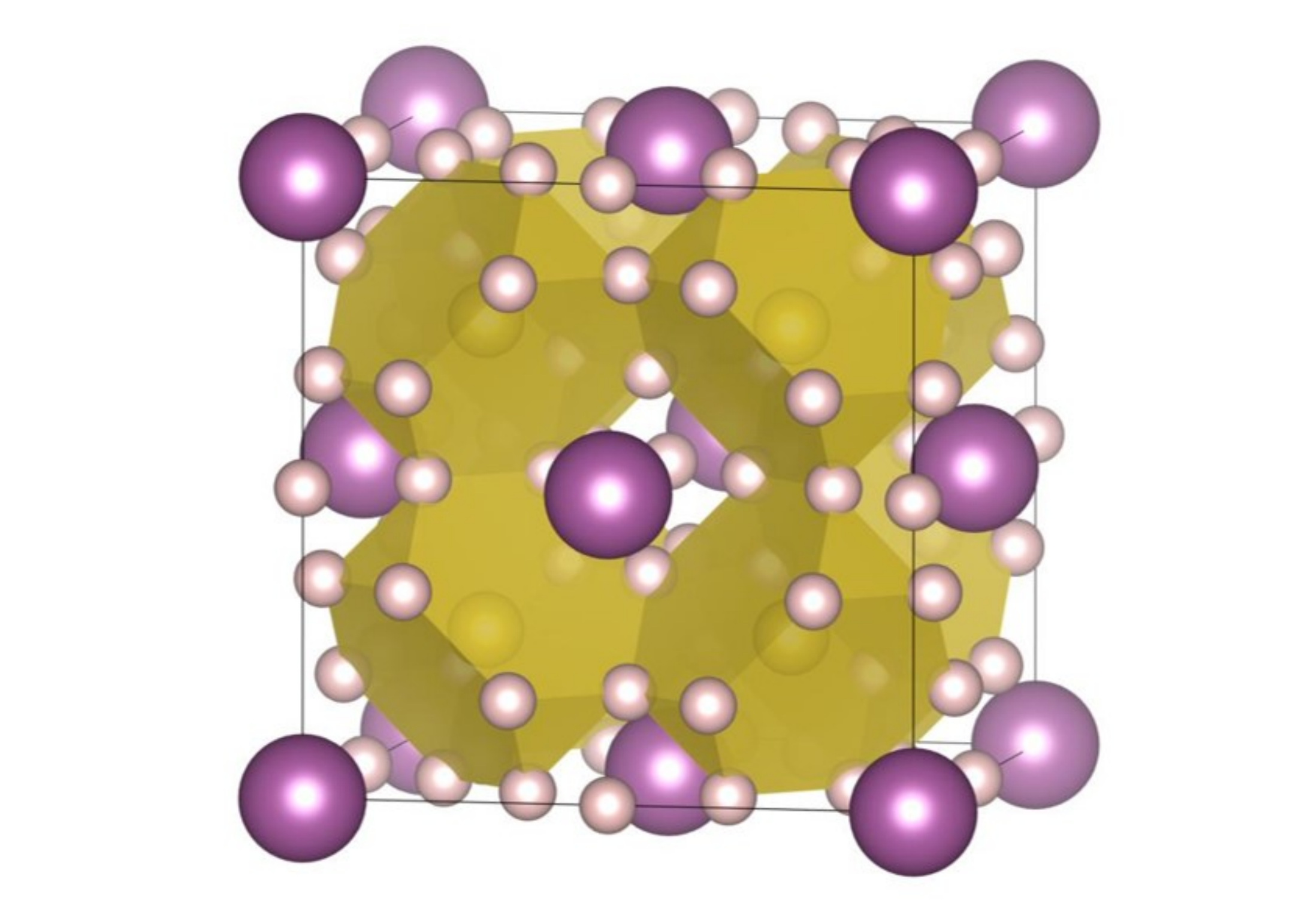}
\caption{The (13-2-1) structure after the slight distortion is corrected. The symmetrized (13-2-1) structure has the $Fm\bar{3}m$ symmetry, where large spheres constitute the FCC structure without contact. A medium sphere is placed in an expanded tetrahedral site. Medium and large spheres are surrounded by small spheres, and the surrounding structures are the semi-regular polyhedrons: the truncated tetrahedron and the truncated octahedron, respectively. Besides, one small sphere is placed on an expanded octahedral site, and it is also be surrounded by other small spheres. The surrounding structure is also the semi-regular polyhedron: the cuboctahedron.}
\label{fig:13-2-1}
\end{figure}

\begin{table*}
\caption{Correspondence between DTSPs and crystals}
\begin{ruledtabular}
\begin{tabular}{cccc}
Packing type & Crystal structure type & space group & Material example \\ \midrule
FCCTC & $\mathrm{AlCu}_2 \mathrm{Mn}$ & $Fm\bar{3}m$ & $\mathrm{ZrNi}_2 \mathrm{In}$, $\mathrm{LiAg}_2 \mathrm{In}$, $\mathrm{MnCo}_2 \mathrm{Si}$ \\
(4-4-2) & $\mathrm{Cu}_2 \mathrm{GaSr}_2$ & $R \bar{3}m$ & $\mathrm{Cu}_2 \mathrm{GaSr}_2$ \\
(4-4-2) & ternary $\mathrm{Cu}_3 \mathrm{Ti}$ & $Pmmn$ & $\mathrm{Li}_2 \mathrm{BaSi}$, $\mathrm{Ni}_2 \mathrm{BaGe}$, $\mathrm{Ni}_2 \mathrm{TiCu}$ \\
(2-2-1) & $\mathrm{ThCr}_2 \mathrm{Si}_2$ & $I4/mmm$ & $\mathrm{EuFe}_2 \mathrm{As}_2$, $\mathrm{CeOs}_2 \mathrm{Si}_2$, $\mathrm{CaRh}_2 \mathrm{P}_2$ \\
(6-2-2) & $\mathrm{BaNiO}_3$ & $P6_3/mmc$ & $\mathrm{CsMgO}_3$, $\mathrm{LaCrGe}_3$, $\mathrm{NdVGe}_3$ \\
(6-4-4) & $\mathrm{Ca}_2 \mathrm{Ni}_3 \mathrm{Ge}_2$ & $Pmma$ & $\mathrm{Ca}_2 \mathrm{Ni}_3 \mathrm{Ge}_2$ \\
\end{tabular}
\label{table:Torquato-densest-packing-and-crystal}
\end{ruledtabular}
\end{table*}

As discussed in Ref.~\cite{PhysRevE.103.023307}, atoms in some kinds of materials such as ion-bonded materials, intermetallic compounds, and materials under high pressure, can be approximated as spheres. In fact, the previous study~\cite{PhysRevE.103.023307} shows that a considerable number of crystals can be understood as DBSPs. For example, the crystal structure of $\mathrm{LaH_{10}}$, which is a superhydride material synthesized under high pressure~\cite{doi:10.1002/anie.201709970}, can be understood as the $\mathrm{XY_{10}}$ structure~\cite{PhysRevE.85.021130,PhysRevE.103.023307}, and the crystal structure of $\mathrm{UB}_4$ corresponds to the (16-4) structure~\cite{PhysRevE.103.023307}.

In this study, we investigated the correspondence of the DTSPs with the crystal structures with respect to the space group. The Spglib~\cite{togo2018textttspglib} is used for determining the space groups of the DTSPs. Note that the structural distortions in DTSPs are corrected by the Spglib. However, we have found only one correspondence between DTSP and crystal structure registered in ICSD \cite{ICSD}: FCCTC which 
corresponds to the $\mathrm{AlCu}_2 \mathrm{Mn}$ structure. Although the correspondence of the DTSPs with the crystal structures seems 
to be exceptional, we discuss possible correspondences below, which might be realized in future experiments.

In recent years, a considerable number of super hydrides under high pressure such as $\mathrm{LaH_{10}}$ are predicted by crystal structure prediction algorithms such as the evolutionary algorithm~\cite{GLASS2006713, doi:10.1063/1.2210932, Oganov_2008, LYAKHOV20101623, LYAKHOV20131172} and the particle-swarm optimization method~\cite{PhysRevB.82.094116, WANG20122063, Wang_2015, WANG2016406}. Those methods search the stable crystal structures by zooming in the promising regions of the energy/enthalpy surface, for example, it is known that the symmetry constraints enhance the efficiency of finding the most stable structure~\cite{LYAKHOV20131172, PhysRevB.82.094116, WANG20122063, WANG2016406}. In addition to the rules of thumb, we can expect that the DSPs can be used as the structure prototypes in searching the materials especially under high pressure. 

We have found that some of DTSPs are highly symmetric when their structural distortions are corrected. For example, the (13-2-1) structure has the $Fm\bar{3}m$ symmetry when the slight distortion is corrected. Large spheres constitute the FCC structure without contact, and medium spheres are placed in the tetrahedral sites, as shown in Fig.~\ref{fig:13-2-1}. The medium and large spheres are surrounded by small spheres. Besides, one small sphere is placed on the octahedral sites, and the small spheres are also surrounded by other small spheres. Small spheres surrounding small, medium, and large spheres constitute the semi-regular polyhedrons. In the $\mathrm{LaH_{10}}$, one $\mathrm{La}$ atom is surrounded by 32 $\mathrm{H}$ atoms~\cite{doi:10.1002/anie.201709970}, and besides, $\mathrm{LaH}_{10}$ has superconductivity above 260 K under high pressure~\cite{PhysRevLett.122.027001}. This similarity indicates that the (13-2-1) structure may be realized by some superhydrides under high pressure that have relatively high $T_c$ superconductivity. In our result, the (13-2-1) structure is the DTSP at the radius ratios of $0.45:0.60:1.00$ and $0.45:0.65:1.00$, and the radius ratios indicate that many kinds of combinations of atoms can be candidate for such superhydride materials. Perhaps, crystals prefer the (12-2-1) structure in which small spheres in octahedral sites are excluded from the (13-2-1) structure. The (12-2-1) structure also has the $Fm\bar{3}m$ symmetry. Otherwise, crystals prefer a quaternary structure in which small spheres in octahedral sites are replaced by other spheres.

In addition, the (4-2-1) and (4-3-1) structures have the $P6/mmm$ symmetry when the slight distortion is corrected. Similarly, the (4-4-2) structure has the $Cmcm$ symmetry, and the (9-6-3) structure has the $R3m$ symmetry when their slight distortion is corrected. Besides, the (13-3-1) structure has the $Pm\bar{3}m$ symmetry when the slight distortion is corrected. These five DTSPs are highly symmetric, so we expect that they might be realized by some materials under high pressure. 

Finally, as discussed in the next subsection, some of semi-DTSPs (SDTSPs) can be associated with the crystal structures as summarized in Table~\ref{table:Torquato-densest-packing-and-crystal}. 
The (4-4-2) structures shown in Figs.~\ref{fig:SDTSPs-4}(c) and \ref{fig:SDTSPs-4}(d) corresponds to the $\mathrm{Cu}_2 \mathrm{GaSr}$ and the ternary $\mathrm{Cu}_3 \mathrm{Ti}$, respectively. 
Note that the ternary $\mathrm{Cu}_3 \mathrm{Ti}$ structure is a crystal in which one of three $\mathrm{Cu}$ atoms is replaced by another element, and an example of the structure is $\mathrm{Li}_2 \mathrm{BaSi}$.
Additionally, the (2-2-1) structure shown in Fig.~\ref{fig:SDTSPs-4}(e), the (6-4-4) structure shown in Fig.~\ref{fig:SDTSPs-4}(f), 
and the (6-2-2) structure shown in Fig.~\ref{fig:SDTSPs-2}(b) correspond to
the $\mathrm{ThCr}_2 \mathrm{Si}_2$,
$\mathrm{Ca}_2 \mathrm{Ni}_3 \mathrm{Ge}_2$,
and 
$\mathrm{BaNiO}_3$ structures, respectively.
The correspondences also indicate that the (S)DTSPs can be used as structural prototypes for crystals.

\subsection{Semi-DTSPs}
\label{associate_dtsps}

\begin{figure}
\centering
\includegraphics[width=\columnwidth]{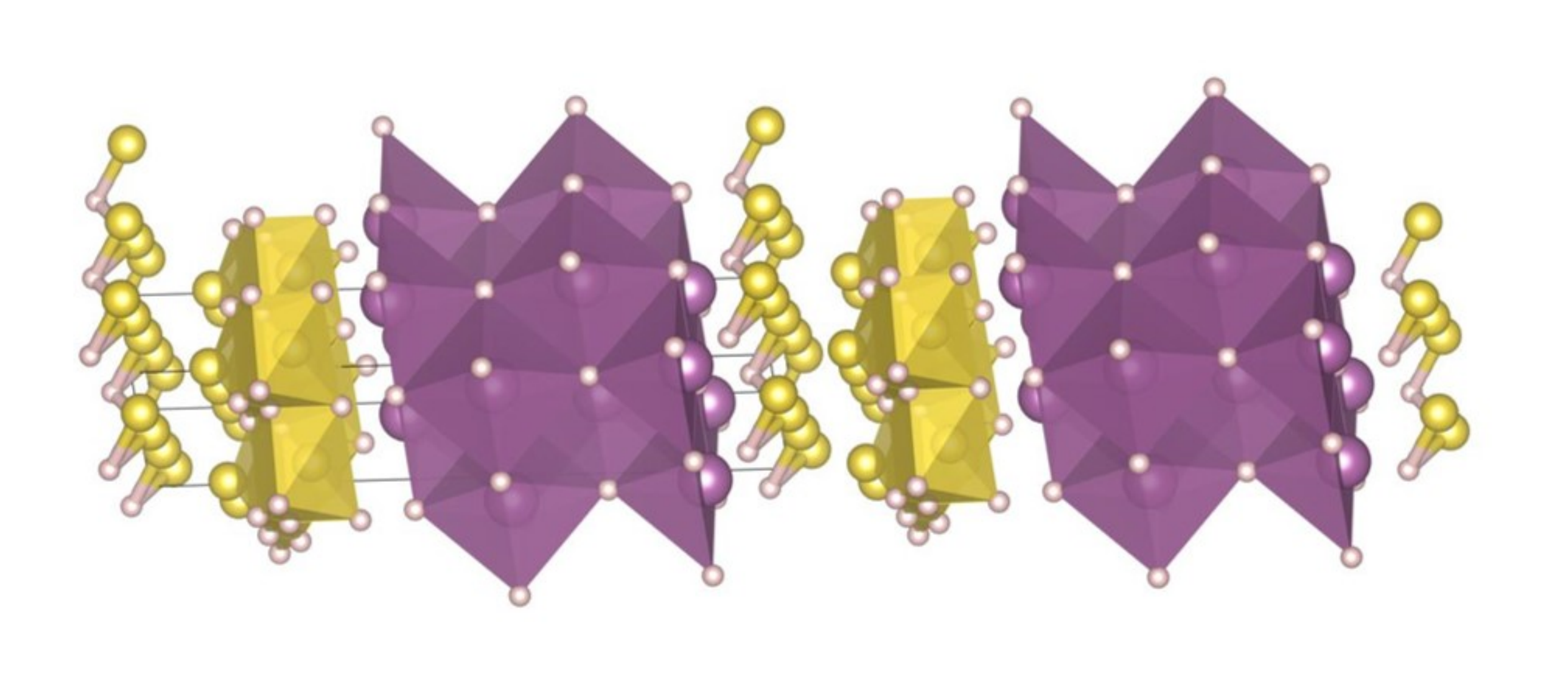}
\caption{The (10-6-4) structure that is the SDTSP at the radius ratio of $0.40:0.70:1.00$. The SDTSPs have long unit cells, in which a few kinds of local structures are combined separately. The separations per cell are the same kind of that of the monophase DBSPs that are discussed in Ref.~\cite{PhysRevE.103.023307}. It is unlikely for such disordered structures to appear on the phase diagrams as the DTSPs.}
\label{fig:SDTSPs-bad}
\end{figure}
\begin{figure}
\centering
\includegraphics[width=\columnwidth]{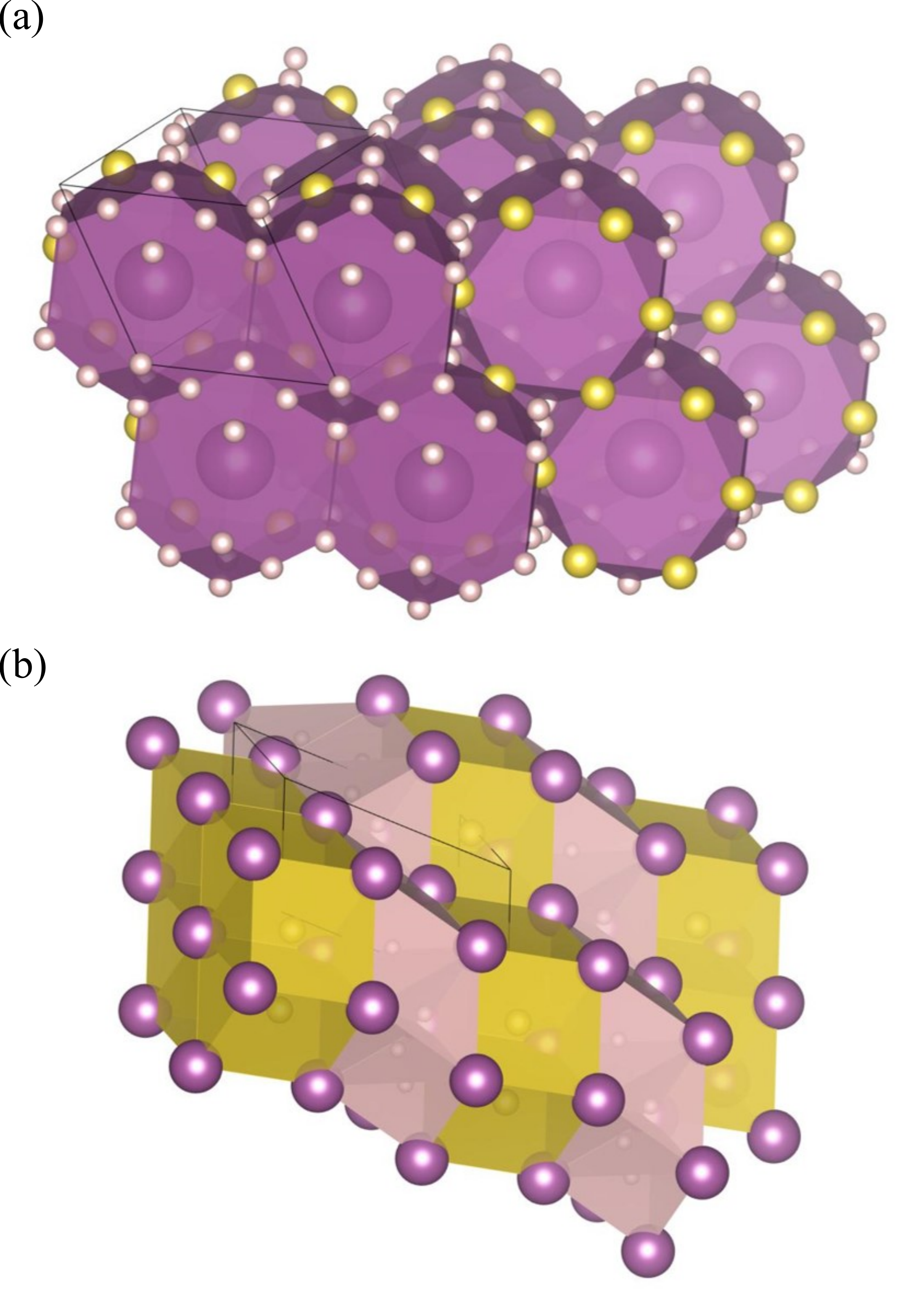}
\caption{(a) The (10-3-2) structure that is the SDTSP at the radius ratios of $0.30:0.40:1.00$, $0.30:0.45:1.00$, and $0.30:0.50:1.00$. (b) The (6-2-2) structure that is the SDTSP at the radius ratios of $0.30:0.50:1.00$ and $0.30:0.55:1.00$.}
\label{fig:SDTSPs-1}
\end{figure}
\begin{figure}
\centering
\includegraphics[width=\columnwidth]{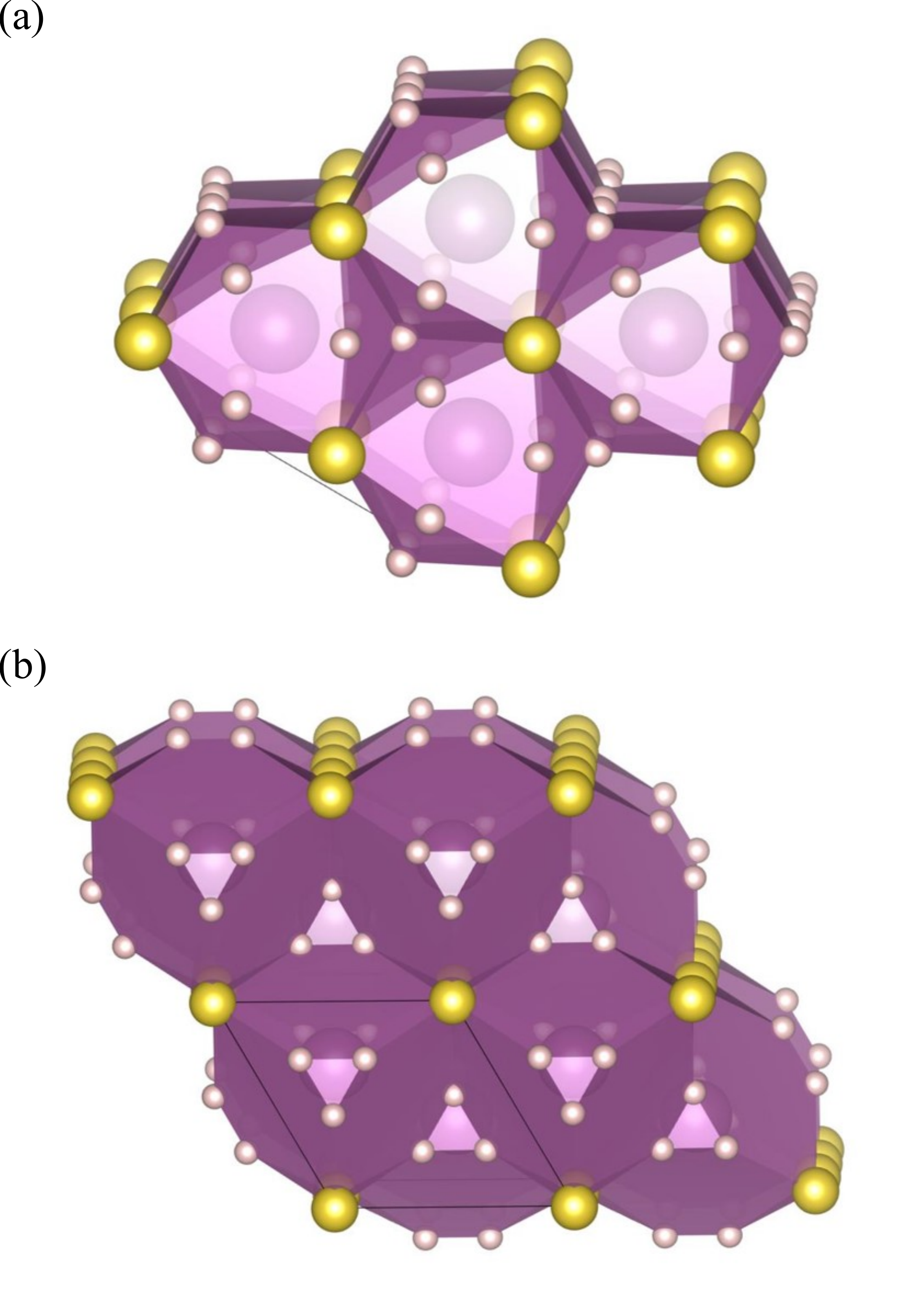}
\caption{The SDTSPs at the radius ratio of $0.35:0.65:1.00$. (a) The (5-1-1) structure. The structure has the $P \bar{6}m2$ symmetry when the slight distortion is corrected. (b) The (6-2-2) structure. The structure corresponds to the $\mathrm{BaSiO}_3$ structure. The $\mathrm{BaSiO}_3$ structure has the $P6_{3}/mmc$ symmetry.}
\label{fig:SDTSPs-2}
\end{figure}
\begin{figure}
\centering
\includegraphics[width=\columnwidth]{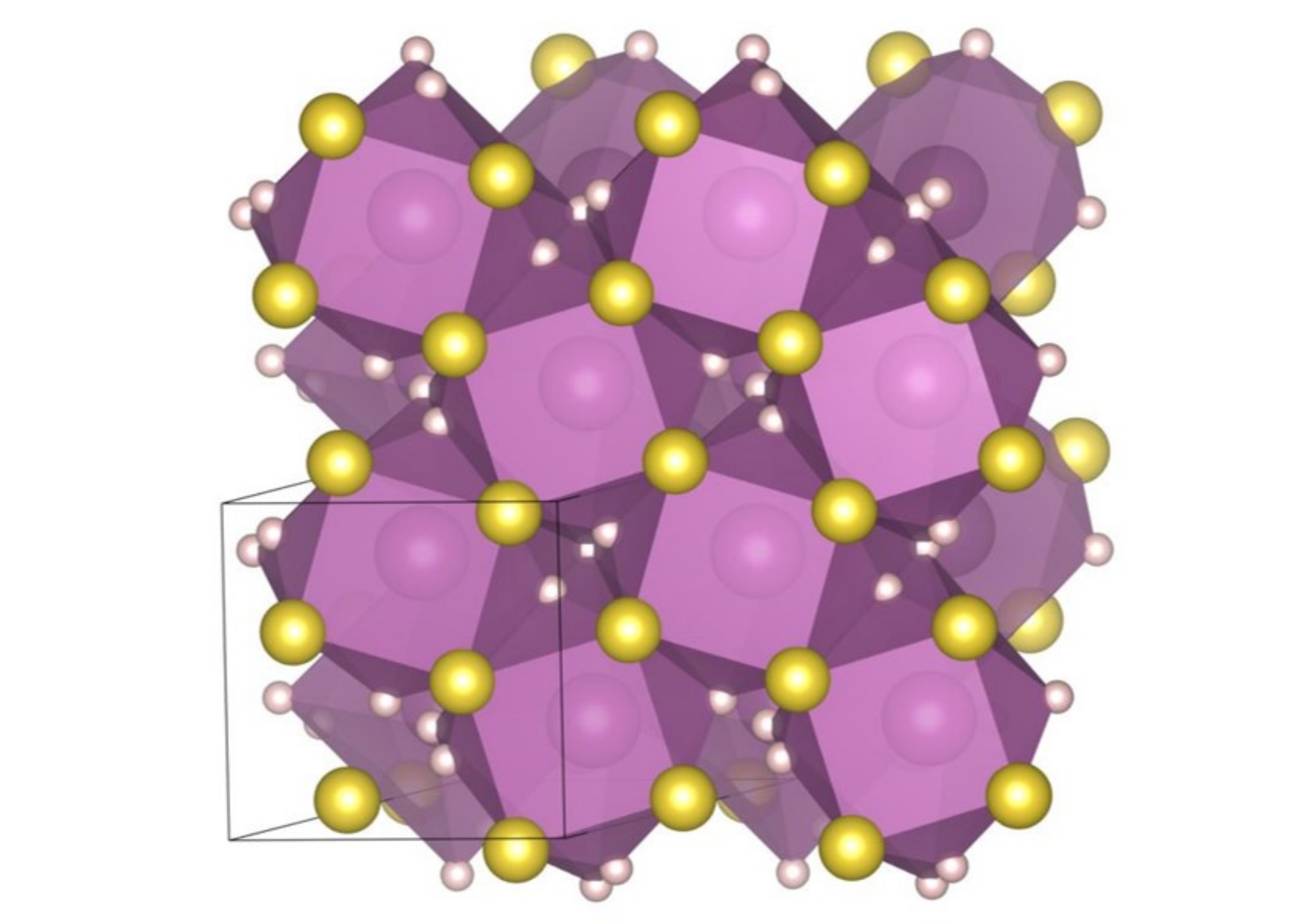}
\caption{The (12-4-4) structure that is the SDTSP at the radius ratio of $0.35:0.70:1.00$. The structure on a plane constituted by medium spheres is almost the same as that by large spheres in the (16-4) structure~\cite{PhysRevE.103.023307} that is one of the DBSPs.}
\label{fig:SDTSPs-3}
\end{figure}
\begin{figure*}
\centering
\includegraphics[width=2\columnwidth]{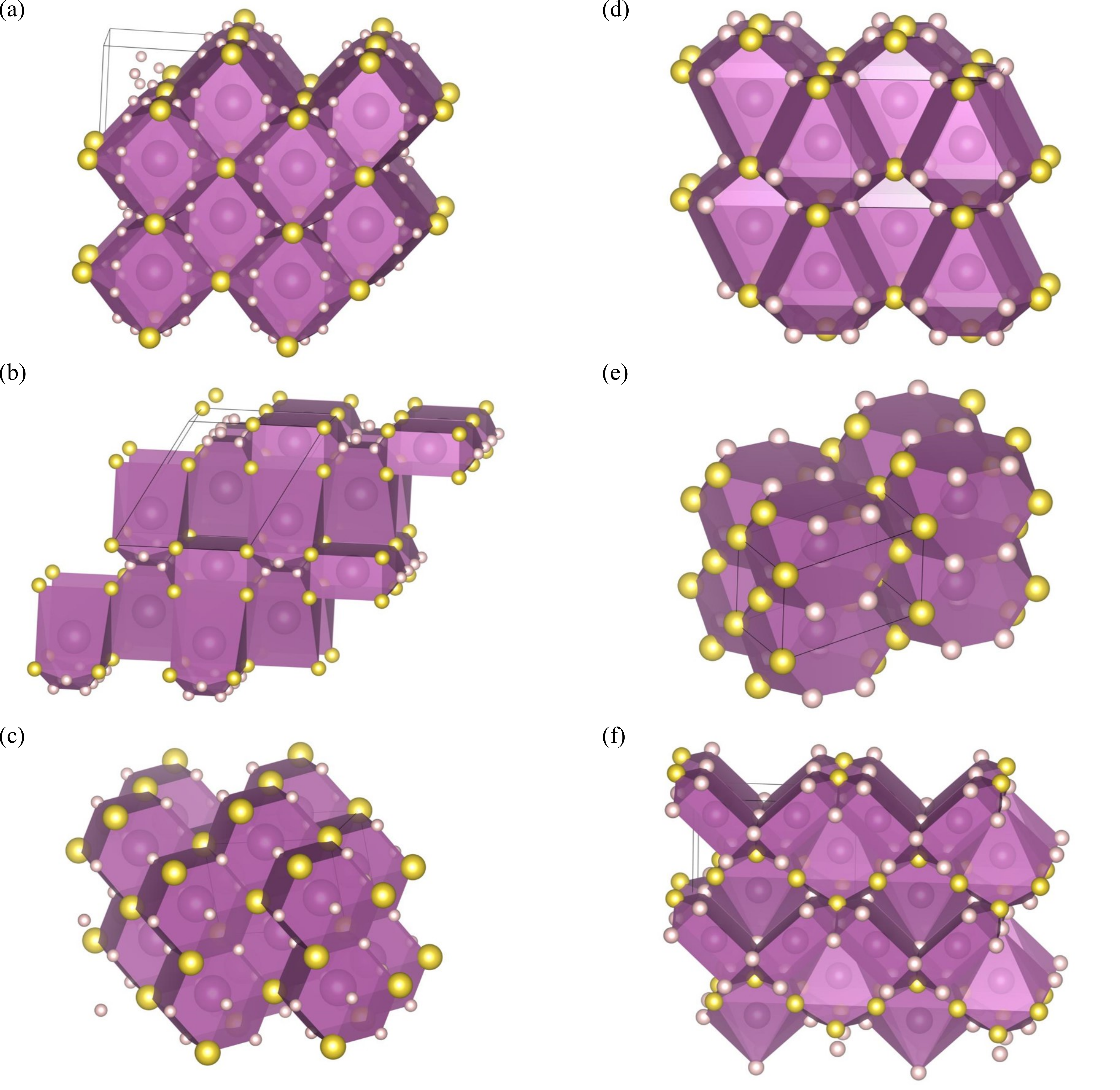}
\caption{The SDTSPs that have high symmetries. (a) The (12-2-2) structure that is the SDTSP at the radius ratio of $0.30:0.60:1.00$. The structure has the $Pmmn$ symmetry when the slight distortion is corrected. (b) The (5-4-3) structure that is the SDTSP at the radius ratio of $0.35:0.45:1.00$. The structure has the $Imm2$ symmetry when the slight distortion is corrected. (c) The (4-2-2) structure that is the SDTSP at the radius ratio of $0.45:0.65:1.00$. The structure has the $R \bar{3}m$ symmetry when the slight distortion is corrected, and it corresponds to the the crystal structure of the $\mathrm{Cu}_2 \mathrm{GaSr}$. (d) The (4-2-2) structure that is the SDTSP at the radius ratio of $0.50:0.65:1.00$. The structure has the $Pmmn$ symmetry when the slight distortion is corrected, and it corresponds to the ternary $\mathrm{Cu}_3 \mathrm{Ti}$ structure that is one of ternary crystal structures: one of crystals which consist of the $\mathrm{Cu}_3 \mathrm{Ti}$ structure is $\mathrm{Li}_2 \mathrm{BaSi}$. (e) The (2-2-1) structure that is the SDTSP at the radius ratio of $0.55:0.70:1.00$. The structure has the $I4/mmm$ symmetry when the slight distortion is corrected, and it can be associated with the $\mathrm{ThCr}_2 \mathrm{Si}_2$ structure. (f) The (6-4-4) structure that is the SDTSP at the radius ratio of $0.60:0.70:1.00$. The structure has the $Pmma$ symmetry when the slight distortion is corrected, and it corresponds to the the crystal structure of the $\mathrm{Ca}_2 \mathrm{Ni}_3 \mathrm{Ge}_2$.}
\label{fig:SDTSPs-4}
\end{figure*}
\begin{figure*}
\centering
\includegraphics[width=2\columnwidth]{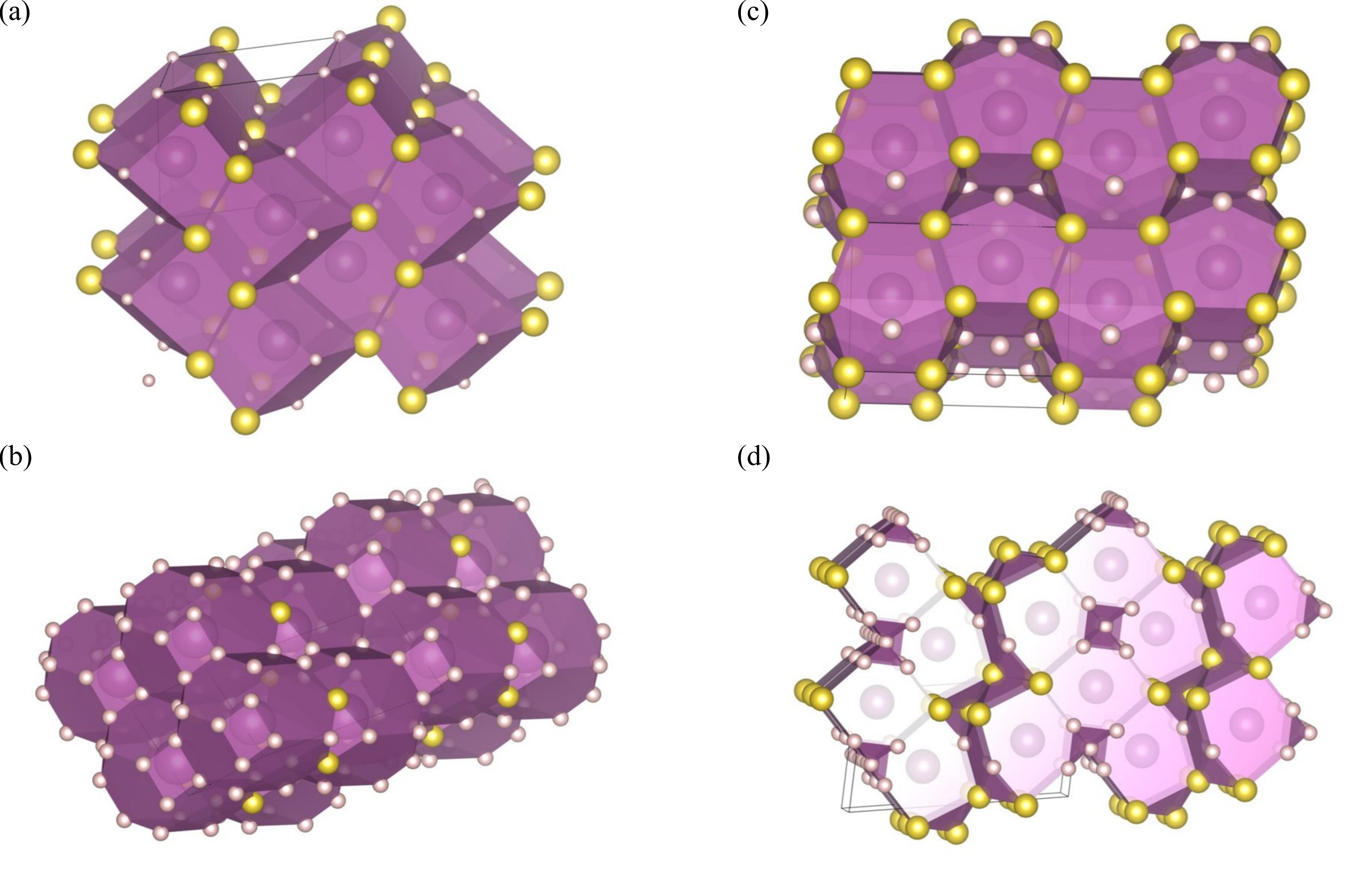}
\caption{(a) The (4-2-2) structure that is the SDTSP at the radius ratio of $0.30:0.70:1.00$. (b) The (12-2-2) structure that is the SDTSP at the radius ratio of $0.35:0.50:1.00$. (c) The (6-4-2) structure that is the SDTSP at the radius ratio of $0.45:0.70:1.00$. (d) The (8-4-3) structure that is the SDTSP at the radius ratio of $0.45:0.70:1.00$.}
\label{fig:SDTSP-other}
\end{figure*}

So far, we focus on the DTSPs, which are denser than any phase separations including DBSPs. In this subsection, we discuss the semi-DTSPs (SDTSPs) that are denser than any phase separation consisting of only the FCC structures and/or the other ternary packings. In other words, to discover the interesting dense ternary packings, we exclude the DBSPs when we calculate the densest phase separations. 
As a result, we have discovered 65 SDTSPs in total, and found that some of them are highly well-ordered.
There is a possibility that the SDTSPs can be DTSPs at suitable radius ratios which have not been investigated in this study.
Actually, for the binary system, some DBSPs appear in a narrow range of the binary phase diagram, e.g. the (9-4) structure appears in a very narrow region, as shown in Ref.~\cite{PhysRevE.103.023307}. Therefore, it is reasonable to imagine that some DTSPs also appear in a small range of radius ratio. In our result, the number of the SDTSPs at each radius ratio is generally less than five, and some of the SDTSPs have long unit cells, in which a few kinds of local structures are combined separately as shown in Fig.~\ref{fig:SDTSPs-bad}. 
The separation of local structures is similar to the feature found in the monophase DBSPs that are discussed in Ref.~\cite{PhysRevE.103.023307}. However, it is unlikely for such disordered structures to appear on the phase diagrams as the DTSPs. On the other hand, some SDTSPs are well-ordered. We expect that they might be DTSPs at suitable radius ratios, and realized by materials especially under pressure, as discussed in Sec.~\ref{dtsps_and_crystal_structures}. In this subsection, we show some of interesting SDTSPs: the well-ordered SDTSPs.

First, we introduce the (10-3-2) and (6-2-2) structures. The (10-3-2) structure is the SDTSP at the radius ratios of $0.30:0.40:1.00$, $0.30:0.45:1.00$, and $0.30:0.50:1.00$, and the (6-2-2) structure is the SDTSP at the radius ratios of $0.30:0.50:1.00$ and $0.30:0.55:1.00$. The two structures shown in Fig.~\ref{fig:SDTSPs-1} are the SDTSPs in a broad range of radius ratio, compared with the other SDTSPs. Besides, their local structures are similar to that of the DTSPs. In the (10-3-2) structure, medium spheres constitute a kagome lattice as in the (4-3-1) structure. However, unlike in the (4-3-1) structure, one of the hexagonal side of the hexagonal prisms consisting of medium spheres are replaced by six small spheres, and accordingly, the kagome lattice comprised by medium spheres are layered with sliding. On the other hand, in the (6-2-2) structure, medium spheres are placed in the triangular prisms constituted by large spheres, and small spheres are placed in the square pyramid constituted by large spheres. The feature is the same as that of the (2-2-2) structure that is the DTSP at the radius ratio of $0.30:0.50:1.00$. Considering in a comprehensive way, it is highly probable for those two SDTSPs to be the DTSPs at suitable radius ratios 
which have not been investigated in this study.

Second, we discuss the two SDTSPs at the radius ratio of $0.35:0.65:1.00$: the (5-1-1) and (6-2-2) structures. These two structures are shown in Fig.~\ref{fig:SDTSPs-2}. In the (5-1-1) structure, large spheres are placed in triangular prisms comprised by medium spheres. In addition, there is one small spheres near the center of each edge of the triangles comprised by medium spheres, and there are two small spheres in a small triangular prism comprised by the small spheres placed on the triangles comprised by medium spheres. The (5-1-1) structure has the $P \bar{6}m2$ symmetry when the slight distortion is corrected. On the other hand, the (6-2-2) structure can be associated with the $\mathrm{BaSiO}_3$ structure that is a crystal structure having the $P6_{3}/mmc$ symmetry. In the (6-2-2) structure, large spheres constitute the HCP structure without contacts, and a large sphere is surrounded by six medium and 12 small spheres. 
It is also likely for these highly well-ordered SDTSPs to be the DTSPs at suitable radius ratios 
which have not been investigated in this study.

Third, we describe the (12-4-4) structure that is the SDTSP at the radius ratio of $0.35:0.70:1.00$. The structure on a plane constituted by medium spheres is almost the same as that by large spheres in the (16-4) structure~\cite{PhysRevE.103.023307}. Due to the existence of large spheres in the (12-4-4) structure, the unit cell is longer than that of the (16-4) structure. As well as the well-ordered SDTSPs such as the (6-2-2) structure, the (12-4-4) structure may also be DTSPs at some radius ratios, and furthermore the geometric feature indicates that unknown DTSPs, which can be derived from the DBSPs comprised by large unit cells such as the (16-4) structure, may be discovered if the maximum size of unit cells is set to be larger than those used in this study.

Besides, we have discovered some SDTSPs that have high symmetries. The (12-2-2) structure shown in Fig.~\ref{fig:SDTSPs-4}(a) is the SDTSP
at the radius ratio of $0.30:0.60:1.00$. The structure has the $Pmmn$ symmetry when the the slight distortion is corrected. 
The (5-4-3) structure shown in Fig.~\ref{fig:SDTSPs-4}(b) is the SDTSP at the radius ratio of $0.35:0.45:1.00$. 
The structure has the $Imm2$ symmetry when the slight distortion is corrected. 
The (4-2-2) structure shown in Fig.~\ref{fig:SDTSPs-4}(c) is the SDTSP at the radius ratio of $0.45:0.65:1.00$. 
The structure has the $R \bar{3}m$ symmetry when the slight distortion is corrected, and corresponds to the crystal structure 
of $\mathrm{Cu}_2 \mathrm{GaSr}$. 
The (4-2-2) structure shown in Fig.~\ref{fig:SDTSPs-4}(d) is the SDTSP at the radius ratio of $0.50:0.65:1.00$. 
The structure has the $Pmmn$ symmetry when the slight distortion is corrected, and corresponds to 
the ternary $\mathrm{Cu}_3 \mathrm{Ti}$ structure. 
The (2-2-1) structure shown in Fig.~\ref{fig:SDTSPs-4}(e) is the SDTSP at the radius ratio of $0.55:0.70:1.00$ . 
The structure has the $I4/mmm$ symmetry when the slight distortion is corrected, 
and can be associated with the $\mathrm{ThCr}_2 \mathrm{Si}_2$ structure. 
The (6-4-4) structure shown in Fig.~\ref{fig:SDTSPs-4}(f) is the SDTSP at the radius ratio of $0.60:0.70:1.00$. 
The structure has the $Pmma$ symmetry when the slight distortion is corrected, and corresponds to the the crystal structure 
of $\mathrm{Ca}_2 \mathrm{Ni}_3 \mathrm{Ge}_2$. 
These six structures are highly symmetric, so it is highly probable for them to be the DTSPs at suitable radius ratios
which have not been investigated in this study. 
Since the four SDTSPs among the six SDTSPs with high symmetries discussed above can be associated with real crystal structures 
as summarized in Table~\ref{table:Torquato-densest-packing-and-crystal}, 
it is natural to consider that discovering DTSPs and SDTSPs with high symmetries may lead to discovering of novel crystal structures. 

Finally, we show some examples of the other SDTSPs which have relatively unique geometric features. The (4-2-2) structure shown in Fig.~\ref{fig:SDTSP-other}(a) is the SDTSP at the radius ratio of $0.30:0.70:1.00$. The (6-4-2) structure shown in Fig.~\ref{fig:SDTSP-other}(c) is the SDTSP at the radius ratio of $0.45:0.70:1.00$. The (8-4-3) structure shown in Fig.~\ref{fig:SDTSP-other}(d) is the SDTSP at the radius ratio of $0.45:0.70:1.00$. The three structures are examples of structures in which large spheres are surrounded by small and medium spheres and the surrounding structure is characteristic from symmetry's point of view. On the other hand, the (12-2-2) structure shown in Fig.~\ref{fig:SDTSP-other}(b) is the SDTSP at the radius ratio of $0.35:0.50:1.00$. This structure is an example of the SDTSPs that can be regarded as clathrate structures. The four structures shown in Fig.~\ref{fig:SDTSP-other} are not highly symmetric, but their structures are characteristic and might be the DTSPs at suitable radius ratios which have not been investigated in this study.  

In summary, we have discovered a considerable number of the SDTSPs and some of them are highly well-ordered. Some structures have high symmetries and appear in a broad range of radius ratios, and some of other structures consist of unique local structures. Besides, structures with high symmetries can be associated with crystal structures. The well-ordered SDTSPs identified in this study may be the DTSPs at suitable radius ratios. These characteristic structures seem to be constituted by relatively large three spheres such as the radius ratio of $0.45:0.65:1.00$. 

\subsection{Public release}

Three-dimensional data of the DTSPs and the \textbf{SAMLAI} package, in which our methods are implemented, are available
online~\cite{samlai}. The distribution of the program package and the source codes follow the practice of the GNU General Public
License version 3 (GPLv3).

\subsection{Effectiveness of our exploration method}

The structural diversity found in the discovered DTSPs and SDTSPs suggests that our piling-up method can find a wide variety of DTSPs. 
Besides, as discussed in Ref.~\cite{PhysRevE.103.023307}, the iterative balance method can find the densest packing fraction 
with optimal distortions. The computational accuracy of packing fraction is important to discover the putative DTSPs, 
because the densest phase separations generally tend to have competitive phase separations, and 
as discussed in Sec.~\ref{sec:geometric_features_of_dtsps_with_much_smaller_spheres}, at some radius ratios, 
several DTSPs have competitive to each other with respect to the packing fraction. 
As discussed in Sec.~\ref{sec:improved_piling-up_method}, the improved piling-up method directly generates multilayered structures. 
In many cases, initial structures generated by the improved piling-up method are close to packing structures and 
the structural optimization does not significantly transform the initial structures. 
Since we try to unbiasedly distribute the initial structures in the configuration space for a given composition and radius ratio 
as much as possible, it is expected that the exhaustive exploration can reach the DSP structure if the number of the trials is 
large enough.
In fact, we confirmed that the improved method can easily find DTSPs compared to the previous method discussed 
in Ref.~\cite{PhysRevE.103.023307}.
The (6-6)-type structure is one of the most difficult structures to be found by the previous method, in which the relative 
angle of two lattice vectors is about $60^{\circ}$. The fact implies that the previous method does not properly explore
structures having two lattice vectors with the relative angle of $60^{\circ}$. On the other hand, we see that 
the improved method can easily find the (6-6)-type structures compared to the previous method.

Although our method successfully discovered diverse DTSPs, we report a failure in finding (14-6-6) structure, which is one of DTSPs 
at the radius ratio of $0.25:0.45:1.00$. 
The (14-6-6) structure can be derived from the (6-6) structure as discussed in Sec.~\ref{sec:phase_diagrams_and_discovered_dtsps}, 
where the 12 expanded tetrahedral sites are occupied by one small sphere, respectively, and two large boundaries, which are 
two of six boundary surfaces between two octahedrons constituted by large spheres, are occupied by one small sphere, respectively. 
As the first trial, we found the (8-6-6) structure, which can be derived from the (6-6) structure, was found by the exhaustive search 
for DTSPs, where some of the tetrahedral sites and boundary surfaces are empty. 
Since there are still vacant tetrahedral sites and vacant boundary surfaces, which can be occupied by small spheres,  
in the (8-6-6) structure, 
we decided to explore the DTSPs at additional compositions, and successfully found the (12-6-6) structure by our random search method, 
where some of the tetrahedral sites and boundary surfaces are still empty. 
So, we estimated that the (14-6-6) structure, in which the 12 expanded tetrahedral sites are occupied by one small sphere, respectively, 
and two large boundaries are occupied by one small sphere, respectively, should have a larger packing fraction than 
that of the (12-6-6) structure, and performed the exhaustive search for the composition. 
However, our calculation could not discover the (14-6-6) structure, although the number of generated structures was 50,000,000. 
Finally, we manually prepared a seed structure of the (14-6-6) structure, and found after the optimization
that the (14-6-6) structure is a DTSP. 
The case of the (14-6-6) structure suggests that there seem to exist structures which cannot be easily found
even by the improved method with the large number of trials. 
We consider that there might be room for further improvement of the random search method by analyzing such a special case, 
while the easiness and difficulty in finding a dense packing structure must be related to the structure of 
the configurational hyperspace.

\section{Conclusion}
\label{sec:conclusion}

In the present research, we have exhaustively explored the densest ternary sphere packings (DTSPs) under periodic boundary conditions at 45 kinds of radius ratios and 237 kinds of compositions. 
Accordingly, we have successfully constructed the ternary phase diagrams with discovering of 37 putative DTSPs.

To explore the DTSPs, we have improved the piling-up method~\cite{PhysRevE.103.023307} to generate initial structures and the structural optimization scheme based on the iterative balance method~\cite{PhysRevE.103.023307}. The improved piling-up method  directly generates multilayered initial structures which are close to packing structures. Since the structural optimization does not significantly transform the initial structures unless there are large overlaps between spheres in generated layers, we can assume that if the number of generated structures are large enough, we can be sure that we successfully explore the DSPs exhaustively in the configuration space. The improved optimization scheme properly rejects sparse structures, and accordingly computational cost for exploration is reduced. The methods for exploring DSPs are implemented in our open-source program package \textbf{SAMLAI} (Structure search Alchemy for MateriaL Artificial Invention), and the package is available online~\cite{samlai}.

Since any composition ratio can be achieved by the combination of the densest FCC structures and/or the DBSPs, it is not apparent whether there is at least one DTSP in the densest phase separation at any radius and composition ratio. In fact, the discovered putative DTSPs have particularly high packing fractions. When the small spheres have much smaller radius ratio than large spheres such as $0.20$ or $0.25$, there are two kinds of DTSPs. One can be derived from the DBSPs, and the other can be regarded as clathrate structures. Since very small spheres can be placed in small voids constituted by medium and/or large spheres, in some cases with expanding the structural frameworks, small voids in DBSPs can be occupied by small spheres. Besides, very small radius ratio does not disturb for medium and large spheres to constitute a network structure, so some of the DTSPs with very small spheres prefer clathrate structures so as to increase packing fractions.

On the other hands, as the radius of small spheres is getting large, several kinds of unique DTSPs appear on the phase diagrams. Some of the DTSPs contain cluster structures comprised by small and medium spheres and the geometries are ordered well. Besides, we have discovered a highly well-ordered clathrate structure named the (13-2-1) structure, which consists of three kinds of relatively large spheres. If the slight distortion is corrected, the (13-2-1) structure has the $Fm\bar{3}m$ symmetry with large spheres constituting the FCC structure without contact despite the large radii of three kinds of spheres such that the radius ratio is $0.45:0.65:1.00$. Surprisingly, a medium sphere is placed in an expanded tetrahedral site. Medium and large spheres are surrounded by small spheres, and the surrounding structures are the semi-regular polyhedrons: the truncated tetrahedron and the truncated octahedron, respectively. Besides, one small sphere is placed on an expanded octahedral site, and it is also surrounded by other small spheres. The surrounding structure is also the semi-regular polyhedron: the cuboctahedron. The discovery of these two kinds of DTSPs implies that unknown and well-ordered DTSPs might be discovered if the maximum number of spheres per unit cells is set to be larger than those used in this study.

In recent years, a considerable number of super hydrides under high pressure such as $\mathrm{LaH_{10}}$~\cite{doi:10.1002/anie.201709970} are synthesized. $\mathrm{LaH}_{10}$ has superconductivity above 260 K under high pressure~\cite{PhysRevLett.122.027001}. In the $\mathrm{LaH_{10}}$~\cite{doi:10.1002/anie.201709970}, one $\mathrm{La}$ atom is surrounded by 32 $\mathrm{H}$ atoms. The similarity indicates that the (13-2-1) structure may be realized by some materials under high pressure, which may have high $T_c$ superconductivity. Besides, our result shows that the (13-2-1) structure is the DTSP at the radius ratios of $0.45:0.60:1.00$ and $0.45:0.65:1.00$, and the radius ratios indicate that many combinations of atoms can be chosen to realize such super hydride materials.

Not only the (13-2-1) structure but also the other DTSPs may be realized by some materials especially under high pressure. Some of the DTSPs such as the (4-2-1), (4-3-1), (4-4-2), (9-6-3), and (13-3-1) structures have high symmetries. Besides, some of DTSPs can be regarded as clathrate structures, and they may also be realized by super hydride materials.

Furthermore, we have discussed the semi-DTSPs (SDTSPs) that are denser than any phase separation consisting of only the FCC structures and/or the other ternary packings: It means that we exclude the DBSPs when we identified the densest phase separations.
As a result, we have discovered 65 SDTSPs, and found that some of SDTSPs are highly well-ordered, characterized by 
a space group of high symmetry. 
The well-ordered SDTSPs identified in this study may be the DTSPs at suitable radius ratios
which have not been investigated in this study. 
These unique structures tend to be constituted by relatively large three spheres such that the radius ratio is $0.45:0.65:1.00$. 
The trend indicates that more DTSPs, consisting of unknown local structures, may be discovered if an exhaustive search for the DTSPs 
are executed at a finer interval for the radius ratio in promising regions such that radii of small and medium spheres are relatively large.

Finally, we have found that a considerable number of SDTSPs with high symmetries can be associated with crystals. 
We expect that discovering not only DTSPs and but also SDTSPs with high symmetries may lead to discovering of novel 
crystal structures. 

\begin{acknowledgments}
We would like to thank Dr. Mitsuaki Kawamura and Dr. Masahiro Fukuda for kindly supporting computational issues.
R. K. is financially supported by the Quantum Science and Technology Fellowship Program (Q-STEP), the University Fellowship 
program for Science and Technology Innovations.
The part of calculations was performed by the supercomputer, Ohtaka, in ISSP, the Univ. of Tokyo.  
\end{acknowledgments}

\bibliographystyle{apsrev4-2}
\bibliography{submit}

\end{document}